\documentclass[aps,prd,showkeys,superscriptaddress,singlecolumn,nofootinbib,floatfix]{revtex4-1}

\pdfoutput=1

\usepackage{amsmath}
\usepackage{amssymb}
\usepackage{amsthm}
\usepackage{mathrsfs}
\usepackage{graphicx}
\usepackage{epstopdf}
\usepackage{fancyhdr}
\usepackage{array}
\usepackage[all]{xy}
\usepackage{eufrak}
\usepackage{euscript}
\usepackage{enumerate}
\usepackage{slashed}
\usepackage{hyperref}
\usepackage{subfigure} 
\usepackage{epstopdf} 

\hypersetup{pdftex,colorlinks=true,linkcolor=blue,citecolor=blue,menucolor=black,urlcolor=blue,filecolor=blue}

\begin{document}

\title{Non-hydrodynamic quasinormal modes and equilibration of a baryon dense holographic QGP with a critical point}

\author{Romulo Rougemont}
\email{rrougemont@iip.ufrn.br}
\affiliation{International Institute of Physics, Federal University of Rio Grande do Norte,
Campus Universit\'{a}rio - Lagoa Nova, CEP 59078-970, Natal, Rio Grande do Norte, Brazil}

\author{Renato Critelli}
\email{renato.critelli@usp.br}
\affiliation{Instituto de F\'{i}sica, Universidade de S\~{a}o Paulo, Rua do Mat\~{a}o, 1371, Butant\~{a}, CEP 05508-090, S\~{a}o Paulo, S\~{a}o Paulo, Brazil}

\author{Jorge Noronha}
\email{noronha@if.usp.br}
\affiliation{Instituto de F\'{i}sica, Universidade de S\~{a}o Paulo, Rua do Mat\~{a}o, 1371, Butant\~{a}, CEP 05508-090, S\~{a}o Paulo, S\~{a}o Paulo, Brazil}

\begin{abstract}
We compute the homogeneous limit of non-hydrodynamic quasinormal modes (QNM's) of a phenomenologically realistic Einstein-Maxwell-Dilaton (EMD) holographic model for the Quark-Gluon Plasma (QGP) that is able to: i) {\it quantitatively} describe state-of-the-art lattice results for the QCD equation of state and higher order baryon susceptibilities with $2+1$ flavors and physical quark masses up to highest values of the baryon chemical potential currently reached in lattice simulations; ii) describe the nearly perfect fluidity of the strongly coupled QGP produced in ultrarelativistic heavy ion collisions; iii) give a very good description of the bulk viscosity extracted via some recent Bayesian analyzes of hydrodynamical descriptions of heavy ion experimental data. This EMD model has been recently used to predict the location of the QCD critical point in the QCD phase diagram, which was found to be within the reach of upcoming low energy heavy ion collisions. The lowest quasinormal modes of the $SO(3)$ rotationally invariant quintuplet, triplet, and singlet channels evaluated in the present work provide upper bounds for characteristic equilibration times describing how fast the dense medium returns to thermal equilibrium after being subjected to small disturbances. We find that the equilibration times in the different channels come closer to each other at high temperatures, although being well separated at the critical point. Moreover, in most cases, these equilibration times decrease with increasing baryon chemical potential while keeping temperature fixed.
\end{abstract}



\maketitle
\tableofcontents

\section{Introduction}
\label{sec:intro}

For more than a decade ultrarelativistic heavy ion collisions \cite{Arsene:2004fa,Adcox:2004mh,Back:2004je,Adams:2005dq,Aad:2013xma} have been used to probe many different aspects of QCD under such extreme conditions which defy a complete description from first principles or even from a single effective model or theoretical framework. In fact, the spacetime evolution of the rapidly expanding fireball produced in such collisions involves out-of-equilibrium real time phenomena in a strongly coupled non-conformal quark-gluon plasma (QGP) \cite{Gyulassy:2004zy,Heinz:2013th,Shuryak:2014zxa}, a crossover transition towards a hadron dominated phase \cite{Aoki:2006we,Borsanyi:2016ksw}, and finally the subsequent freeze-out and hadronic decay later stages.

The QGP formed in these collisions possesses a small shear viscosity to entropy density ratio ($\eta/s$) and a nontrivial profile for the bulk viscosity to entropy density ratio ($\zeta/s$) \cite{Schenke:2010rr,Song:2010mg,Gale:2012rq,Noronha-Hostler:2013gga,Noronha-Hostler:2014dqa,Ryu:2015vwa,Bernhard:2016tnd,Bernhard:2017vql,Bernhard:2018hnz,Ryu:2017qzn}, while the relationship between the local energy density and pressure of this rapidly out-of-equilibrium evolving system is well described in practice (see, e.g., \cite{Pratt:2015zsa,Monnai:2017cbv,Alba:2017hhe}) by lattice QCD results computed in equilibrium. These three key ingredients are present in the bottom-up Einstein-Maxwell-Dilaton (EMD) holographic model of Ref.\ \cite{Critelli:2017oub} which has been recently employed to provide a phenomenologically plausible prediction for the location of the long sought QCD critical end point (CEP) \cite{Stephanov:1998dy,Stephanov:1999zu,Rischke:2003mt,Stephanov:2011pb}. Indeed, the EMD model of Ref.\ \cite{Critelli:2017oub} is uniquely able to {\it simultaneously} \cite{Critelli:2017oub,Rougemont:2017tlu}:\footnote{The hydrodynamic viscosities predicted by the refined EMD model of Ref.\ \cite{Critelli:2017oub} are close to the results already published in Ref.\ \cite{Rougemont:2017tlu}, which used the previous version \cite{Rougemont:2015wca} of the model's parameters.}

\begin{enumerate}

\item {\it Predict and match at the quantitative level} state-of-the-art lattice results \cite{Bazavov:2017dus,Bellwied:2015lba} for the QCD equation of state and higher order baryon susceptibilities with $2+1$ flavors and physical quark masses at finite baryon density up to the highest values of baryon chemical potential ($\mu_B \sim 600$ MeV) currently reached in lattice simulations;

\item Describe the nearly perfect fluidity of the strongly coupled QGP encoded in the small value of $\eta/s$ \cite{Ryu:2015vwa,Bernhard:2016tnd};\footnote{This point is the only feature shared with other holographic models as it stems from the general result derived in \cite{Kovtun:2004de}.}

\item Describe the dynamical effects associated with the non-conformal nature of the QGP encoded in a nontrivial $\zeta/s$ profile which peaks in the crossover region \cite{Bernhard:2017vql,Bernhard:2018hnz}, in accordance with phenomenological expectations \cite{Karsch:2007jc,NoronhaHostler:2008ju}.
\end{enumerate}

The CEP location in the plane of temperature and baryon chemical potential was estimated to lie at $(T^{\textrm{CEP}},\mu_B^{\textrm{CEP}})\sim(89,724)$ MeV. Moreover, in Ref.\ \cite{Critelli:2017oub} a chemical freeze-out analysis was also pursued providing the following estimate for the CEP location in terms of the center of mass energy of the collisions: $\sqrt{s_{\textrm{NN}}^{\textrm{CEP}}}\sim 2.5$ --- $4.1$ GeV. This prediction will soon be put to test in upcoming low energy heavy ion collisions, such as the fixed target experiments at RHIC \cite{Meehan:2017cum} and later the 
Compressed Baryonic Matter (CBM) experiment at FAIR/GSI \cite{Ablyazimov:2017guv}.

By disturbing the equilibrated charged black hole backgrounds which are static solutions of the EMD holographic model, and solving the linearized equations of motion for the corresponding perturbations with some adequate Dirichlet boundary conditions, one obtains the spectra of quasinormal modes (QNM's) of the theory associated with the response of the medium to these disturbances \cite{Kovtun:2005ev,Berti:2009kk}. In more general situations comprising far-from-equilibrium dynamics, the QNM's describe the linear part of the decaying perturbations of the medium as it approaches thermal equilibrium. Consequently, as discussed e.g. in Ref.\ \cite{Horowitz:1999jd}, the longest-lived non-hydrodynamic QNM's\footnote{Non-hydrodynamic QNM's correspond to collective excitation modes of the system that possess nonzero frequencies even in the homogeneous regime of perturbations with zero wavenumber. Those are very different from the so-called hydrodynamic modes, such as a sound wave, whose excitation frequency satisfies the condition $\lim_{k\to 0}\omega_{\textrm{sound}}(k)=0$.} with lowest imaginary part (in absolute value) provide upper bounds for characteristic equilibration times of the late linear stage of the full nonlinear evolution of the medium. In fact, it has been explicitly shown e.g. in Ref.\ \cite{Critelli:2017euk} that the lowest QNM's of a top-down conformal holographic plasma with a critical point quantitatively describe the late time behavior of the full nonlinear far-from-equilibrium evolution of the system. On the other hand, earlier nonlinear stages are beyond the scope of these characteristic equilibration times extracted from the QNM analysis and, thus, in general one cannot read e.g. the isotropization and thermalization times of the medium without performing numerical simulations and following the full nonlinear evolution of the system.

Here we initiate the study of how the presence of a critical point affects the non-equilibrium behavior of strongly coupled (non-conformal) baryon dense holographic models. In this work, we investigate the behavior of the lowest, homogeneous non-hydrodynamic QNM's of the bottom-up EMD model of Ref.\ \cite{Critelli:2017oub} up to the critical region, from which we extract upper bounds for characteristic equilibration times of the hot and baryon dense QGP specifically in the regime of linearized homogeneous perturbations. As it will be shown in the course of the next sections, these equilibration times of the plasma in different channels are generally reduced by increasing the baryon chemical potential keeping the temperature fixed.

We also remark that, as it will be discussed in more details in the conclusions, the dynamical universality class \cite{Hohenberg:1977ym} of large $N_c$ theories, like the present holographic model, is likely to be of type B \cite{Natsuume:2010bs}, while QCD is expected to be in the type H dynamical universality class \cite{Son:2004iv}. Therefore, one must be careful and keep this distinction in mind when applying holographic models to obtain insights for dynamical critical phenomena in real QCD.

\emph{Definitions}: We use a mostly plus metric signature and natural units i.e. $\hbar = c = k_B = 1$.

\section{EMD holography and thermodynamics}
\label{sec:EMD}

In this section we give a very brief review of the basic properties of the EMD holographic model of Ref.\ \cite{Critelli:2017oub} and its main thermodynamic results. For detailed discussions including technical details on numerics we refer the interested reader to consult the aforementioned work.

The construction of bottom-up dilatonic holographic models seeded with some phenomenological inputs in order to mimic the behavior of the QGP at finite temperature has been originally proposed in the seminal work of Ref.\ \cite{Gubser:2008ny} and extended to nonzero baryon densities in \cite{DeWolfe:2010he,DeWolfe:2011ts}. The dilaton field is used to dynamically break conformal symmetry in a very specific way dictated by the phenomenological inputs used to engineer the holographic model. Gauge/gravity \cite{Maldacena:1997re,Gubser:1998bc,Witten:1998qj,Witten:1998zw} models constructed in this way may be then used to provide estimates for a large variety of physical observables in the QGP \cite{Gubser:2008yx,DeWolfe:2010he,DeWolfe:2011ts,Finazzo:2013efa,Finazzo:2014cna,Rougemont:2015wca,Rougemont:2015ona,Finazzo:2015xwa,Rougemont:2017tlu,Critelli:2017oub,Rougemont:2015oea,Finazzo:2016mhm,Critelli:2016cvq}.

The bulk action of the EMD model is given by
\begin{align}
S= \frac{1}{2\kappa_5^2}\int d^5x\,\sqrt{-g}\left[R-\frac{(\partial_\mu\phi)^2}{2}-V(\phi) -\frac{f(\phi)F_{\mu\nu}^2}{4}\right],
\label{eq:EMDaction}
\end{align}
where $\kappa_5^2\equiv 8\pi G_5$ is the five dimensional Newton's constant. The action \eqref{eq:EMDaction} is accompanied by some boundary terms which will play no role for the calculations to be carried out here and, therefore, we omit them. In \eqref{eq:EMDaction}, the dilaton potential $V(\phi)$ and the Maxwell-dilaton coupling $f(\phi)$ are free functions of the bottom-up EMD construction. There are also two free parameters given by the gravitational constant $\kappa_5^2$ and the radius $L$ of the five dimensional asymptotically Anti-de Sitter (AdS) spacetime. Without any loss of generality, we set $L=1$ and consider instead a scaling factor $\Lambda$ corresponding to a fixed energy scale employed to convert observables with mass dimension $p$ evaluated on the gravity side of the holographic duality, which are naturally computed in units of $L^{-p}$, to physical units of MeV$^p$ on the gauge theory side of the holographic correspondence.

The free functions and parameters of the bottom-up EMD model were dynamically fixed in Ref.\ \cite{Critelli:2017oub} by matching the holographic equation of state and second order baryon susceptibility evaluated at $\mu_B=0$ to the corresponding QCD results obtained in state-of-the-art lattice simulations \cite{Bellwied:2015lba,Borsanyi:2011sw,Borsanyi:2013bia},
\begin{align}
V(\phi)&=-12\cosh(0.63\,\phi)+0.65\,\phi^2-0.05\,\phi^4+0.003\,\phi^6,\nonumber\\
\kappa_5^2&=8\pi G_5=8\pi(0.46), \quad \Lambda=1058.83\,\textrm{MeV},\nonumber\\
f(\phi)&=\frac{\textrm{sech}\left(-0.27\,\phi + 0.4\,\phi^2\right) + 1.7\,\textrm{sech}(100\,\phi)}{2.7}.
\label{eq:EMDfits}
\end{align}
We remark that results for any other observable, apart from the ones used to fix the EMD parameters in Eq.\ \eqref{eq:EMDfits}, follow as legitimate predictions of the EMD holographic model.

\begin{figure*}
\begin{center}
\includegraphics[width=0.85\textwidth]{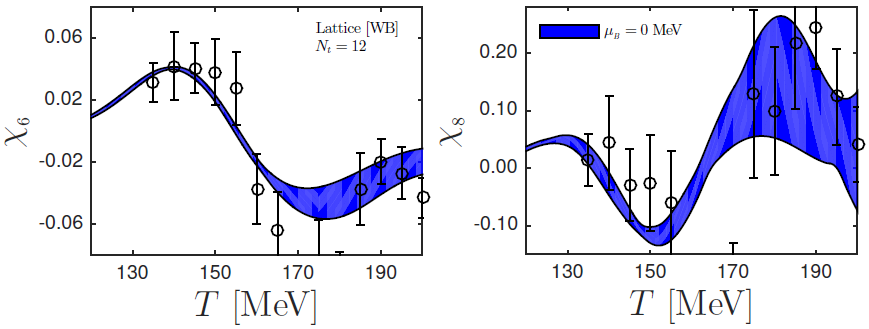}
\end{center}
\caption{{\small (Color online) Sixth and eighth order dimensionless baryon susceptibilities $\chi_n\equiv \partial^n(P/T^4)/\partial(\mu_B/T)^n$ predicted by the EMD model in Ref.\ \cite{Critelli:2017oub} compared to the very recent lattice QCD results from Ref.\ \cite{Borsanyi:2018grb}.}
\label{fig:thermo}}
\end{figure*}

The static and homogeneous charged black hole backgrounds describing isotropic and translationally invariant thermal states at finite density in the gauge theory are obtained by numerically solving the EMD equations of motion, where the Ansatz for the EMD fields may be written as follows \cite{DeWolfe:2010he},
\begin{align}
ds^2=e^{2A(r)} \left[ -h(r)dt^2+d\vec{x}^2 \right] +\frac{dr^2}{h(r)}, \quad
\phi=\phi(r), \quad A=A_\mu dx^\mu=\Phi(r)dt. \label{eq:EMDansatz}
\end{align}
The black hole horizon is located at the value of the radial coordinate corresponding to the largest zero of $h(r_H)=0$, while the boundary of the asymptotically AdS$_5$ spacetime lies at $r\rightarrow\infty$. The set of coupled second order differential equations of motion for the EMD fields may be numerically solved following the method discussed in Ref.\ \cite{Critelli:2017oub} by choosing values for a pair of initial conditions corresponding to the value of the dilaton field at the horizon ($\phi_0$) and the value of the derivative of the Maxwell field at the horizon ($\Phi_1$). Each value chosen for the pair of initial conditions $(\phi_0,\Phi_1)$ generates a numerical charged black hole background associated with a definite thermal state at finite baryon density in the gauge theory. By constructing an ensemble of charged black holes, one populates the $(T,\mu_B)$ phase diagram and may then proceed to evaluate a wealth of different physical observables.

In Fig. \ref{fig:thermo} we show the comparisons between the holographic EMD predictions for the sixth and eighth order baryon susceptibilities \cite{Critelli:2017oub} and the very recent lattice QCD results of Ref. \cite{Borsanyi:2018grb}.\footnote{Even though the lattice simulations of Ref. \cite{Borsanyi:2018grb} are done with $2+1+1$ flavors, in the range of temperatures considered in Fig. \ref{fig:thermo} the results agree with lattice QCD simulations with $2+1$ flavors \cite{Borsanyi:2016ksw}.} One notes that the EMD predictions are in good quantitative agreement with first principle lattice QCD results. As far as we know, when it comes to the baryon susceptibilities, the only other model currently available in the literature which has also achieved a high level of quantitative agreement with lattice results is the Cluster Expansion Model (CEM) of Ref.\ \cite{Vovchenko:2017gkg}. Moreover, in Ref. \cite{Critelli:2017oub} the EMD predictions for the pressure and the baryon charge density at finite temperature and baryon chemical potential were compared to the lattice results of Ref. \cite{Bazavov:2017dus}, where it was obtained an excellent quantitative agreement with state-of-the-art lattice QCD simulations.

By analyzing the behavior of the baryon susceptibilities in the EMD model beyond the region of the phase diagram currently probed by lattice simulations, in Ref.\ \cite{Critelli:2017oub} we predicted the QCD CEP to be located at $(T^{\textrm{CEP}},\mu_B^{\textrm{CEP}})\sim(89,724)$ MeV. As mentioned in Sec.\ \ref{sec:intro}, a chemical freeze-out analysis was also carried out in Ref.\ \cite{Critelli:2017oub} and the following window for the critical region in terms of the center of mass energy of the collisions was found: $\sqrt{s_{\textrm{NN}}^{\textrm{CEP}}}\sim 2.5$ --- $4.1$ GeV. Though these values of $\sqrt{s_{\textrm{NN}}}$ are lower than the upcoming Beam Energy Scan II at RHIC, planned fixed-target experiments also at RHIC \cite{Meehan:2017cum} and future low energy heavy ion collisions at FAIR/GSI \cite{Ablyazimov:2017guv} should be able to investigate the baryon rich system formed in these low energies.

We close this section by remarking that our motivation here to use the bottom-up EMD holographic model of Ref.\ \cite{Critelli:2017oub} is that it provides a very practical way to estimate how a strongly coupled QGP at finite temperature behaves at moderately large baryon density in and out of equilibrium. This is still beyond the scope of first principle lattice QCD calculations and, thus, our knowledge about the QGP in this regime must rely on effective models that display some of the QGP's well-known essential features, such as its nearly perfect fluidity in the crossover region. Moreover, in this type of framework, it is mandatory to have at least the thermodynamics of the system (in the regime amenable to the lattice) under investigation properly described, a nontrivial feature displayed by the model of Ref.\ \cite{Critelli:2017oub} as discussed above. Also, we note that we are assuming that in the range of the $(T,\mu_B)$ plane we are interested in the system is still sufficiently strongly coupled and behaves as a nearly perfect fluid. This is clearly not the case at very large temperatures (where a weakly coupled description must hold \cite{Haque:2014rua}) or at very low temperatures in the hadron dominated phase.

Furthermore, we also remember that, as detailed discussed in Refs. \cite{Gubser:2008ny,Gubser:2008yx,DeWolfe:2010he,DeWolfe:2011ts,Rougemont:2015wca,Critelli:2017oub}, in the kind of bottom-up EMD holographic model used here, the flavor dynamics at the boundary is effectively encoded in the bulk dilaton potential $V(\phi)$ and Maxwell-dilaton coupling function $f(\phi)$. The phenomenological applicability of such approach must be then empirically checked by comparing the predictions of the EMD model with the corresponding phenomenology intended to be described. In what regards thermodynamics and hydrodynamics of the strongly coupled QGP with $2+1$ flavors and physical quark masses, as emphasized by items 1 --- 3 of Section \ref{sec:intro}, the EMD model of Ref. \cite{Critelli:2017oub} has so far an unmatched degree of simultaneous quantitative agreement with many different QCD results, providing a strong empirical evidence of the phenomenological applicability of this kind of effective holographic approach.

\section{Quasinormal modes and equilibration times}
\label{sec:QNMs}

Quasinormal modes \cite{Horowitz:1999jd,Kokkotas:1999bd,Kovtun:2005ev,Berti:2009kk,Konoplya:2011qq,Buchel:2015saa} in some non-conformal dilatonic holographic models have been analyzed, for instance, in Refs.\ \cite{Janik:2015waa,Janik:2015iry,Janik:2016btb,Attems:2016ugt,Gursoy:2016ggq,Demircik:2016nhr,Betzios:2017dol}. In the context of EMD holography at finite baryon chemical potential, some results for the $SO(3)$ quintuplet channel (to be discussed below) but still far from the critical regime have been discussed by some of us in Ref.\ \cite{Rougemont:2015wca} using a previous version of the EMD model.

As discussed in detail in Ref. \cite{DeWolfe:2011ts}, for an EMD model at finite temperature and chemical potential in the homogeneous regime of zero wavenumber disturbances, the physically relevant gauge and diffeomorphism invariant linearized perturbations of the system are organized into different representations of the $SO(3)$ rotational symmetry: the quintuplet, triplet, and singlet channels. By considering a wave-plane profile for these perturbations, one can derive linearized equations of motion, whose solutions expanded asymptotically close to the boundary typically possess a leading non-normalizable mode and a subleading normalizable mode for each perturbation. According to the holographic dictionary, the leading modes act as sources for local gauge invariant operators of the dual quantum field theory (QFT) at the boundary (e.g, the energy-momentum tensor, vector currents, and scalar fields), while the subleading modes are related to the expectation values of these operators. By setting the subleading modes to zero as a boundary condition and by also imposing the causal in-falling wave condition at the black hole horizon in the interior of the bulk, one obtains particular solutions of the linearized equations of motion for the perturbations of the bulk fields. By plugging them back into the action one can calculate the thermal two-point retarded correlators of the dual QFT. Kubo formulas relating these thermal retarded correlators to transport coefficients of the dual QFT may be derived using linear response theory within the framework of the gradient expansion of the dual QFT operators in terms of the hydrodynamic variables of the system (e.g., the local energy and charge densities and the local fluid flow velocity), as done for instance in Ref. \cite{Baier:2007ix}.

Also according to the holographic dictionary \cite{Son:2002sd}, the thermal two-point retarded correlators of the dual QFT may be written as (minus) the ratio between the subleading and the leading modes subjected to the in-falling wave condition at the horizon. Consequently, if one now sets the leading modes to zero as a boundary condition, one obtains the poles of these thermal retarded correlators, which correspond to physical collective excitations in the dual QFT. But taking the leading modes to zero as a boundary condition, implying that the (in-falling) perturbations must vanish at the boundary, is exactly what defines the eigenvalue problem for the quasinormal modes in asymptotically AdS geometries \cite{Kovtun:2005ev}. Therefore, by means of the holographic dictionary, the quasinormal modes of black holes in the bulk correspond to poles of thermal retarded correlators of the dual QFT at the boundary.

We remark that in the in-falling Eddington-Finkelstein (EF) coordinates, which we shall use in this section, the causal condition for obtaining in-falling modes at the black hole horizon is simply expressed by requiring that the solutions are regular at the horizon. In this case, one may evaluate the shear viscosity by making use of a Kubo formula that relates this transport coefficient to the retarded 2-point Green's function of the non-normalizable perturbation in the $SO(3)$ quintuplet channel \cite{DeWolfe:2011ts}. Similarly, by working with the perturbations of the $SO(3)$ triplet and singlet channels, one may obtain through the use of Kubo formulas the baryon conductivity \cite{DeWolfe:2011ts,Rougemont:2015ona} and the bulk viscosity \cite{DeWolfe:2011ts,Rougemont:2017tlu} of the medium, respectively.

The spectra of QNM's of the theory is obtained by solving the linearized equations of motion for the perturbations requiring regularity at the horizon while imposing the Dirichlet condition that these perturbations vanish at the boundary. This means that such equations will admit solutions only for a discrete set of complex eigenfrequencies $\omega$, which are the QNM's of the theory. In general, the QNM's $\omega$ depend on the wavenumber $k$ of the perturbations and also on the properties of the black hole background, with the latter translating in a dependence on $T$ and $\mu_B$. In the present work we are only interested in analyzing the upper bounds for characteristic equilibration times of the medium in the regime of linear perturbations and, for this sake, it suffices to consider only the lowest non-hydrodynamic QNM's in the homogeneous regime with $k=0$, which we shall explore in the course of the next sections.

We remember that the hydrodynamic QNM's may be used to obtain the hydrodynamic transport coefficients of the system in an alternative way to the more direct and straightforward method of Kubo formulas previously mentioned. This has been discussed, for instance, in Refs. \cite{Policastro:2002se,Policastro:2002tn,Janik:2016btb,Heller:2013fn}.

On the other hand, the non-hydrodynamic QNM's, as discussed for instance in Ref. \cite{Horowitz:1999jd}, may be used to obtain upper bounds for characteristic equilibration times of the linearly disturbed black holes. This is so since the lowest non-hydrodynamic modes are the latest to be damped in the homogeneous zero wavenumber limit. By the holographic correspondence, these correspond to characteristic equilibration times of the dual plasma slightly driven out of equilibrium.\footnote{Far-from-equilibrium dynamics, in turn, requires the analysis of the full nonlinear evolution of the system, which will not be treated in the present work.} Moreover, as discussed in the seminal work of Ref. \cite{Heller:2013fn}, higher order hydrodynamic transport coefficients may be linked to the lowest non-hydrodynamic QNM's of the system using a Borel resummation of the asymptotic gradient expansion for hydrodynamics. Consequently, the non-hydrodynamic QNM's, even though not describable within hydrodynamics, leave their fingerprints in the behavior of high order hydrodynamic transport coefficients.

The gauge and diffeomorphism invariant EMD linearized perturbations in the homogeneous regime of disturbances with zero wavenumber, and their corresponding equations of motion, were originally obtained in Ref. \cite{DeWolfe:2011ts}. In the $SO(3)$ quintuplet channel the relevant perturbation is given by $\chi\equiv h_{ij}$, where $h_{\mu\nu}$ is the perturbation of the metric field and $h_{ij}$ is any of the five traceless spatial components of the graviton. This case will be analyzed in Section \ref{sec:5et}. In the $SO(3)$ triplet channel the relevant perturbation $a\equiv a_i$ represents any of the three spatial components of the perturbation $a_\mu$ of the Maxwell field - this shall be studied in Section \ref{sec:3et}. In the $SO(3)$ singlet channel the relevant perturbation is given by $\mathcal{S}\equiv\varphi-\left(\phi'/2A'\right)\left[(h_{xx}+h_{yy}+h_{zz})/3\right]$, where $\varphi$ is the perturbation of the dilaton field $\phi$. We note that the background dilaton field couples the dilaton perturbation to the spatial trace of the graviton through this $\mathcal{S}$-perturbation. The explicit form of the linearly perturbed EMD fields reads as follows \cite{DeWolfe:2011ts},
\begin{align}
ds_{\textrm{(pert)}}^2&=\left(g_{\mu\nu}^{(0)}+\textrm{Re}\left[e^{2A(r)}h_{\mu\nu}(r)e^{-i\omega t}\right]\right)dx^\mu dx^\nu, \quad
\phi_{\textrm{(pert)}}=\phi(r)+\textrm{Re}\left[\varphi(r)e^{-i\omega t}\right],\nonumber\\
A_{\textrm{(pert)}}&=A_{\mu\,\textrm{(pert)}}\,dx^\mu=\Phi(r)dt +\textrm{Re}\left[a_\mu(r)e^{-i\omega t}\right]dx^\mu, \label{eq:EMDpert}
\end{align}
where $g_{\mu\nu}^{(0)}$ is the undisturbed background metric given in Eq. \eqref{eq:EMDansatz}.

Moreover, the $\mathcal{S}$-perturbation inherits the same ultraviolet asymptotics of the background dilaton field \cite{DeWolfe:2011ts,Janik:2016btb,Critelli:2017euk}, which is the reason why it has been dubbed in Ref.\ \cite{Critelli:2017euk} the ``dilaton channel''. Furthermore, the $SO(3)$ quintuplet channel may be also called the ``external scalar channel'' since its equation of motion, to be discussed in what follows, equals that of a massless external scalar field placed on the EMD backgrounds \cite{Rougemont:2015wca,DeWolfe:2011ts,Finazzo:2016psx}. Similarly, the $SO(3)$ triplet channel may be also referred as the ``vector diffusion'' channel, since this channel is related to the charge conductivity and the diffusion coefficient \cite{DeWolfe:2011ts,Rougemont:2015ona}.

As discussed in Ref.\ \cite{Critelli:2017euk}, in a more general far-from-equilibrium homogeneous (but highly anisotropic and nonlinear) configuration, each of these channels are associated with physical observables describing the full time evolution of the medium. Namely, the $SO(3)$ quintuplet, triplet, and singlet channels are associated with the pressure anisotropy, the charge density, and the scalar condensate (dual to the bulk dilaton field), respectively. Under fairly general circumstances, the approach of each of these observables toward thermal equilibrium may provide many different characteristic relaxation time scales for the medium \cite{Attems:2017zam}. The approach of the pressure anisotropy toward zero gives us the so-called ``isotropization time'' of the system, while the last equilibration time of the medium is to be naturally identified as the actual ``thermalization time'' of the system. 

As explicitly shown in Ref.\ \cite{Critelli:2017euk} for a top-down conformal EMD model with a critical point describing an $\mathcal{N}=4$ Super Yang-Mills plasma at finite density, the late time behavior of the pressure anisotropy is quantitatively described by linearized oscillations given by the lowest QNM of the $SO(3)$ quintuplet channel \cite{Finazzo:2016psx}. Moreover, in that case, the scalar condensate was found to be always the last observable to equilibrate and, as such, the thermalization time was associated with the approach of the scalar condensate to its equilibrium value, which turned out to be quantitatively described by the lowest QNM of the $SO(3)$ singlet channel. However, one must keep in mind that the isotropization and thermalization times of the system can only be computed by numerically simulating the full nonlinear evolution of the medium, i.e. they cannot be obtained by considering just the characteristic equilibration times associated with the late linear oscillations of the system described in the QNM analysis.

In the present work, even though we are still not investigating far-from-equilibrium properties of the phenomenologically realistic EMD model of Ref.\ \cite{Critelli:2017oub}, from previous considerations in the literature we expect that the behavior of the lowest QNM's in the $SO(3)$ quintuplet, triplet, and singlet channels will provide correct descriptions of the late time dynamics of the pressure anisotropy, baryon charge density, and the scalar condensate, respectively. In this regard, the characteristic equilibration times of the medium, which we shall obtain here by analyzing the lowest homogeneous non-conformal QNM's of each channel in the EMD model up to the critical region, should be more properly seen as ``lower bounds'' for the full relaxation times of the medium, since they do not take into account the time spent by the system in earlier nonlinear stages. On the other hand, when considering just small perturbations around thermal equilibrium, the characteristic equilibration times extracted from the QNM analysis indeed provide upper bounds for how fast the disturbed system returns to equilibrium.

Before we proceed to the evaluation of the QNM's in each of the aforementioned channels, let us first specify the in-falling EF coordinates by the relation given below,
\begin{align}
dv = dt + \sqrt{-\frac{g_{rr}}{g_{tt}}}\,dr = dt + \frac{e^{-A}}{h}\,dr,
\end{align}
where $v$ is the EF time. With this, one may easily translate the equations of motion for the homogeneous EMD perturbations derived in Ref.\ \cite{DeWolfe:2011ts} in domain-wall coordinates to the EF coordinates. This will be done in the following sections. Some technical details on the grids employed to obtain the lowest QNM's in each channel are deferred to Appendix \ref{sec:num}.

\subsection{$SO(3)$ quintuplet channel}
\label{sec:5et}

The equation of motion for the $\chi$-perturbation of the quintuplet channel in EF coordinates reads \cite{Rougemont:2015wca},
\begin{align}
\chi '(r) \left(4 A'(r)-\frac{2 i \omega  e^{-A(r)}}{h(r)}+\frac{h'(r)}{h(r)}\right)-\frac{3 i \omega  e^{-A(r)} \chi (r) A'(r)}{h(r)}+\chi ''(r) = 0.
\label{eq:eom5et}
\end{align}

\begin{figure*}
\begin{center}
\begin{tabular}{c}
\includegraphics[width=0.45\textwidth]{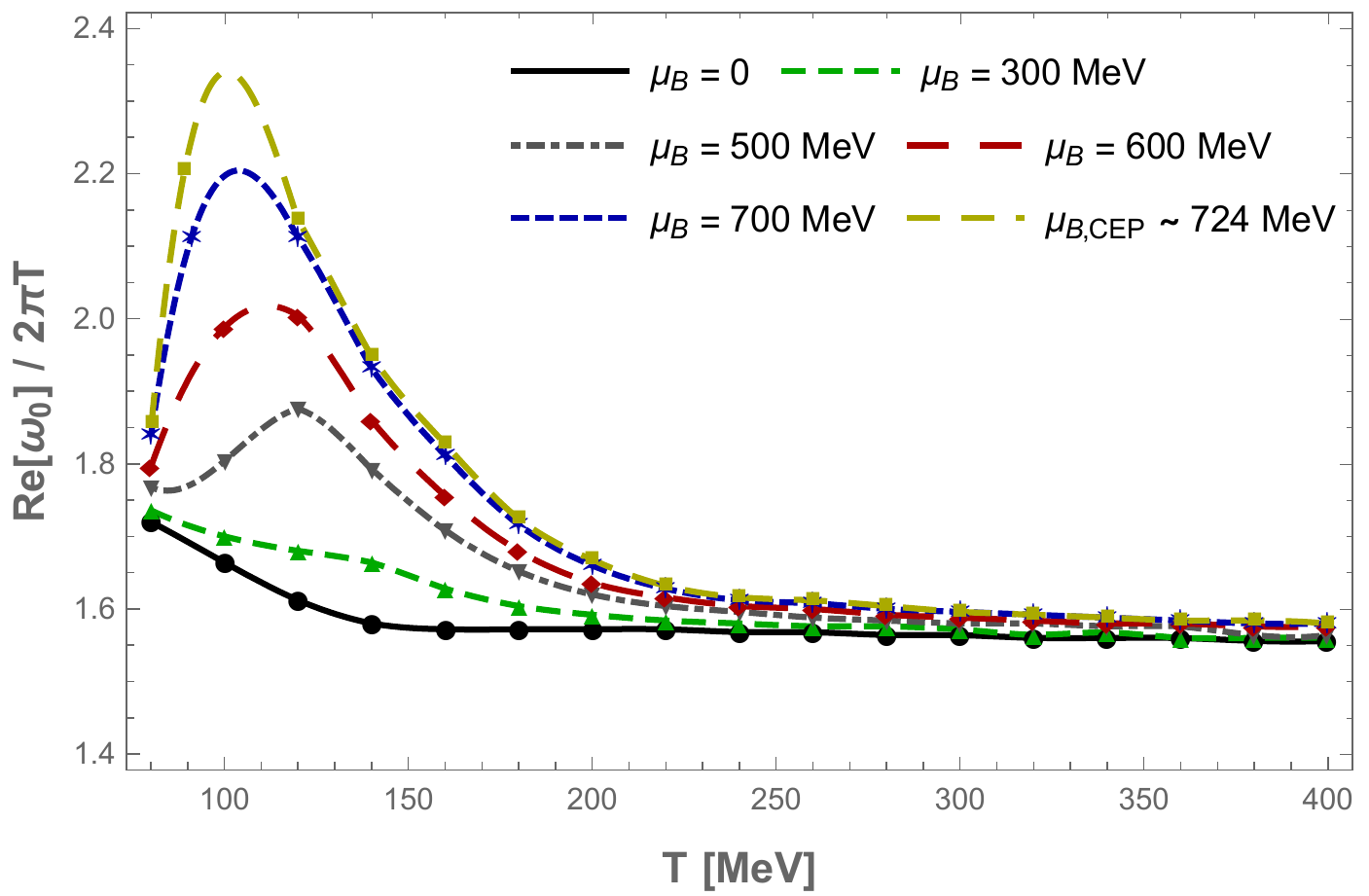} 
\end{tabular}
\begin{tabular}{c}
\includegraphics[width=0.45\textwidth]{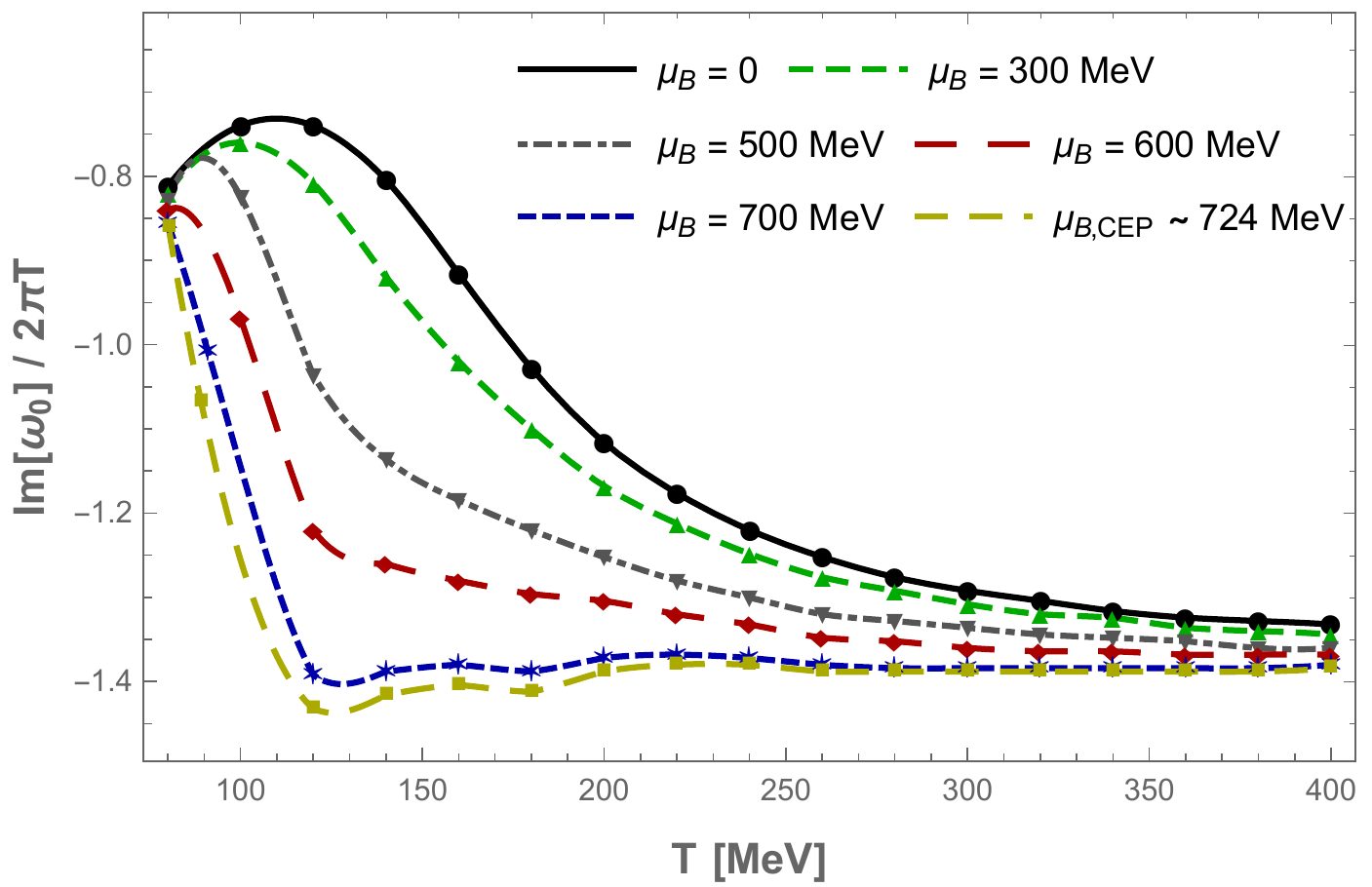} 
\end{tabular}
\begin{tabular}{c}
\includegraphics[width=0.45\textwidth]{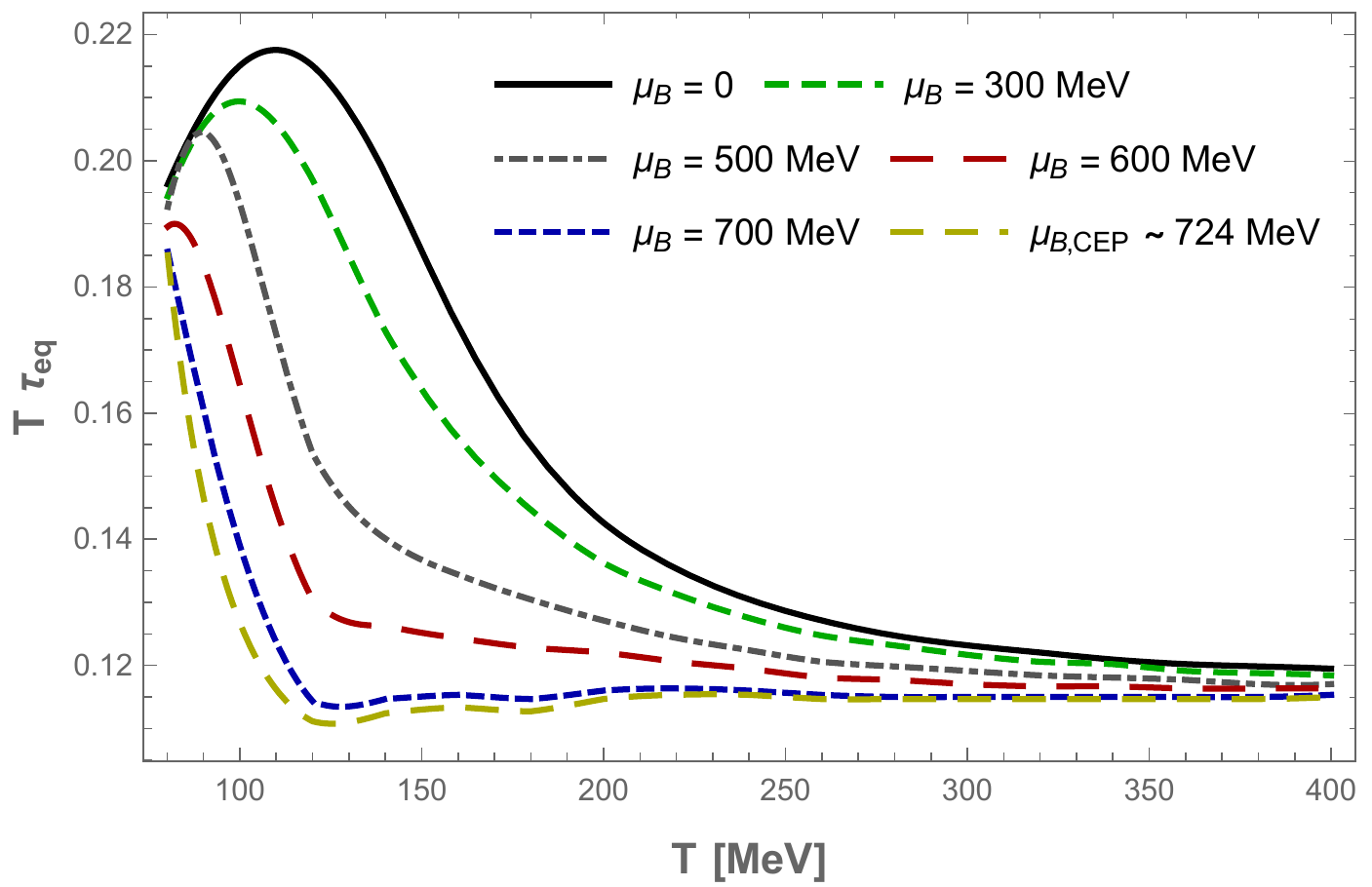} 
\end{tabular}
\begin{tabular}{c}
\includegraphics[width=0.45\textwidth]{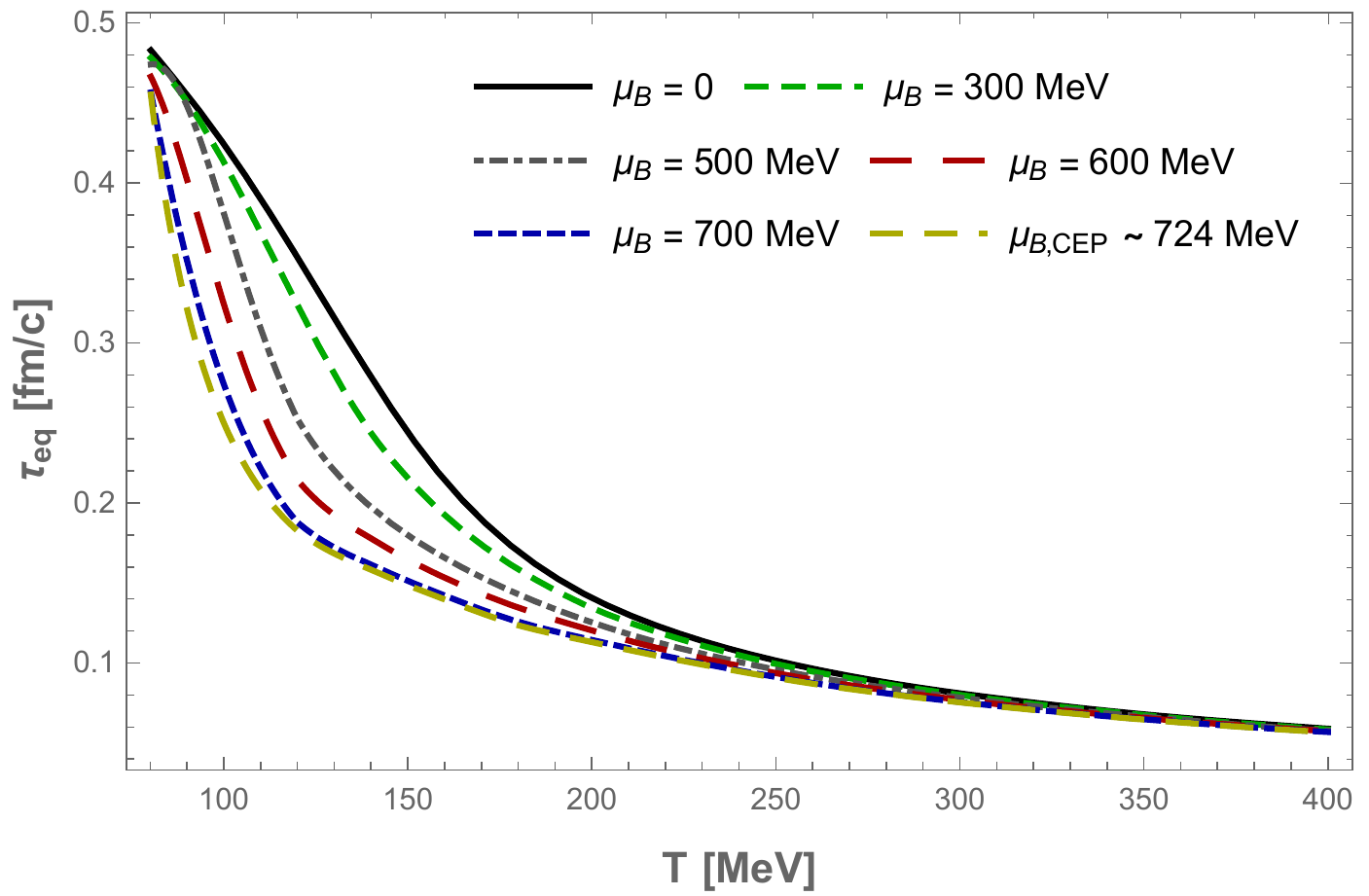} 
\end{tabular}
\end{center}
\caption{{\small (Color online) Absolute value of the real part of the lowest QNM in the $SO(3)$ quintuplet channel (top left), imaginary part of the lowest QNM (top right), the dimensionless combination given by the temperature times the equilibration time (bottom left), and the equilibration time measured in units of fm/c (bottom right) as functions of temperature for different values of the baryon chemical potential. In the upper panels we plot both the calculated points and the interpolated curves between them.}
\label{fig:5et}}
\end{figure*}

This is a linear second order differential equation which must be solved with appropriate boundary conditions in order to find the corresponding QNM's. As discussed before, the boundary conditions to be imposed are that the solutions must be regular at the black hole horizon and vanish at the boundary.

In order to numerically integrate the equation of motion \eqref{eq:eom5et} from the horizon up to the boundary\footnote{Eq.\ \eqref{eq:eom5et} is expressed in the {\it numerical coordinates} discussed in detail in Ref.\ \cite{Critelli:2017oub}. As also discussed in that reference, the final results are to be written in the {\it standard coordinates} and, as shown in Ref.\ \cite{Finazzo:2015xwa}, the dimensionless combination $\omega/(2\pi T)$ expressed in the standard coordinates is simply given by twice the value of the quasinormal eigenfrequency obtained in the numerical coordinates.} we need to specify the values of the $\chi$-perturbation and its derivative at the horizon. One may simply set $\chi(r_H)=1$, while the expression for its derivative may be obtained by Taylor-expanding the $\chi$-perturbation and the background around the horizon and plugging these expansions back into Eq.\ \eqref{eq:eom5et}, which then becomes an algebraic equation for $\chi'(r_H)$ whose solution is given by \cite{Rougemont:2015wca}
\begin{align}
\chi'(r_H) = -\frac{3 A_1 \omega }{2 \omega +i},
\label{eq:chi1num}
\end{align}
where $A_1$ is the value of the derivative of the background field $A(r)$ evaluated at the horizon (this is itself a function of the initial conditions $(\phi_0,\Phi_1)$ \cite{Critelli:2017oub}).

Now that we have the initial conditions $(\chi(r_H),\chi'(r_H))$ required to initialize the numerical integration of Eq.\ \eqref{eq:eom5et}, we take the following steps \cite{Rougemont:2015wca}:

\begin{enumerate}[i.]

\item We construct a grid of backgrounds on top of which we shall obtain the QNM's as functions of $(T,\mu_B)$. The background grid considered in this work is discussed in Appendix \ref{sec:num};

\item We construct on each point within the aforementioned background grid a rough grid of complex frequencies $\omega$ which are seeded to the differential equation \eqref{eq:eom5et} to be numerically integrated on each point within this $\omega$-grid. The rough $\omega$-grid we consider in this work has a numerical step-size of 0.03 between consecutive points both in the real and imaginary directions of the complex $\omega$-plane; 

\item Once the previous step is accomplished we apply a shooting method upon the $\omega$-grid generated for each point of the background grid. In this shooting method we begin by picking some value for $|\chi|$ evaluated at the boundary of the background spacetime and then we lower its value until we can clearly identify isolated clusters of complex eigenfrequencies with small values of the perturbation at the boundary. Since we are only interested in computing the lowest non-hydrodynamic QNM $\omega_0$, we pay attention to the cluster with lowest imaginary part (in absolute value) and keep decreasing the boundary value of $|\chi|$ until only one point remains within that cluster;

\item Next we take these rough estimates of $\omega_0$ for each background point as the centers of finer $\omega$-grids constructed with a numerical step-size of 0.002 between consecutive points both in the real and imaginary directions of the complex $\omega$-plane. We solve again Eq.\ \eqref{eq:eom5et} on top of these finer $\omega$-grids and, for each point on the background grid, we store the eigenfrequency associated with the lowest value of $|\chi|$ at the boundary.

\end{enumerate}

In principle, one may repeat this process of $\omega$-grid refinement until the desired numerical accuracy is reached. Moreover, by enlarging the $\omega$-grid size one may also identify excited QNM's besides the lowest one\footnote{In this work, we used rough $\omega$-grids with real part spanning the interval $[-2,2]$ and imaginary part spanning the interval [-2,0.1].}. The same shooting method for obtaining the QNM's, which has been previously employed in Ref.\ \cite{Rougemont:2015wca}, will also be used in the next sections.

The results for the lowest QNM $\omega_0$ of the quintuplet channel up to the critical region are shown in Fig.\ \ref{fig:5et}, where we also plot the results for the corresponding equilibration time given by minus the inverse of the imaginary part of $\omega_0$. As a basic consistency check of the numerics, we note that at high $T$ the real and imaginary parts of $\omega_0$ do return to the correct ultraviolet conformal results \cite{Janik:2015waa,Rougemont:2015wca}. One also observes that at fixed temperatures the equilibration time in this channel always decreases with increasing the baryon chemical potential.

From the bottom plot in Fig.\ \ref{fig:5et} one can immediately read some reference values for the equilibration time of the quintuplet channel, e.g. at $(T,\mu_B)\sim\{(400,0),(145,0),(89,724)\}$ MeV the corresponding equilibration times are, respectively, $\tau_{\textrm{eq}}\sim\{0.06,0.26,0.33\}$ fm/c, with the last point corresponding to the equilibration time at the CEP. By comparing this result with those from the next two sections, we shall see that the equilibration time in the quintuplet channel is the shortest relaxation time of the system.

\subsection{$SO(3)$ triplet channel}
\label{sec:3et}

The equation of motion for the $a$-perturbation of the triplet channel in EF coordinates reads
\begin{widetext}
\begin{align}
&a''(r)+a'(r) \left(2 A'(r)+\frac{h'(r)-2 i \omega  e^{-A(r)}}{h(r)}+\frac{f'(\phi ) \phi '(r)}{f(\phi )}\right)+\nonumber\\
&\frac{a(r) e^{-2 A(r)} \left(-f^2(\phi ) \Phi '(r)^2-i \omega 
   e^{A(r)} \left(f(\phi ) A'(r)+f'(\phi ) \phi '(r)\right)\right)}{f(\phi ) h(r)} = 0.
\label{eq:eom3et}
\end{align}
\end{widetext}

\begin{figure*}
\begin{center}
\begin{tabular}{c}
\includegraphics[width=0.45\textwidth]{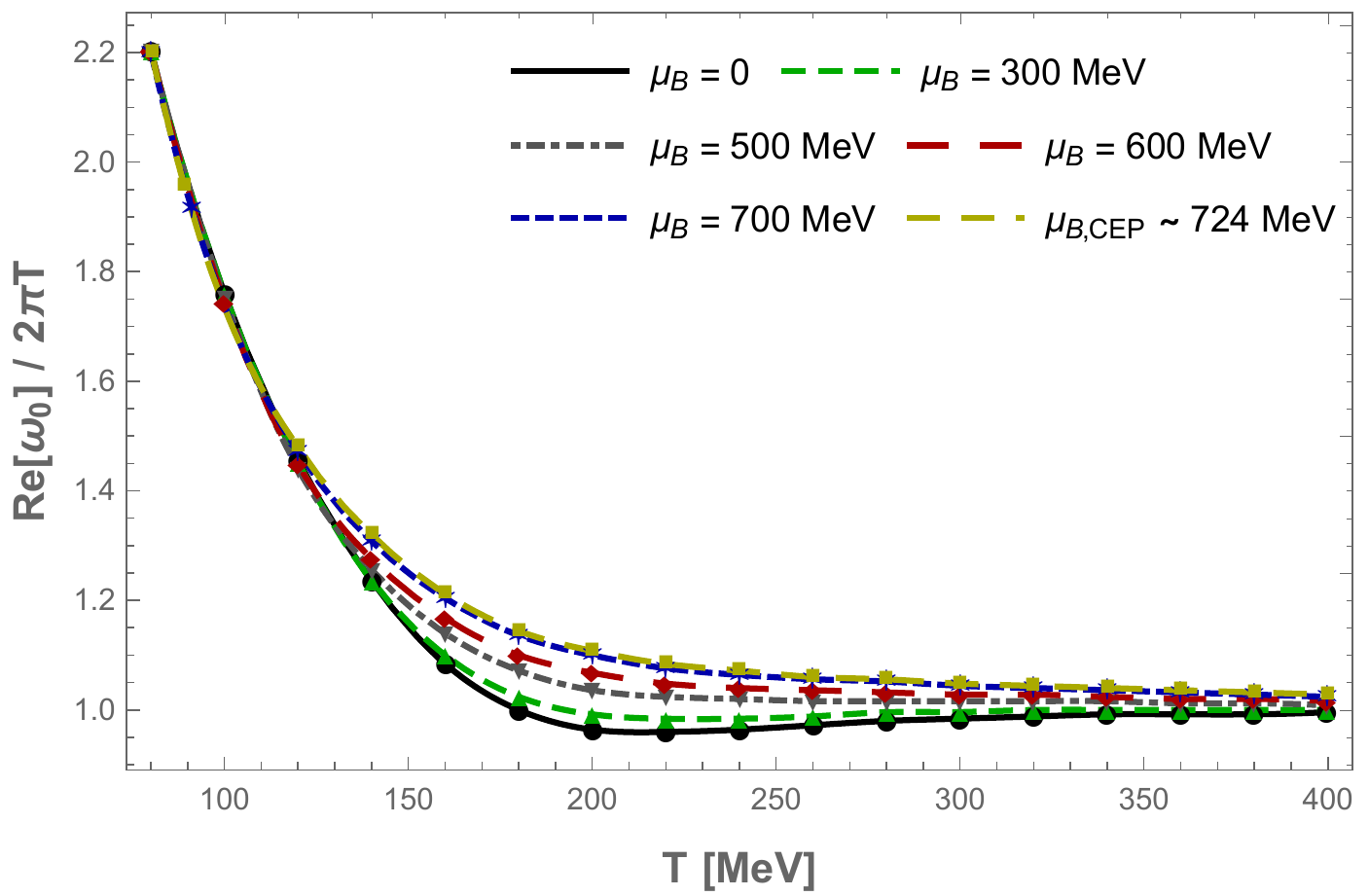} 
\end{tabular}
\begin{tabular}{c}
\includegraphics[width=0.45\textwidth]{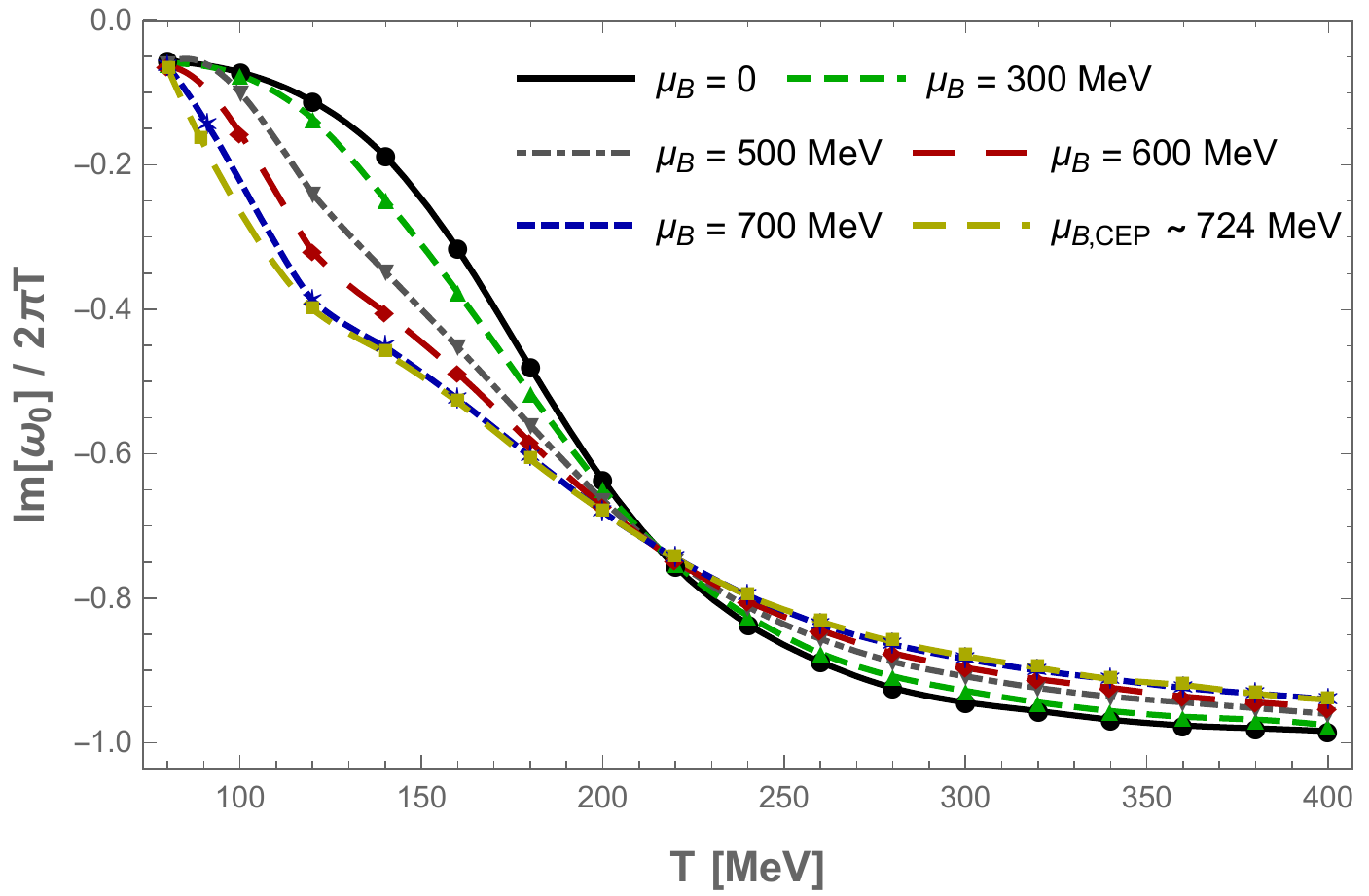} 
\end{tabular}
\begin{tabular}{c}
\includegraphics[width=0.45\textwidth]{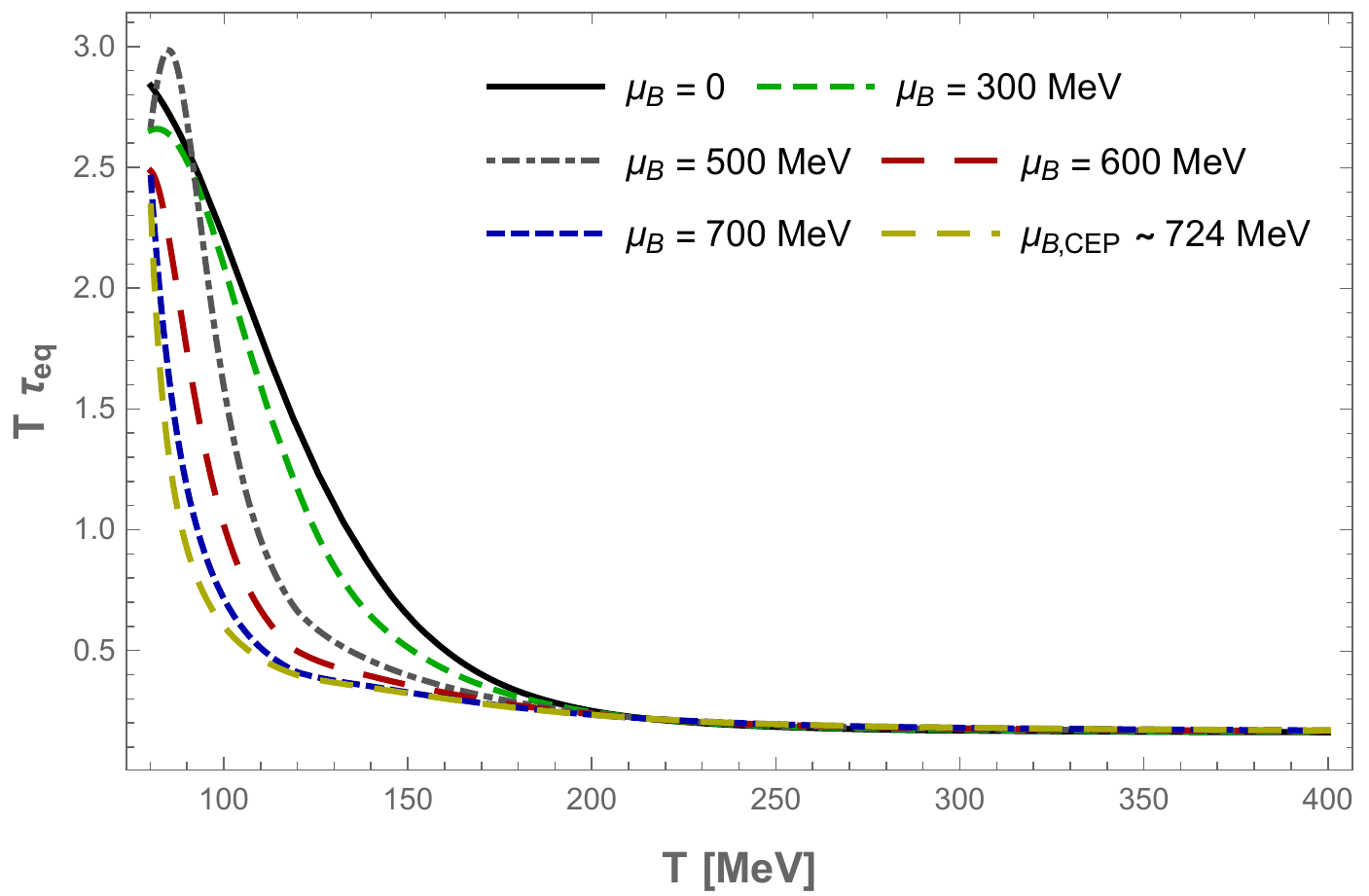} 
\end{tabular}
\begin{tabular}{c}
\includegraphics[width=0.45\textwidth]{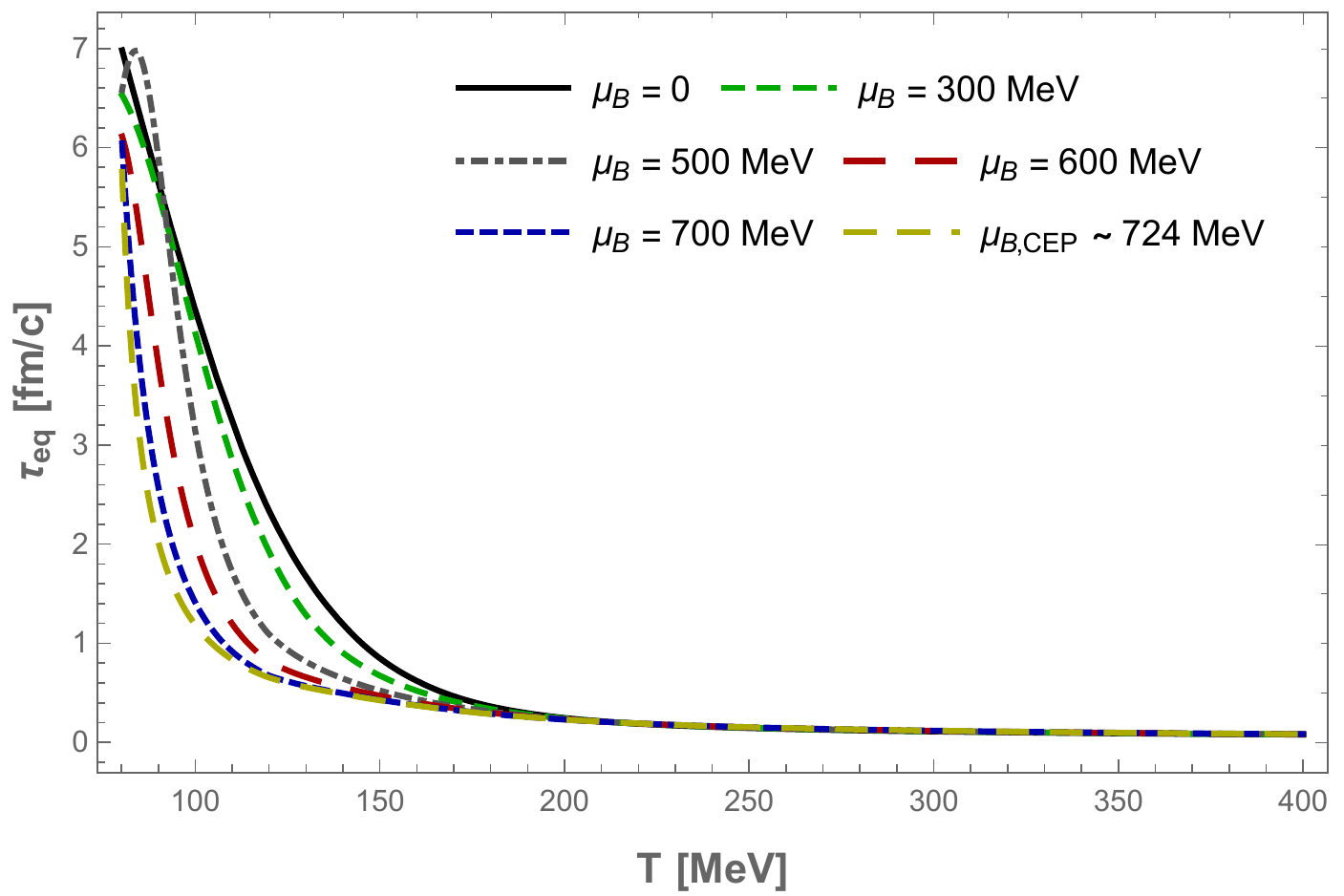} 
\end{tabular}
\end{center}
\caption{{\small (Color online) Absolute value of the real part of the lowest QNM in the $SO(3)$ triplet channel (top left), imaginary part of the lowest QNM (top right), the dimensionless combination given by the temperature times the equilibration time (bottom left), and the equilibration time measured in units of fm/c (bottom right) as functions of temperature for different values of the baryon chemical potential. In the upper panels we plot both the calculated points and the interpolated curves between them.}
\label{fig:3et}}
\end{figure*}

\begin{figure*}
\begin{center}
\begin{tabular}{c}
\includegraphics[width=0.45\textwidth]{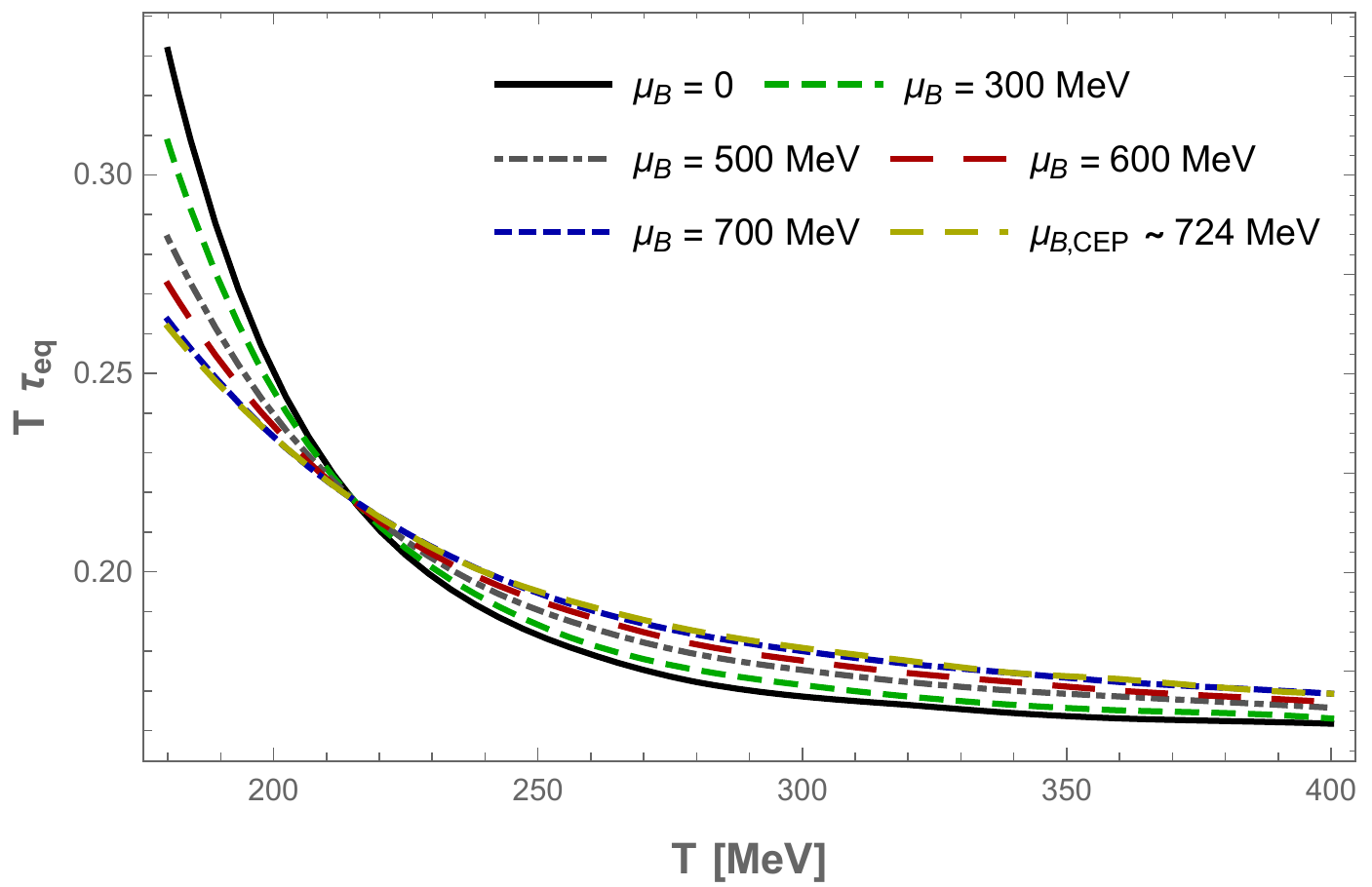} 
\end{tabular}
\begin{tabular}{c}
\includegraphics[width=0.45\textwidth]{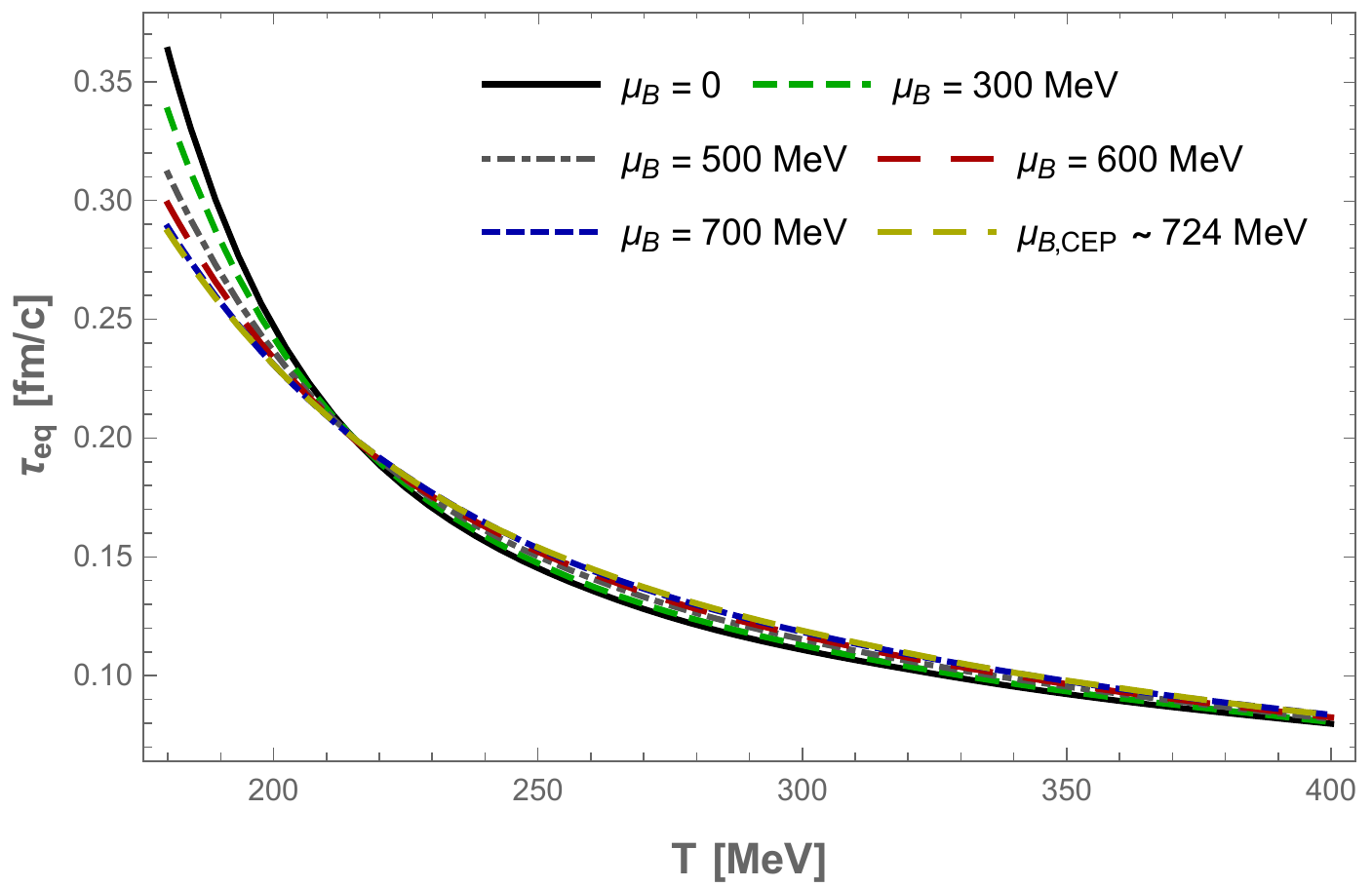} 
\end{tabular}
\begin{tabular}{c}
\includegraphics[width=0.45\textwidth]{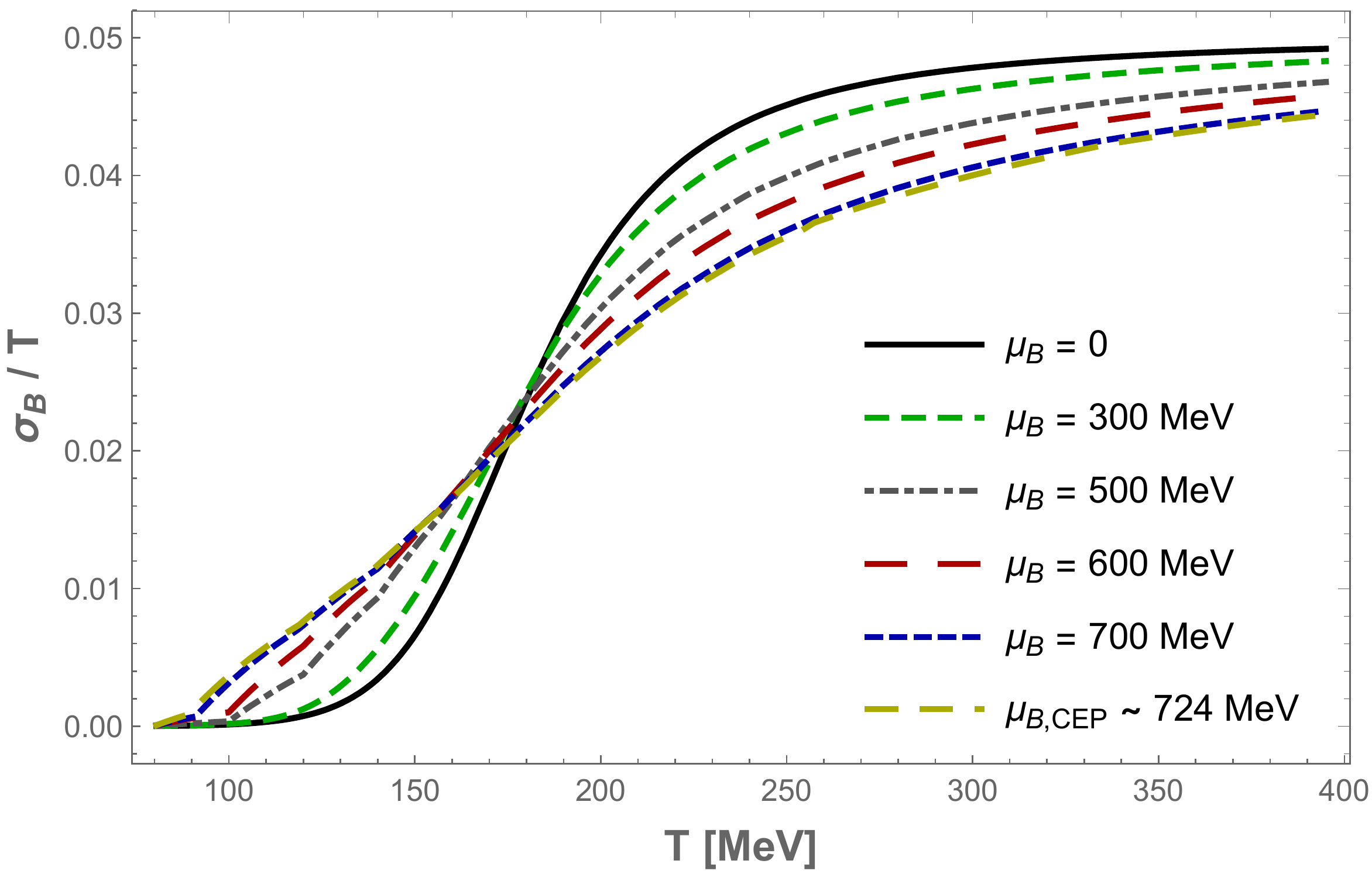} 
\end{tabular}
\end{center}
\caption{{\small (Color online) In the upper panels we zoom in the region containing the crossing point for the equilibration time in the $SO(3)$ triplet channel. In the bottom panel we plot the result for the baryon DC conductivity, which also displays a crossing point (although at a lower temperature).}
\label{fig:3et_zoom}}
\end{figure*}

As before, one may simply set $a(r_H)=1$ and work out its derivative at the horizon,
\begin{widetext}
\begin{align}
a'(r_H) &= \left( 170 \text{sech}\left(100 \phi _0\right) \left(-27 \omega  \left(A_1-100 \phi _1 \tanh \left(100 \phi _0\right)\right)+20 i \Phi _1^2 \text{sech}\left(\frac{1}{100}
   \left(27-40 \phi _0\right) \phi _0\right)\right)\right.\nonumber\\
&+\text{sech}\left(\frac{1}{100} \left(27-40 \phi _0\right) \phi _0\right) \left(-27 \omega  \left(100 A_1+\left(80 \phi
   _0-27\right) \phi _1 \tanh \left(\frac{1}{100} \left(27-40 \phi _0\right) \phi _0\right)\right)\right.\nonumber\\
&\left.\left. +1000 i \Phi _1^2 \text{sech}\left(\frac{1}{100} \left(27-40 \phi _0\right) \phi
   _0\right)\right)+2890 i \Phi _1^2 \text{sech}^2\left(100 \phi _0\right)\right) / \left(270 (2 \omega +i) \left(17 \text{sech}\left(100 \phi _0\right)\right.\right.\nonumber\\
&\left.\left.+10 \text{sech}\left(\frac{1}{100}
   \left(27-40 \phi _0\right) \phi _0\right)\right)\right),
\label{eq:a1num}
\end{align}
\end{widetext}
where $\phi_1$ is the value of the derivative of the dilaton field evaluated at the horizon (this is itself a function of the initial conditions $(\phi_0,\Phi_1)$ \cite{Critelli:2017oub}). The general steps used to numerically obtain the QNM's are the same discussed in the previous section.

In Fig.\ \ref{fig:3et} we show the results for the lowest non-hydrodynamical QNM in the triplet channel, as well as the corresponding equilibration times. As a basic consistency check of the numerics, one notes that at high $T$ the real and imaginary parts of $\omega_0$ go to their correct ultraviolet conformal results, which are known analytically for the vector diffusion channel \cite{Kovtun:2005ev}. In the present case, one notes that for $T\lesssim 215$ MeV the equilibration time is reduced by increasing the baryon chemical potential at fixed temperature but at $T\sim 215$ MeV all the curves with different fixed $\mu_B$ meet at the same crossing point, while for $T\gtrsim 215$ MeV the equilibration time slightly increases with increasing baryon chemical potential at fixed temperatures. This is more clearly illustrated in Fig.\ \ref{fig:3et_zoom}, where we also plot the results for the baryon DC conductivity in the refined EMD model of Ref.\ \cite{Critelli:2017oub}\footnote{This is very similar to the baryon conductivity calculated in Ref.\ \cite{Rougemont:2015ona} using the previous version of the EMD model \cite{Rougemont:2015wca}.}. Intriguingly, we observe that the baryon conductivity, which is the transport coefficient extracted from the $SO(3)$ triplet channel, also displays a crossing point, although at a lower temperature than the equilibration time of this channel.

From Fig.\ \ref{fig:3et} one reads some reference values for the equilibration time in the triplet channel, e.g. at $(T,\mu_B)\sim\{(400,0),(145,0),(89,724)\}$ MeV the corresponding equilibration times are, respectively, $\tau_{\textrm{eq}}\sim\{0.08,1.00,2.15\}$ fm/c, with the last point corresponding to the CEP. At $(T,\mu_B)\sim (145,0)$ MeV and at the CEP the equilibration times of the triplet channel are the longest in comparison with the results from the other two channels.

\subsection{$SO(3)$ singlet channel}
\label{sec:1et}
 
The equation of motion for the $\mathcal{S}$-perturbation of the singlet channel in EF coordinates reads
\begin{widetext}
\begin{align}
&\left( e^{-2 A(r)} \mathcal{S}(r) \left(-18 A'(r)^2 f'(\phi )^2 \Phi '(r)^2+f(\phi ) \left(3 A'(r)^2 \left(8 e^{2 A(r)} h(r) \phi '(r)^2-6 e^{2 A(r)} V''(\phi )+3 f''(\phi ) \Phi
   '(r)^2\right)\right.\right.\right.\nonumber\\
&\left.\left.\left. +6 A'(r) \phi '(r) \left(e^{2 A(r)} \left(h'(r) \phi '(r)-2 V'(\phi )\right)+f'(\phi ) \Phi '(r)^2\right)-54 i \omega  e^{A(r)} A'(r)^3-e^{2 A(r)} h(r) \phi
   '(r)^4\right)\right)\right) / \left(18 f(\phi ) h(r) A'(r)^2\right)\nonumber\\
& +\frac{\mathcal{S}'(r) \left(4 h(r) A'(r)-2 i \omega  e^{-A(r)}+h'(r)\right)}{h(r)}+\mathcal{S}''(r) = 0.
\label{eq:eom1et}
\end{align}
\end{widetext}

As in the previous sections, one may set $\mathcal{S}(r_H)=1$ and work out its derivative at the horizon,
\begin{widetext}
\begin{align}
\mathcal{S}'(r_H) &= \left(2430000 i A_1 \omega -\left( 40 \phi _1 \left(2000 \Phi _1^2 \phi _0 \tanh \left(\frac{1}{100} \left(27-40 \phi _0\right) \phi _0\right) \text{sech}\left(\frac{1}{100}
   \left(27-40 \phi _0\right) \phi _0\right)\right.\right.\right.\nonumber\\
& -425000 \Phi _1^2 \tanh \left(100 \phi _0\right) \text{sech}\left(100 \phi _0\right)-675 \Phi _1^2 \tanh \left(\frac{1}{100}
   \left(27-40 \phi _0\right) \phi _0\right) \text{sech}\left(\frac{1}{100} \left(27-40 \phi _0\right) \phi _0\right)\nonumber\\
&\left.\left. -243 \phi _0^5+2700 \phi _0^3-17550 \phi _0+6750 \phi
   _1+102060 \sinh \left(\frac{63 \phi _0}{100}\right)\right)\right) / (A_1)\nonumber\\
& +\frac{300 \Phi _1^2 \left(17000 \tanh \left(100 \phi _0\right) \text{sech}\left(100 \phi _0\right)+\left(27-80
   \phi _0\right) \tanh \left(\frac{1}{100} \left(27-40 \phi _0\right) \phi _0\right) \text{sech}\left(\frac{1}{100} \left(27-40 \phi _0\right) \phi _0\right)\right){}^2}{17
   \text{sech}\left(100 \phi _0\right)+10 \text{sech}\left(\frac{1}{100} \left(27-40 \phi _0\right) \phi _0\right)}\nonumber\\
& -15 \Phi _1^2 \left(85000000 \left(\cosh \left(200 \phi
   _0\right)-3\right) \text{sech}^3\left(100 \phi _0\right)+\frac{1}{2} \text{sech}^3\left(\frac{1}{100} \left(27-40 \phi _0\right) \phi _0\right) \left(-3 \left(27-80 \phi
   _0\right){}^2\right.\right.\nonumber\\
&\left.\left. +8000 \sinh \left(\frac{1}{50} \left(27-40 \phi _0\right) \phi _0\right)+\left(27-80 \phi _0\right){}^2 \cosh \left(\frac{1}{50} \left(27-40 \phi _0\right) \phi
   _0\right)\right)\right)\nonumber\\
&\left. +324 \left(25 \left(9 \phi _0^4-60 \phi _0^2+130\right)-11907 \cosh \left(\frac{63 \phi _0}{100}\right)\right)\right) / \left(810000 (1-2 i \omega )\right).
\label{eq:S1num}
\end{align}
\end{widetext}
These are the initial conditions used to numerically integrate the equation of motion \eqref{eq:eom1et} following the shooting procedure discussed before.

\begin{figure*}
\begin{center}
\begin{tabular}{c}
\includegraphics[width=0.45\textwidth]{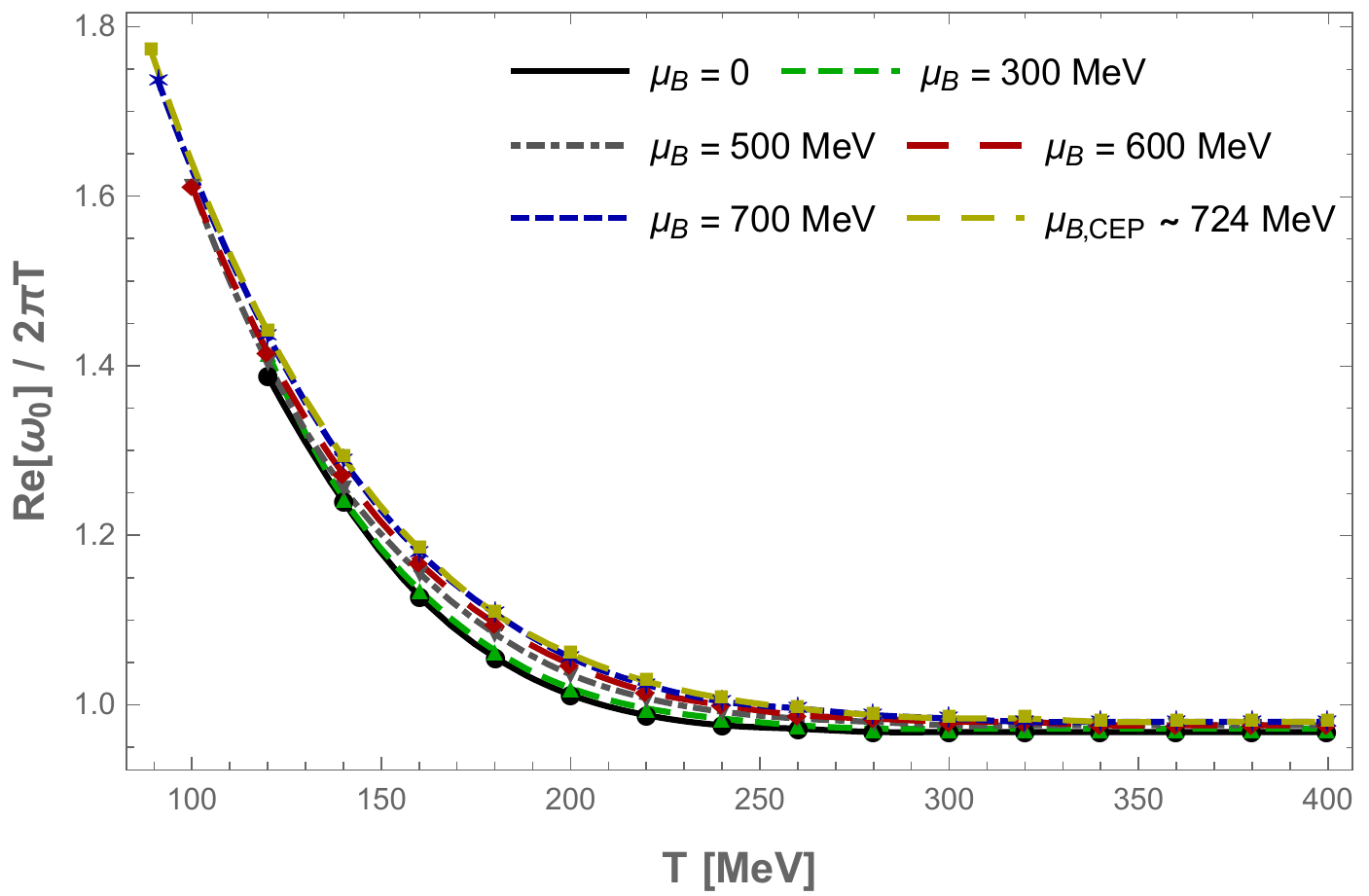} 
\end{tabular}
\begin{tabular}{c}
\includegraphics[width=0.45\textwidth]{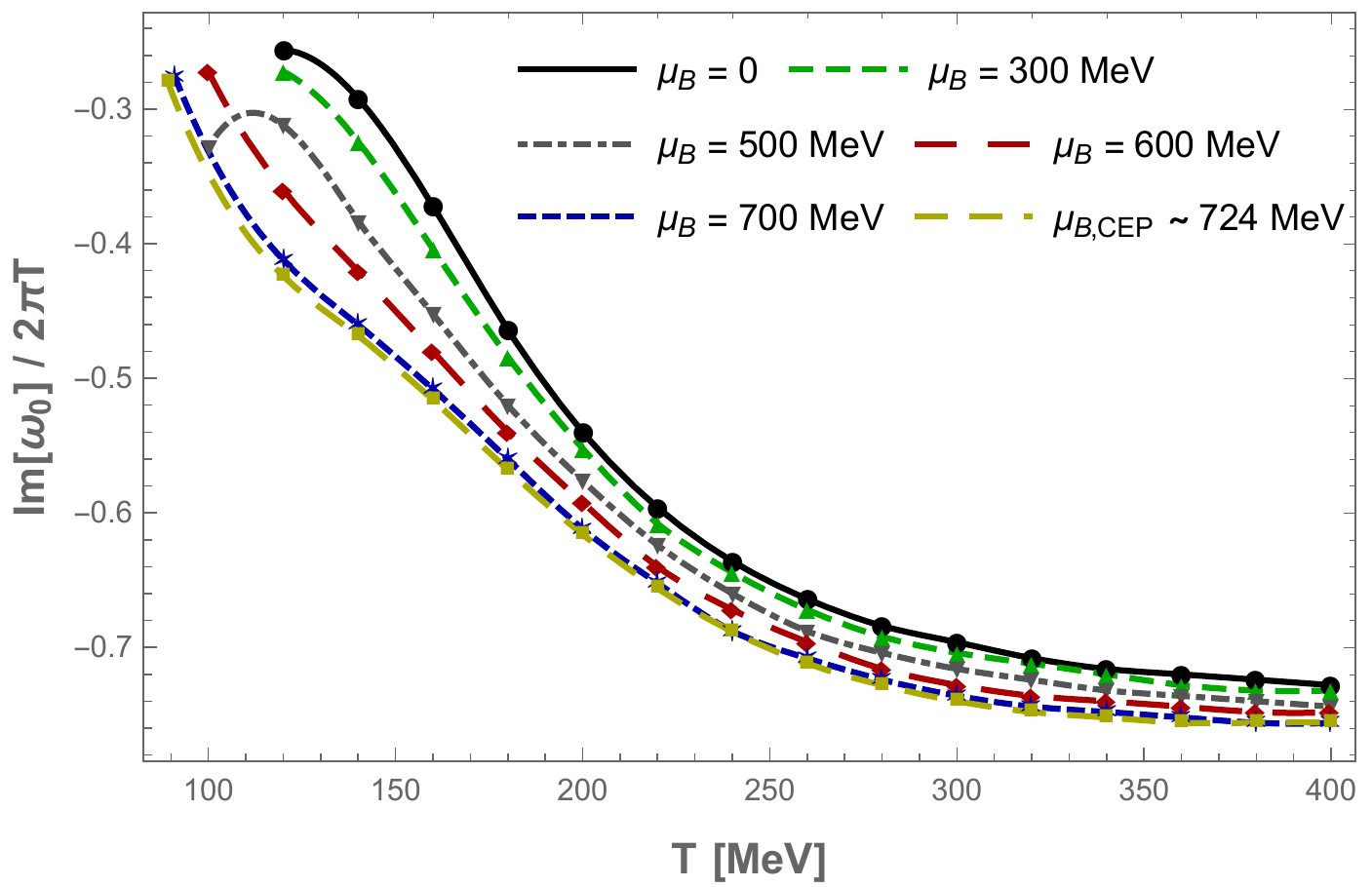} 
\end{tabular}
\begin{tabular}{c}
\includegraphics[width=0.45\textwidth]{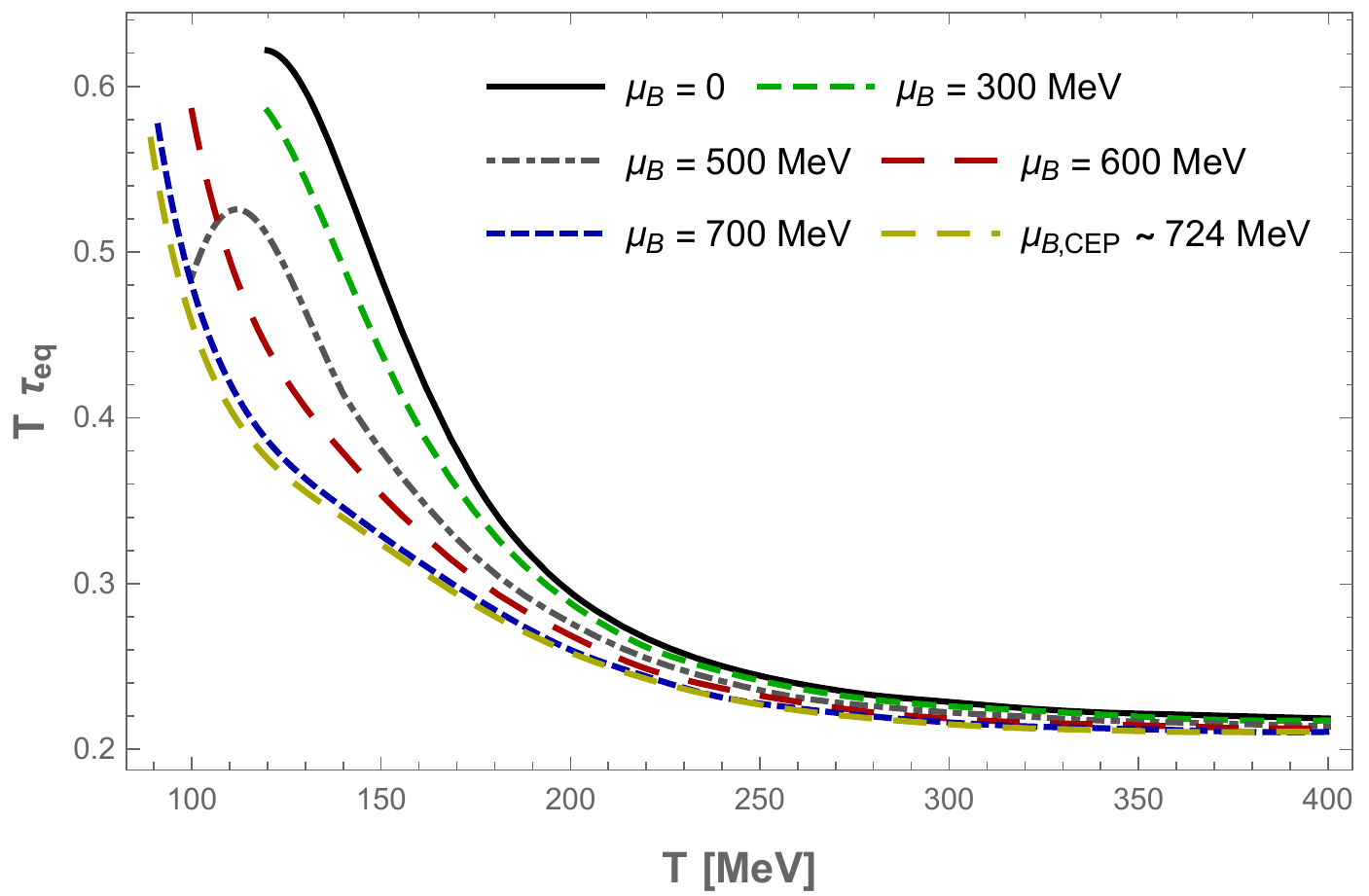} 
\end{tabular}
\begin{tabular}{c}
\includegraphics[width=0.45\textwidth]{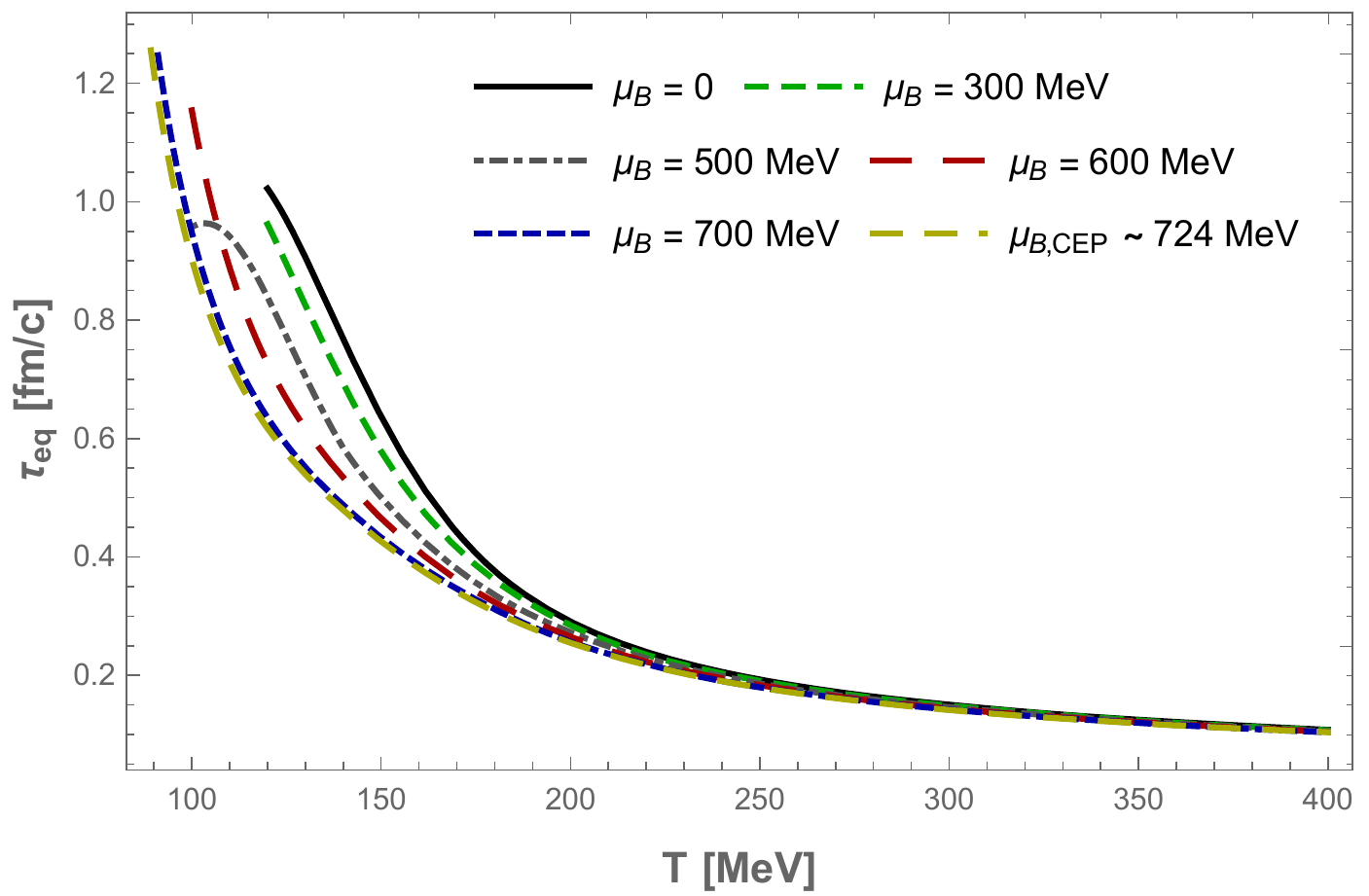} 
\end{tabular}
\end{center}
\caption{{\small (Color online) Absolute value of the real part of the lowest QNM in the $SO(3)$ singlet channel (top left), imaginary part of the lowest QNM (top right), the dimensionless combination given by the temperature times the equilibration time (bottom left), and the equilibration time measured in units of fm/c (bottom right) as functions of temperature for different values of the baryon chemical potential. In the upper panels we plot both the calculated points and the interpolated curves between them.}
\label{fig:1et}}
\end{figure*}

In Fig.\ \ref{fig:1et} we display the results for the lowest non-hydrodynamical QNM of the singlet channel and the corresponding equilibration times. As discussed in Refs.\ \cite{DeWolfe:2011ts,Janik:2016btb,Critelli:2017euk}, the $\mathcal{S}$-perturbation inherits the same ultraviolet asymptotics of the background dilaton field and, therefore, the ultraviolet conformal behavior of the QNM's in the dilaton channel depends on the scaling dimension $\Delta$ of the gauge theory operator dual to the dilaton. In the case of the top-down conformal EMD model with finite chemical potential and a critical point investigated in Ref.\ \cite{Critelli:2017euk} $\Delta = 2$, while in the present phenomenological EMD model at finite baryon density $\Delta\approx 2.73$ \cite{Critelli:2017oub}. Consequently, the conformal behavior attained in the ultraviolet by QNM's in the dilaton channels of both EMD models are different.

One notes from Fig.\ \ref{fig:1et} that the equilibration time in the dilaton channel generally decreases with increasing $\mu_B$ at fixed $T$. Also from this figure one can read some reference values for the equilibration time in the singlet channel, e.g. at $(T,\mu_B)\sim\{(400,0),(145,0),(89,724)\}$ MeV the corresponding equilibration times are, respectively, $\tau_{\textrm{eq}}\sim\{0.11,0.70,1.26\}$ fm/c, with the last point corresponding to the CEP. By comparison with the other two channels, the singlet channel gives the longest equilibration time of the medium for the first point above.

\section{Conclusions and outlook}
\label{sec:conclusion}

In this work we investigated the homogeneous limit of the lowest non-hydrodynamic QNM's of the EMD model of Ref.\ \cite{Critelli:2017oub} at finite temperature and baryon density up to the critical region. We then computed the associated equilibration times in the $SO(3)$ rotationally invariant quintuplet, triplet, and singlet channels. The equilibration times in the different channels come closer to each other at high temperatures, although being well separated at the critical point. In most cases, these equilibration times decrease with increasing baryon chemical potential while keeping temperature fixed.

Some considerations on dynamic universality classes \cite{Hohenberg:1977ym} are in order here. As discussed in Ref. \cite{Natsuume:2010bs}, in holographic models defined in the large $N_c$ limit, one expects the dynamical universality class to be of type B due to the $N_c^2$ suppression of convective transport compared to diffusive transport. This is different from what is expected to take place in $N_c=3$ QCD, whose dynamical universality class was argued to be of type H in Ref. \cite{Son:2004iv}. In type H theories the convective transport is dominant near criticality since it diverges with the correlation length. On the other hand, within the landscape of large $N_c$ theories, where holographic models belong to, convective transport is a subleading effect in $N_c$ and cannot be seen in this limit. In particular, for type B theories the shear and bulk viscosities, and also the conductivities remain finite at the critical point. This is exactly what happens in the EMD model (see for instance the bottom plot in Fig. \ref{fig:3et_zoom} with the result for the baryon conductivity). This is interesting in its own since, for instance, the divergence of the shear viscosity at the critical point in type H theories indicates that a hydrodynamic description in such theories must break down at least close to the critical region. On the other hand, this is not the case for holographic models, where $\eta/s=1/4\pi$. And at least for the characteristic equilibration times of spatially homogeneous QNM's in the EMD model, we have concluded that no divergences appear at criticality.

Possible comparisons with critical slowing down \cite{Stephanov:2017ghc} would require an analysis of an inhomogeneous setting, which is beyond the scope of the present work. However, we draw the attention to the fact that recently in Ref. \cite{Critelli:2018osu}, by analysing the Bjorken flow \cite{Bjorken:1982qr} of a conformal top-down holographic model at finite temperature and chemical potential, we concluded that the hydrodynamization time of that system significantly increases as one approaches the critical point of the model. It will be interesting to investigate in the near future the Bjorken flow of the present EMD model.

We note that even though the universality classes of QCD and large $N_c$ theories may be different, since the QGP produced in heavy ion collisions has a very small size and a very short lifetime the system cannot access the divergence of the correlation length. This raises the question to which extent type B and type H behaviors can be actually distinguished under experimental conditions, as originally inquired in Ref. \cite{Natsuume:2010bs}.

We must also remark that, besides the aforementioned points, the homogeneous QNM's evaluated in the present work are not expected to give a precise description of characteristic equilibration times of the QGP under the conditions actually realized in heavy ion collisions since the latter corresponds to a medium that rapidly expands in spatial directions and, in the present analysis, we considered a non-expanding homogeneous plasma which is slightly out-of-equilibrium.


The first sequel of the present work which we intend to pursue in the near future comprises an investigation of the far-from-equilibrium homogeneous isotropization dynamics of the EMD model. This would allow us to check whether the QNM's obtained here in the $SO(3)$ quintuplet, triplet, and singlet channels indeed describe the late time behavior of the pressure anisotropy, baryon charge density, and the scalar condensate, respectively, of the full nonlinear system of equations. Moreover, in this setting we shall also be able to evaluate the homogeneous isotropization and thermalization times of the EMD system near the critical region.

Future projects also comprise numerical simulations in the EMD model of more realistic expanding scenarios involving for instance the holographic boost-invariant Bjorken flow \cite{Chesler:2009cy,Heller:2011ju,Jankowski:2014lna,Romatschke:2017vte,Spalinski:2017mel,Critelli:2018osu} and also collisions of holographic shock waves \cite{Chesler:2010bi,Casalderrey-Solana:2013aba,vanderSchee:2013pia,Chesler:2015bba,Chesler:2016ceu,Attems:2016tby,Casalderrey-Solana:2016xfq,Attems:2017zam}. The latter may provide new insight into the complex interplay between the early time non-equilibrium dynamics and the late time hydrodynamic behavior of the strongly coupled baryon rich QGP near a critical point, which could be formed in upcoming low energy heavy ion collisions at RHIC and FAIR.

Another interesting question concerns the QNM's of the anisotropic EMD model at finite temperature and magnetic field proposed in Ref.\ \cite{Finazzo:2016mhm}, whose thermodynamics is in excellent agreement with lattice QCD results. This setup, in particular, may be of great relevance to high energy peripheral heavy ion collisions where very intense magnetic fields of order $eB\lesssim 0.3$ GeV$^2$ may be produced at the earliest stages of such non-central collisions \cite{Skokov:2009qp}.

\begin{figure*}
\begin{center}
\begin{tabular}{c}
\includegraphics[width=0.48\textwidth]{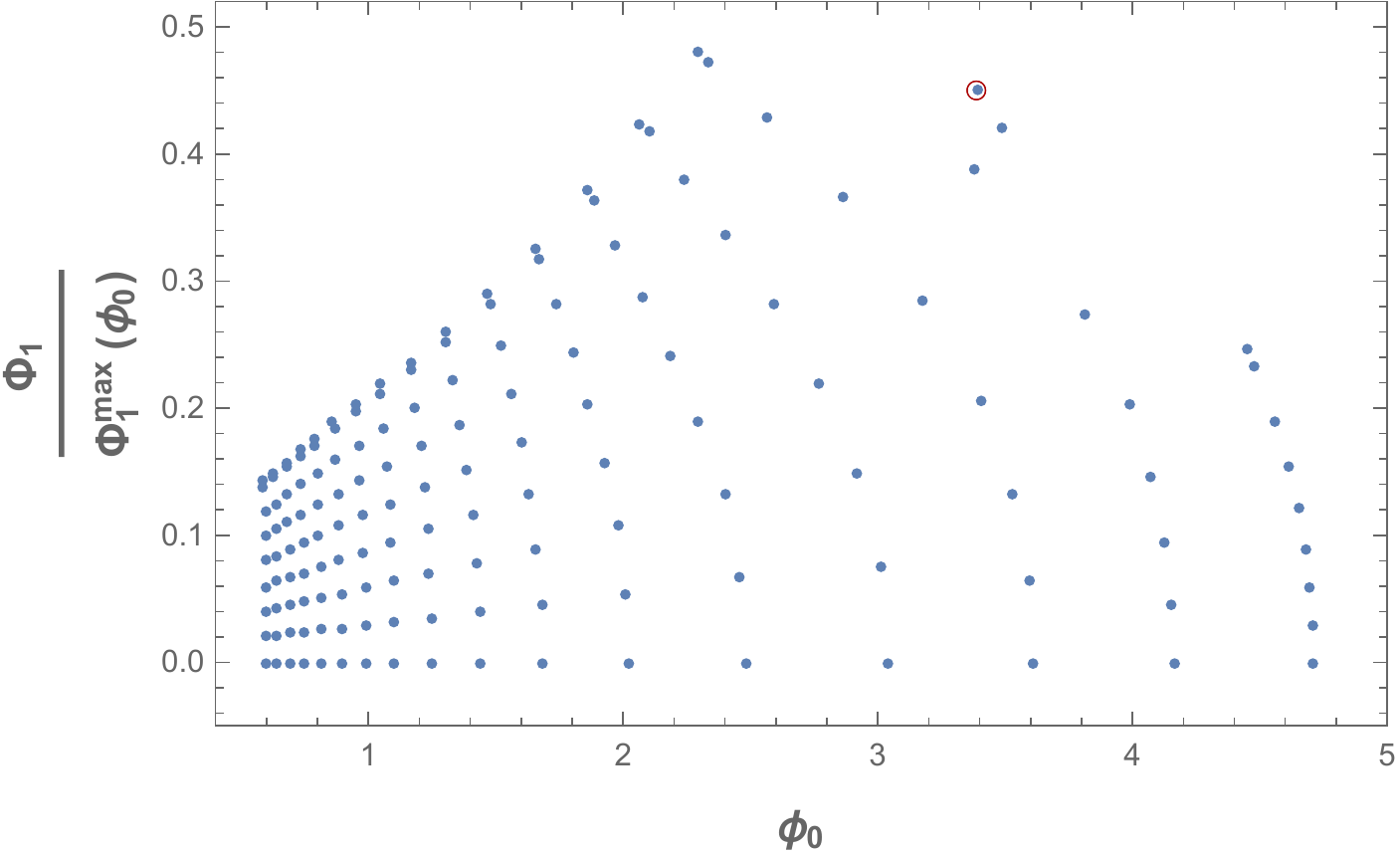} 
\end{tabular}
\begin{tabular}{c}
\includegraphics[width=0.45\textwidth]{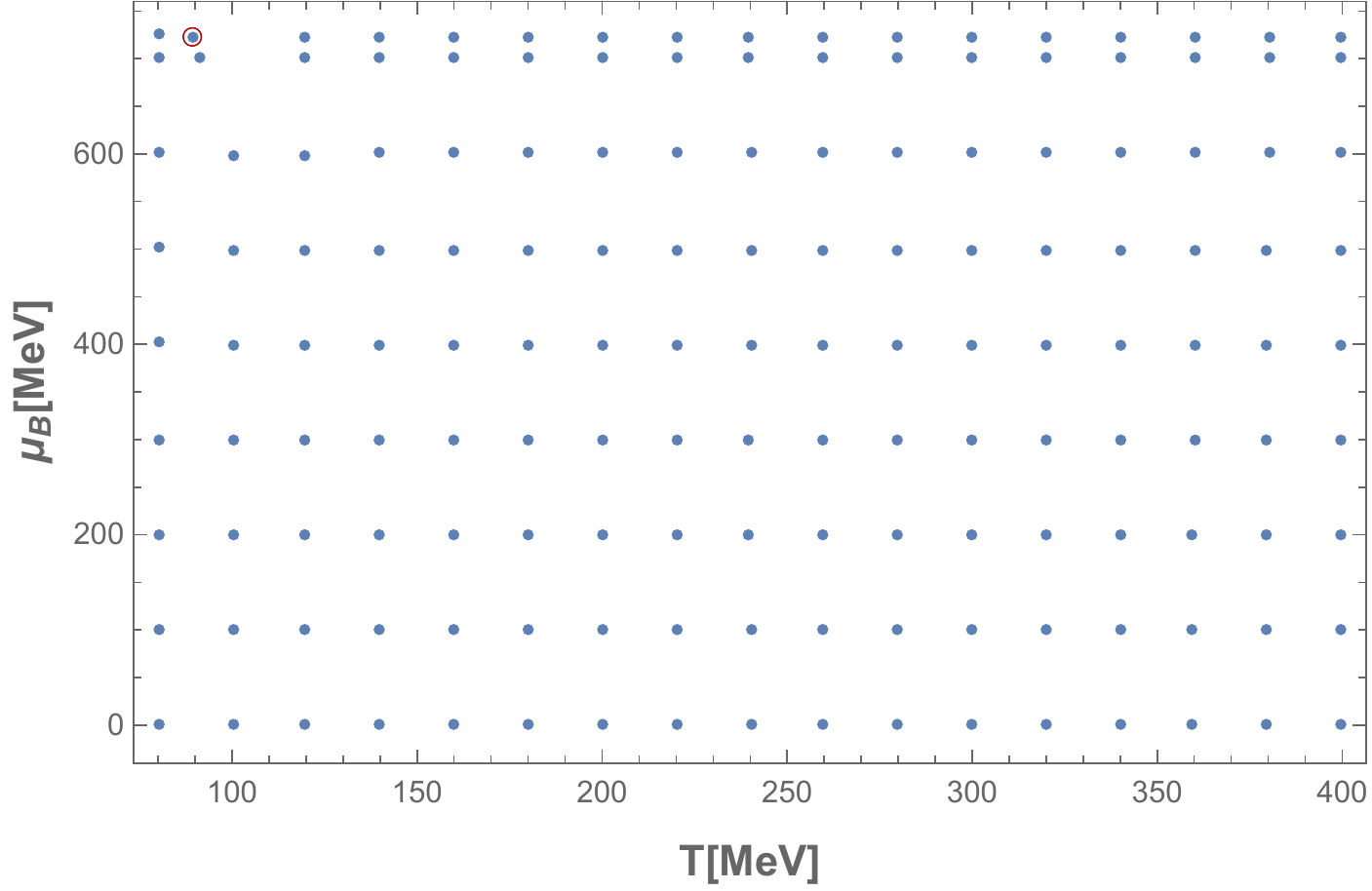} 
\end{tabular}
\end{center}
\caption{{\small (Color online) Grid of backgrounds generated on the plane of initial conditions (left) and in the $(T,\mu_B)$ plane (right). The background corresponding to the critical point in the phase diagram is highlighted by the red circle in both panels.}
\label{fig:grid}}
\end{figure*}

\begin{acknowledgments}
We thank I.~Portillo for the plots with the latest lattice QCD results on $\chi_6$ and $\chi_8$. R.R. acknowledges financial support by Funda\c c\~ao Norte Riograndense de Pesquisa e Cultura (FUNPEC). R.C. was supported by Funda\c c\~ao de Amparo \`a Pesquisa do Estado de S\~ao Paulo (FAPESP) grant no. 2016/09263-2. J.N. thanks FAPESP for financial support under grants no. 2015/50266-2 and  2017/05685-2 and also Conselho Nacional de Desenvolvimento Cient\'ifico e Tecnol\'ogico (CNPq).
\end{acknowledgments}

\appendix
\section{Further details about the numerical work}
\label{sec:num}

In this appendix we give some further details about the numerical evaluation of the non-hydrodynamic QNM's of the EMD model carried out in the present work.

The grid of numerical black hole backgrounds we used to compute the QNM's as functions of $(T,\mu_B)$ is shown in Fig.\ \ref{fig:grid}. This grid is almost regular in the $(T,\mu_B)$ plane, except for some ``missing points'' (e.g., $(T,\mu_B)=(100,700)$ MeV) which were not generated in the irregular shaped original background grid\footnote{By constructing a grid with regular shape in the plane of initial conditions $(\phi_0,\Phi_1)$ we obtain a grid with irregular shape in the physical $(T,\mu_B)$ plane and vice-versa, see e.g. Fig.\ 8 of Ref.\ \cite{Critelli:2017oub}.} \cite{Critelli:2017oub} due to numerical difficulties we face in solving the coupled set of differential equations of motion for the EMD fields with some initial conditions $(\phi_0,\Phi_1)$. We took a dense grid of points in this original irregular shaped grid and applied a filtering process upon it to eliminate unstable and metastable solutions close to the phase transition region corresponding to the CEP where competing branches of black hole solutions exist. We selected the stable solutions by looking at the backgrounds with maximum entropy between the competing branches. Using these stable backgrounds we then managed to invert the map $(\phi_0,\Phi_1)\to(T,\mu_B)$ (which could not be inverted before the filtering process due to the fact that this map is not 1-to-1 in the presence of competing branches of black hole solutions) and constructed the regular shaped grid displayed in Fig.\ \ref{fig:grid}. We further applied the Jacobian test of Ref.\ \cite{DeWolfe:2010he} to confirm that no unstable solutions were present after the inversion described above.

A final observation concerns the $\omega$-grid used in the $SO(3)$ singlet channel. For some backgrounds at low temperatures we were not able to identify the isolated clusters discussed in Section \ref{sec:5et} via the shooting technique. Therefore, no points were calculated in these cases, which is the reason why some of the curves in Fig. \ref{fig:1et} end at different values of $T$ coming from higher temperatures.

\bibliographystyle{apsrev4-1}
\bibliography{BH_Bib} 

\begin{thebibliography}{97}%
\makeatletter
\providecommand \@ifxundefined [1]{%
 \@ifx{#1\undefined}
}%
\providecommand \@ifnum [1]{%
 \ifnum #1\expandafter \@firstoftwo
 \else \expandafter \@secondoftwo
 \fi
}%
\providecommand \@ifx [1]{%
 \ifx #1\expandafter \@firstoftwo
 \else \expandafter \@secondoftwo
 \fi
}%
\providecommand \natexlab [1]{#1}%
\providecommand \enquote  [1]{``#1''}%
\providecommand \bibnamefont  [1]{#1}%
\providecommand \bibfnamefont [1]{#1}%
\providecommand \citenamefont [1]{#1}%
\providecommand \href@noop [0]{\@secondoftwo}%
\providecommand \href [0]{\begingroup \@sanitize@url \@href}%
\providecommand \@href[1]{\@@startlink{#1}\@@href}%
\providecommand \@@href[1]{\endgroup#1\@@endlink}%
\providecommand \@sanitize@url [0]{\catcode `\\12\catcode `\$12\catcode
  `\&12\catcode `\#12\catcode `\^12\catcode `\_12\catcode `\%12\relax}%
\providecommand \@@startlink[1]{}%
\providecommand \@@endlink[0]{}%
\providecommand \url  [0]{\begingroup\@sanitize@url \@url }%
\providecommand \@url [1]{\endgroup\@href {#1}{\urlprefix }}%
\providecommand \urlprefix  [0]{URL }%
\providecommand \Eprint [0]{\href }%
\providecommand \doibase [0]{http://dx.doi.org/}%
\providecommand \selectlanguage [0]{\@gobble}%
\providecommand \bibinfo  [0]{\@secondoftwo}%
\providecommand \bibfield  [0]{\@secondoftwo}%
\providecommand \translation [1]{[#1]}%
\providecommand \BibitemOpen [0]{}%
\providecommand \bibitemStop [0]{}%
\providecommand \bibitemNoStop [0]{.\EOS\space}%
\providecommand \EOS [0]{\spacefactor3000\relax}%
\providecommand \BibitemShut  [1]{\csname bibitem#1\endcsname}%
\let\auto@bib@innerbib\@empty
\bibitem [{\citenamefont {Arsene}\ \emph {et~al.}(2005)\citenamefont {Arsene}
  \emph {et~al.}}]{Arsene:2004fa}%
  \BibitemOpen
  \bibfield  {author} {\bibinfo {author} {\bibfnamefont {I.}~\bibnamefont
  {Arsene}} \emph {et~al.} (\bibinfo {collaboration} {BRAHMS}),\ }\href
  {\doibase 10.1016/j.nuclphysa.2005.02.130} {\bibfield  {journal} {\bibinfo
  {journal} {Nucl. Phys.}\ }\textbf {\bibinfo {volume} {A757}},\ \bibinfo
  {pages} {1} (\bibinfo {year} {2005})},\ \Eprint
  {http://arxiv.org/abs/nucl-ex/0410020} {arXiv:nucl-ex/0410020 [nucl-ex]}
  \BibitemShut {NoStop}%
\bibitem [{\citenamefont {Adcox}\ \emph {et~al.}(2005)\citenamefont {Adcox}
  \emph {et~al.}}]{Adcox:2004mh}%
  \BibitemOpen
  \bibfield  {author} {\bibinfo {author} {\bibfnamefont {K.}~\bibnamefont
  {Adcox}} \emph {et~al.} (\bibinfo {collaboration} {PHENIX}),\ }\href
  {\doibase 10.1016/j.nuclphysa.2005.03.086} {\bibfield  {journal} {\bibinfo
  {journal} {Nucl. Phys.}\ }\textbf {\bibinfo {volume} {A757}},\ \bibinfo
  {pages} {184} (\bibinfo {year} {2005})},\ \Eprint
  {http://arxiv.org/abs/nucl-ex/0410003} {arXiv:nucl-ex/0410003 [nucl-ex]}
  \BibitemShut {NoStop}%
\bibitem [{\citenamefont {Back}\ \emph {et~al.}(2005)\citenamefont {Back} \emph
  {et~al.}}]{Back:2004je}%
  \BibitemOpen
  \bibfield  {author} {\bibinfo {author} {\bibfnamefont {B.~B.}\ \bibnamefont
  {Back}} \emph {et~al.},\ }\href {\doibase 10.1016/j.nuclphysa.2005.03.084}
  {\bibfield  {journal} {\bibinfo  {journal} {Nucl. Phys.}\ }\textbf {\bibinfo
  {volume} {A757}},\ \bibinfo {pages} {28} (\bibinfo {year} {2005})},\ \Eprint
  {http://arxiv.org/abs/nucl-ex/0410022} {arXiv:nucl-ex/0410022 [nucl-ex]}
  \BibitemShut {NoStop}%
\bibitem [{\citenamefont {Adams}\ \emph {et~al.}(2005)\citenamefont {Adams}
  \emph {et~al.}}]{Adams:2005dq}%
  \BibitemOpen
  \bibfield  {author} {\bibinfo {author} {\bibfnamefont {J.}~\bibnamefont
  {Adams}} \emph {et~al.} (\bibinfo {collaboration} {STAR}),\ }\href {\doibase
  10.1016/j.nuclphysa.2005.03.085} {\bibfield  {journal} {\bibinfo  {journal}
  {Nucl. Phys.}\ }\textbf {\bibinfo {volume} {A757}},\ \bibinfo {pages} {102}
  (\bibinfo {year} {2005})},\ \Eprint {http://arxiv.org/abs/nucl-ex/0501009}
  {arXiv:nucl-ex/0501009 [nucl-ex]} \BibitemShut {NoStop}%
\bibitem [{\citenamefont {Aad}\ \emph {et~al.}(2013)\citenamefont {Aad} \emph
  {et~al.}}]{Aad:2013xma}%
  \BibitemOpen
  \bibfield  {author} {\bibinfo {author} {\bibfnamefont {G.}~\bibnamefont
  {Aad}} \emph {et~al.} (\bibinfo {collaboration} {ATLAS}),\ }\href {\doibase
  10.1007/JHEP11(2013)183} {\bibfield  {journal} {\bibinfo  {journal} {JHEP}\
  }\textbf {\bibinfo {volume} {11}},\ \bibinfo {pages} {183} (\bibinfo {year}
  {2013})},\ \Eprint {http://arxiv.org/abs/1305.2942} {arXiv:1305.2942
  [hep-ex]} \BibitemShut {NoStop}%
\bibitem [{\citenamefont {Gyulassy}\ and\ \citenamefont
  {McLerran}(2005)}]{Gyulassy:2004zy}%
  \BibitemOpen
  \bibfield  {author} {\bibinfo {author} {\bibfnamefont {M.}~\bibnamefont
  {Gyulassy}}\ and\ \bibinfo {author} {\bibfnamefont {L.}~\bibnamefont
  {McLerran}},\ }\bibfield  {booktitle} {\emph {\bibinfo {booktitle} {{Quark
  gluon plasma. New discoveries at RHIC: A case of strongly interacting quark
  gluon plasma. Proceedings, RBRC Workshop, Brookhaven, Upton, USA, May 14-15,
  2004}}},\ }\href {\doibase 10.1016/j.nuclphysa.2004.10.034} {\bibfield
  {journal} {\bibinfo  {journal} {Nucl. Phys.}\ }\textbf {\bibinfo {volume}
  {A750}},\ \bibinfo {pages} {30} (\bibinfo {year} {2005})},\ \Eprint
  {http://arxiv.org/abs/nucl-th/0405013} {arXiv:nucl-th/0405013 [nucl-th]}
  \BibitemShut {NoStop}%
\bibitem [{\citenamefont {Heinz}\ and\ \citenamefont
  {Snellings}(2013)}]{Heinz:2013th}%
  \BibitemOpen
  \bibfield  {author} {\bibinfo {author} {\bibfnamefont {U.}~\bibnamefont
  {Heinz}}\ and\ \bibinfo {author} {\bibfnamefont {R.}~\bibnamefont
  {Snellings}},\ }\href {\doibase 10.1146/annurev-nucl-102212-170540}
  {\bibfield  {journal} {\bibinfo  {journal} {Ann. Rev. Nucl. Part. Sci.}\
  }\textbf {\bibinfo {volume} {63}},\ \bibinfo {pages} {123} (\bibinfo {year}
  {2013})},\ \Eprint {http://arxiv.org/abs/1301.2826} {arXiv:1301.2826
  [nucl-th]} \BibitemShut {NoStop}%
\bibitem [{\citenamefont {Shuryak}(2017)}]{Shuryak:2014zxa}%
  \BibitemOpen
  \bibfield  {author} {\bibinfo {author} {\bibfnamefont {E.}~\bibnamefont
  {Shuryak}},\ }\href {\doibase 10.1103/RevModPhys.89.035001} {\bibfield
  {journal} {\bibinfo  {journal} {Rev. Mod. Phys.}\ }\textbf {\bibinfo {volume}
  {89}},\ \bibinfo {pages} {035001} (\bibinfo {year} {2017})},\ \Eprint
  {http://arxiv.org/abs/1412.8393} {arXiv:1412.8393 [hep-ph]} \BibitemShut
  {NoStop}%
\bibitem [{\citenamefont {Aoki}\ \emph {et~al.}(2006)\citenamefont {Aoki},
  \citenamefont {Endrodi}, \citenamefont {Fodor}, \citenamefont {Katz},\ and\
  \citenamefont {Szabo}}]{Aoki:2006we}%
  \BibitemOpen
  \bibfield  {author} {\bibinfo {author} {\bibfnamefont {Y.}~\bibnamefont
  {Aoki}}, \bibinfo {author} {\bibfnamefont {G.}~\bibnamefont {Endrodi}},
  \bibinfo {author} {\bibfnamefont {Z.}~\bibnamefont {Fodor}}, \bibinfo
  {author} {\bibfnamefont {S.~D.}\ \bibnamefont {Katz}}, \ and\ \bibinfo
  {author} {\bibfnamefont {K.~K.}\ \bibnamefont {Szabo}},\ }\href {\doibase
  10.1038/nature05120} {\bibfield  {journal} {\bibinfo  {journal} {Nature}\
  }\textbf {\bibinfo {volume} {443}},\ \bibinfo {pages} {675} (\bibinfo {year}
  {2006})},\ \Eprint {http://arxiv.org/abs/hep-lat/0611014}
  {arXiv:hep-lat/0611014 [hep-lat]} \BibitemShut {NoStop}%
\bibitem [{\citenamefont {Borsanyi}\ \emph {et~al.}(2016)\citenamefont
  {Borsanyi} \emph {et~al.}}]{Borsanyi:2016ksw}%
  \BibitemOpen
  \bibfield  {author} {\bibinfo {author} {\bibfnamefont {S.}~\bibnamefont
  {Borsanyi}} \emph {et~al.},\ }\href {\doibase 10.1038/nature20115} {\bibfield
   {journal} {\bibinfo  {journal} {Nature}\ }\textbf {\bibinfo {volume}
  {539}},\ \bibinfo {pages} {69} (\bibinfo {year} {2016})},\ \Eprint
  {http://arxiv.org/abs/1606.07494} {arXiv:1606.07494 [hep-lat]} \BibitemShut
  {NoStop}%
\bibitem [{\citenamefont {Schenke}\ \emph {et~al.}(2011)\citenamefont
  {Schenke}, \citenamefont {Jeon},\ and\ \citenamefont
  {Gale}}]{Schenke:2010rr}%
  \BibitemOpen
  \bibfield  {author} {\bibinfo {author} {\bibfnamefont {B.}~\bibnamefont
  {Schenke}}, \bibinfo {author} {\bibfnamefont {S.}~\bibnamefont {Jeon}}, \
  and\ \bibinfo {author} {\bibfnamefont {C.}~\bibnamefont {Gale}},\ }\href
  {\doibase 10.1103/PhysRevLett.106.042301} {\bibfield  {journal} {\bibinfo
  {journal} {Phys. Rev. Lett.}\ }\textbf {\bibinfo {volume} {106}},\ \bibinfo
  {pages} {042301} (\bibinfo {year} {2011})},\ \Eprint
  {http://arxiv.org/abs/1009.3244} {arXiv:1009.3244 [hep-ph]} \BibitemShut
  {NoStop}%
\bibitem [{\citenamefont {Song}\ \emph {et~al.}(2011)\citenamefont {Song},
  \citenamefont {Bass}, \citenamefont {Heinz}, \citenamefont {Hirano},\ and\
  \citenamefont {Shen}}]{Song:2010mg}%
  \BibitemOpen
  \bibfield  {author} {\bibinfo {author} {\bibfnamefont {H.}~\bibnamefont
  {Song}}, \bibinfo {author} {\bibfnamefont {S.~A.}\ \bibnamefont {Bass}},
  \bibinfo {author} {\bibfnamefont {U.}~\bibnamefont {Heinz}}, \bibinfo
  {author} {\bibfnamefont {T.}~\bibnamefont {Hirano}}, \ and\ \bibinfo {author}
  {\bibfnamefont {C.}~\bibnamefont {Shen}},\ }\href {\doibase
  10.1103/PhysRevLett.106.192301, 10.1103/PhysRevLett.109.139904} {\bibfield
  {journal} {\bibinfo  {journal} {Phys. Rev. Lett.}\ }\textbf {\bibinfo
  {volume} {106}},\ \bibinfo {pages} {192301} (\bibinfo {year} {2011})},\
  \bibinfo {note} {[Erratum: Phys. Rev. Lett.109,139904(2012)]},\ \Eprint
  {http://arxiv.org/abs/1011.2783} {arXiv:1011.2783 [nucl-th]} \BibitemShut
  {NoStop}%
\bibitem [{\citenamefont {Gale}\ \emph {et~al.}(2013)\citenamefont {Gale},
  \citenamefont {Jeon}, \citenamefont {Schenke}, \citenamefont {Tribedy},\ and\
  \citenamefont {Venugopalan}}]{Gale:2012rq}%
  \BibitemOpen
  \bibfield  {author} {\bibinfo {author} {\bibfnamefont {C.}~\bibnamefont
  {Gale}}, \bibinfo {author} {\bibfnamefont {S.}~\bibnamefont {Jeon}}, \bibinfo
  {author} {\bibfnamefont {B.}~\bibnamefont {Schenke}}, \bibinfo {author}
  {\bibfnamefont {P.}~\bibnamefont {Tribedy}}, \ and\ \bibinfo {author}
  {\bibfnamefont {R.}~\bibnamefont {Venugopalan}},\ }\href {\doibase
  10.1103/PhysRevLett.110.012302} {\bibfield  {journal} {\bibinfo  {journal}
  {Phys. Rev. Lett.}\ }\textbf {\bibinfo {volume} {110}},\ \bibinfo {pages}
  {012302} (\bibinfo {year} {2013})},\ \Eprint {http://arxiv.org/abs/1209.6330}
  {arXiv:1209.6330 [nucl-th]} \BibitemShut {NoStop}%
\bibitem [{\citenamefont {Noronha-Hostler}\ \emph {et~al.}(2013)\citenamefont
  {Noronha-Hostler}, \citenamefont {Denicol}, \citenamefont {Noronha},
  \citenamefont {Andrade},\ and\ \citenamefont
  {Grassi}}]{Noronha-Hostler:2013gga}%
  \BibitemOpen
  \bibfield  {author} {\bibinfo {author} {\bibfnamefont {J.}~\bibnamefont
  {Noronha-Hostler}}, \bibinfo {author} {\bibfnamefont {G.~S.}\ \bibnamefont
  {Denicol}}, \bibinfo {author} {\bibfnamefont {J.}~\bibnamefont {Noronha}},
  \bibinfo {author} {\bibfnamefont {R.~P.~G.}\ \bibnamefont {Andrade}}, \ and\
  \bibinfo {author} {\bibfnamefont {F.}~\bibnamefont {Grassi}},\ }\href
  {\doibase 10.1103/PhysRevC.88.044916} {\bibfield  {journal} {\bibinfo
  {journal} {Phys. Rev.}\ }\textbf {\bibinfo {volume} {C88}},\ \bibinfo {pages}
  {044916} (\bibinfo {year} {2013})},\ \Eprint {http://arxiv.org/abs/1305.1981}
  {arXiv:1305.1981 [nucl-th]} \BibitemShut {NoStop}%
\bibitem [{\citenamefont {Noronha-Hostler}\ \emph {et~al.}(2014)\citenamefont
  {Noronha-Hostler}, \citenamefont {Noronha},\ and\ \citenamefont
  {Grassi}}]{Noronha-Hostler:2014dqa}%
  \BibitemOpen
  \bibfield  {author} {\bibinfo {author} {\bibfnamefont {J.}~\bibnamefont
  {Noronha-Hostler}}, \bibinfo {author} {\bibfnamefont {J.}~\bibnamefont
  {Noronha}}, \ and\ \bibinfo {author} {\bibfnamefont {F.}~\bibnamefont
  {Grassi}},\ }\href {\doibase 10.1103/PhysRevC.90.034907} {\bibfield
  {journal} {\bibinfo  {journal} {Phys. Rev.}\ }\textbf {\bibinfo {volume}
  {C90}},\ \bibinfo {pages} {034907} (\bibinfo {year} {2014})},\ \Eprint
  {http://arxiv.org/abs/1406.3333} {arXiv:1406.3333 [nucl-th]} \BibitemShut
  {NoStop}%
\bibitem [{\citenamefont {Ryu}\ \emph {et~al.}(2015)\citenamefont {Ryu},
  \citenamefont {Paquet}, \citenamefont {Shen}, \citenamefont {Denicol},
  \citenamefont {Schenke}, \citenamefont {Jeon},\ and\ \citenamefont
  {Gale}}]{Ryu:2015vwa}%
  \BibitemOpen
  \bibfield  {author} {\bibinfo {author} {\bibfnamefont {S.}~\bibnamefont
  {Ryu}}, \bibinfo {author} {\bibfnamefont {J.~F.}\ \bibnamefont {Paquet}},
  \bibinfo {author} {\bibfnamefont {C.}~\bibnamefont {Shen}}, \bibinfo {author}
  {\bibfnamefont {G.~S.}\ \bibnamefont {Denicol}}, \bibinfo {author}
  {\bibfnamefont {B.}~\bibnamefont {Schenke}}, \bibinfo {author} {\bibfnamefont
  {S.}~\bibnamefont {Jeon}}, \ and\ \bibinfo {author} {\bibfnamefont
  {C.}~\bibnamefont {Gale}},\ }\href {\doibase 10.1103/PhysRevLett.115.132301}
  {\bibfield  {journal} {\bibinfo  {journal} {Phys. Rev. Lett.}\ }\textbf
  {\bibinfo {volume} {115}},\ \bibinfo {pages} {132301} (\bibinfo {year}
  {2015})},\ \Eprint {http://arxiv.org/abs/1502.01675} {arXiv:1502.01675
  [nucl-th]} \BibitemShut {NoStop}%
\bibitem [{\citenamefont {Bernhard}\ \emph {et~al.}(2016)\citenamefont
  {Bernhard}, \citenamefont {Moreland}, \citenamefont {Bass}, \citenamefont
  {Liu},\ and\ \citenamefont {Heinz}}]{Bernhard:2016tnd}%
  \BibitemOpen
  \bibfield  {author} {\bibinfo {author} {\bibfnamefont {J.~E.}\ \bibnamefont
  {Bernhard}}, \bibinfo {author} {\bibfnamefont {J.~S.}\ \bibnamefont
  {Moreland}}, \bibinfo {author} {\bibfnamefont {S.~A.}\ \bibnamefont {Bass}},
  \bibinfo {author} {\bibfnamefont {J.}~\bibnamefont {Liu}}, \ and\ \bibinfo
  {author} {\bibfnamefont {U.}~\bibnamefont {Heinz}},\ }\href {\doibase
  10.1103/PhysRevC.94.024907} {\bibfield  {journal} {\bibinfo  {journal} {Phys.
  Rev.}\ }\textbf {\bibinfo {volume} {C94}},\ \bibinfo {pages} {024907}
  (\bibinfo {year} {2016})},\ \Eprint {http://arxiv.org/abs/1605.03954}
  {arXiv:1605.03954 [nucl-th]} \BibitemShut {NoStop}%
\bibitem [{\citenamefont {Bernhard}\ \emph {et~al.}(2017)\citenamefont
  {Bernhard}, \citenamefont {Moreland},\ and\ \citenamefont
  {Bass}}]{Bernhard:2017vql}%
  \BibitemOpen
  \bibfield  {author} {\bibinfo {author} {\bibfnamefont {J.~E.}\ \bibnamefont
  {Bernhard}}, \bibinfo {author} {\bibfnamefont {J.~S.}\ \bibnamefont
  {Moreland}}, \ and\ \bibinfo {author} {\bibfnamefont {S.~A.}\ \bibnamefont
  {Bass}},\ }\bibfield  {booktitle} {\emph {\bibinfo {booktitle} {{Proceedings,
  26th International Conference on Ultrarelativistic Nucleus-Nucleus Collisions
  (Quark Matter 2017): Chicago,Illinois, USA, February 6-11, 2017}}},\ }\href
  {\doibase 10.1016/j.nuclphysa.2017.05.037} {\bibfield  {journal} {\bibinfo
  {journal} {Nucl. Phys.}\ }\textbf {\bibinfo {volume} {A967}},\ \bibinfo
  {pages} {293} (\bibinfo {year} {2017})},\ \Eprint
  {http://arxiv.org/abs/1704.04462} {arXiv:1704.04462 [nucl-th]} \BibitemShut
  {NoStop}%
\bibitem [{\citenamefont {Bernhard}(4 19)}]{Bernhard:2018hnz}%
  \BibitemOpen
  \bibfield  {author} {\bibinfo {author} {\bibfnamefont {J.~E.}\ \bibnamefont
  {Bernhard}},\ }\emph {\bibinfo {title} {{Bayesian parameter estimation for
  relativistic heavy-ion collisions}}},\ \href
  {https://inspirehep.net/record/1669345/files/1804.06469.pdf} {Ph.D. thesis},\
  \bibinfo  {school} {Duke U.} (\bibinfo {year} {2018-04-19}),\ \Eprint
  {http://arxiv.org/abs/1804.06469} {arXiv:1804.06469 [nucl-th]} \BibitemShut
  {NoStop}%
\bibitem [{\citenamefont {Ryu}\ \emph {et~al.}(2018)\citenamefont {Ryu},
  \citenamefont {Paquet}, \citenamefont {Shen}, \citenamefont {Denicol},
  \citenamefont {Schenke}, \citenamefont {Jeon},\ and\ \citenamefont
  {Gale}}]{Ryu:2017qzn}%
  \BibitemOpen
  \bibfield  {author} {\bibinfo {author} {\bibfnamefont {S.}~\bibnamefont
  {Ryu}}, \bibinfo {author} {\bibfnamefont {J.-F.}\ \bibnamefont {Paquet}},
  \bibinfo {author} {\bibfnamefont {C.}~\bibnamefont {Shen}}, \bibinfo {author}
  {\bibfnamefont {G.}~\bibnamefont {Denicol}}, \bibinfo {author} {\bibfnamefont
  {B.}~\bibnamefont {Schenke}}, \bibinfo {author} {\bibfnamefont
  {S.}~\bibnamefont {Jeon}}, \ and\ \bibinfo {author} {\bibfnamefont
  {C.}~\bibnamefont {Gale}},\ }\href {\doibase 10.1103/PhysRevC.97.034910}
  {\bibfield  {journal} {\bibinfo  {journal} {Phys. Rev.}\ }\textbf {\bibinfo
  {volume} {C97}},\ \bibinfo {pages} {034910} (\bibinfo {year} {2018})},\
  \Eprint {http://arxiv.org/abs/1704.04216} {arXiv:1704.04216 [nucl-th]}
  \BibitemShut {NoStop}%
\bibitem [{\citenamefont {Pratt}\ \emph {et~al.}(2015)\citenamefont {Pratt},
  \citenamefont {Sangaline}, \citenamefont {Sorensen},\ and\ \citenamefont
  {Wang}}]{Pratt:2015zsa}%
  \BibitemOpen
  \bibfield  {author} {\bibinfo {author} {\bibfnamefont {S.}~\bibnamefont
  {Pratt}}, \bibinfo {author} {\bibfnamefont {E.}~\bibnamefont {Sangaline}},
  \bibinfo {author} {\bibfnamefont {P.}~\bibnamefont {Sorensen}}, \ and\
  \bibinfo {author} {\bibfnamefont {H.}~\bibnamefont {Wang}},\ }\href {\doibase
  10.1103/PhysRevLett.114.202301} {\bibfield  {journal} {\bibinfo  {journal}
  {Phys. Rev. Lett.}\ }\textbf {\bibinfo {volume} {114}},\ \bibinfo {pages}
  {202301} (\bibinfo {year} {2015})},\ \Eprint
  {http://arxiv.org/abs/1501.04042} {arXiv:1501.04042 [nucl-th]} \BibitemShut
  {NoStop}%
\bibitem [{\citenamefont {Monnai}\ and\ \citenamefont
  {Ollitrault}(2017)}]{Monnai:2017cbv}%
  \BibitemOpen
  \bibfield  {author} {\bibinfo {author} {\bibfnamefont {A.}~\bibnamefont
  {Monnai}}\ and\ \bibinfo {author} {\bibfnamefont {J.-Y.}\ \bibnamefont
  {Ollitrault}},\ }\href {\doibase 10.1103/PhysRevC.96.044902} {\bibfield
  {journal} {\bibinfo  {journal} {Phys. Rev.}\ }\textbf {\bibinfo {volume}
  {C96}},\ \bibinfo {pages} {044902} (\bibinfo {year} {2017})},\ \Eprint
  {http://arxiv.org/abs/1707.08466} {arXiv:1707.08466 [nucl-th]} \BibitemShut
  {NoStop}%
\bibitem [{\citenamefont {Alba}\ \emph {et~al.}(2017)\citenamefont {Alba},
  \citenamefont {Mantovani~Sarti}, \citenamefont {Noronha}, \citenamefont
  {Noronha-Hostler}, \citenamefont {Parotto}, \citenamefont {Vazquez},\ and\
  \citenamefont {Ratti}}]{Alba:2017hhe}%
  \BibitemOpen
  \bibfield  {author} {\bibinfo {author} {\bibfnamefont {P.}~\bibnamefont
  {Alba}}, \bibinfo {author} {\bibfnamefont {V.}~\bibnamefont
  {Mantovani~Sarti}}, \bibinfo {author} {\bibfnamefont {J.}~\bibnamefont
  {Noronha}}, \bibinfo {author} {\bibfnamefont {J.}~\bibnamefont
  {Noronha-Hostler}}, \bibinfo {author} {\bibfnamefont {P.}~\bibnamefont
  {Parotto}}, \bibinfo {author} {\bibfnamefont {I.~P.}\ \bibnamefont
  {Vazquez}}, \ and\ \bibinfo {author} {\bibfnamefont {C.}~\bibnamefont
  {Ratti}},\ }\href@noop {} {\  (\bibinfo {year} {2017})},\ \Eprint
  {http://arxiv.org/abs/1711.05207} {arXiv:1711.05207 [nucl-th]} \BibitemShut
  {NoStop}%
\bibitem [{\citenamefont {Critelli}\ \emph
  {et~al.}(2017{\natexlab{a}})\citenamefont {Critelli}, \citenamefont
  {Noronha}, \citenamefont {Noronha-Hostler}, \citenamefont {Portillo},
  \citenamefont {Ratti},\ and\ \citenamefont {Rougemont}}]{Critelli:2017oub}%
  \BibitemOpen
  \bibfield  {author} {\bibinfo {author} {\bibfnamefont {R.}~\bibnamefont
  {Critelli}}, \bibinfo {author} {\bibfnamefont {J.}~\bibnamefont {Noronha}},
  \bibinfo {author} {\bibfnamefont {J.}~\bibnamefont {Noronha-Hostler}},
  \bibinfo {author} {\bibfnamefont {I.}~\bibnamefont {Portillo}}, \bibinfo
  {author} {\bibfnamefont {C.}~\bibnamefont {Ratti}}, \ and\ \bibinfo {author}
  {\bibfnamefont {R.}~\bibnamefont {Rougemont}},\ }\href {\doibase
  10.1103/PhysRevD.96.096026} {\bibfield  {journal} {\bibinfo  {journal} {Phys.
  Rev.}\ }\textbf {\bibinfo {volume} {D96}},\ \bibinfo {pages} {096026}
  (\bibinfo {year} {2017}{\natexlab{a}})},\ \Eprint
  {http://arxiv.org/abs/1706.00455} {arXiv:1706.00455 [nucl-th]} \BibitemShut
  {NoStop}%
\bibitem [{\citenamefont {Stephanov}\ \emph {et~al.}(1998)\citenamefont
  {Stephanov}, \citenamefont {Rajagopal},\ and\ \citenamefont
  {Shuryak}}]{Stephanov:1998dy}%
  \BibitemOpen
  \bibfield  {author} {\bibinfo {author} {\bibfnamefont {M.~A.}\ \bibnamefont
  {Stephanov}}, \bibinfo {author} {\bibfnamefont {K.}~\bibnamefont
  {Rajagopal}}, \ and\ \bibinfo {author} {\bibfnamefont {E.~V.}\ \bibnamefont
  {Shuryak}},\ }\href {\doibase 10.1103/PhysRevLett.81.4816} {\bibfield
  {journal} {\bibinfo  {journal} {Phys. Rev. Lett.}\ }\textbf {\bibinfo
  {volume} {81}},\ \bibinfo {pages} {4816} (\bibinfo {year} {1998})},\ \Eprint
  {http://arxiv.org/abs/hep-ph/9806219} {arXiv:hep-ph/9806219 [hep-ph]}
  \BibitemShut {NoStop}%
\bibitem [{\citenamefont {Stephanov}\ \emph {et~al.}(1999)\citenamefont
  {Stephanov}, \citenamefont {Rajagopal},\ and\ \citenamefont
  {Shuryak}}]{Stephanov:1999zu}%
  \BibitemOpen
  \bibfield  {author} {\bibinfo {author} {\bibfnamefont {M.~A.}\ \bibnamefont
  {Stephanov}}, \bibinfo {author} {\bibfnamefont {K.}~\bibnamefont
  {Rajagopal}}, \ and\ \bibinfo {author} {\bibfnamefont {E.~V.}\ \bibnamefont
  {Shuryak}},\ }\href {\doibase 10.1103/PhysRevD.60.114028} {\bibfield
  {journal} {\bibinfo  {journal} {Phys. Rev.}\ }\textbf {\bibinfo {volume}
  {D60}},\ \bibinfo {pages} {114028} (\bibinfo {year} {1999})},\ \Eprint
  {http://arxiv.org/abs/hep-ph/9903292} {arXiv:hep-ph/9903292 [hep-ph]}
  \BibitemShut {NoStop}%
\bibitem [{\citenamefont {Rischke}(2004)}]{Rischke:2003mt}%
  \BibitemOpen
  \bibfield  {author} {\bibinfo {author} {\bibfnamefont {D.~H.}\ \bibnamefont
  {Rischke}},\ }\href {\doibase 10.1016/j.ppnp.2003.09.002} {\bibfield
  {journal} {\bibinfo  {journal} {Prog. Part. Nucl. Phys.}\ }\textbf {\bibinfo
  {volume} {52}},\ \bibinfo {pages} {197} (\bibinfo {year} {2004})},\ \Eprint
  {http://arxiv.org/abs/nucl-th/0305030} {arXiv:nucl-th/0305030 [nucl-th]}
  \BibitemShut {NoStop}%
\bibitem [{\citenamefont {Stephanov}(2011)}]{Stephanov:2011pb}%
  \BibitemOpen
  \bibfield  {author} {\bibinfo {author} {\bibfnamefont {M.~A.}\ \bibnamefont
  {Stephanov}},\ }\href {\doibase 10.1103/PhysRevLett.107.052301} {\bibfield
  {journal} {\bibinfo  {journal} {Phys. Rev. Lett.}\ }\textbf {\bibinfo
  {volume} {107}},\ \bibinfo {pages} {052301} (\bibinfo {year} {2011})},\
  \Eprint {http://arxiv.org/abs/1104.1627} {arXiv:1104.1627 [hep-ph]}
  \BibitemShut {NoStop}%
\bibitem [{\citenamefont {Rougemont}\ \emph {et~al.}(2017)\citenamefont
  {Rougemont}, \citenamefont {Critelli}, \citenamefont {Noronha-Hostler},
  \citenamefont {Noronha},\ and\ \citenamefont {Ratti}}]{Rougemont:2017tlu}%
  \BibitemOpen
  \bibfield  {author} {\bibinfo {author} {\bibfnamefont {R.}~\bibnamefont
  {Rougemont}}, \bibinfo {author} {\bibfnamefont {R.}~\bibnamefont {Critelli}},
  \bibinfo {author} {\bibfnamefont {J.}~\bibnamefont {Noronha-Hostler}},
  \bibinfo {author} {\bibfnamefont {J.}~\bibnamefont {Noronha}}, \ and\
  \bibinfo {author} {\bibfnamefont {C.}~\bibnamefont {Ratti}},\ }\href
  {\doibase 10.1103/PhysRevD.96.014032} {\bibfield  {journal} {\bibinfo
  {journal} {Phys. Rev.}\ }\textbf {\bibinfo {volume} {D96}},\ \bibinfo {pages}
  {014032} (\bibinfo {year} {2017})},\ \Eprint
  {http://arxiv.org/abs/1704.05558} {arXiv:1704.05558 [hep-ph]} \BibitemShut
  {NoStop}%
\bibitem [{\citenamefont {Rougemont}\ \emph
  {et~al.}(2016{\natexlab{a}})\citenamefont {Rougemont}, \citenamefont
  {Ficnar}, \citenamefont {Finazzo},\ and\ \citenamefont
  {Noronha}}]{Rougemont:2015wca}%
  \BibitemOpen
  \bibfield  {author} {\bibinfo {author} {\bibfnamefont {R.}~\bibnamefont
  {Rougemont}}, \bibinfo {author} {\bibfnamefont {A.}~\bibnamefont {Ficnar}},
  \bibinfo {author} {\bibfnamefont {S.}~\bibnamefont {Finazzo}}, \ and\
  \bibinfo {author} {\bibfnamefont {J.}~\bibnamefont {Noronha}},\ }\href
  {\doibase 10.1007/JHEP04(2016)102} {\bibfield  {journal} {\bibinfo  {journal}
  {JHEP}\ }\textbf {\bibinfo {volume} {04}},\ \bibinfo {pages} {102} (\bibinfo
  {year} {2016}{\natexlab{a}})},\ \Eprint {http://arxiv.org/abs/1507.06556}
  {arXiv:1507.06556 [hep-th]} \BibitemShut {NoStop}%
\bibitem [{\citenamefont {Bazavov}\ \emph {et~al.}(2017)\citenamefont {Bazavov}
  \emph {et~al.}}]{Bazavov:2017dus}%
  \BibitemOpen
  \bibfield  {author} {\bibinfo {author} {\bibfnamefont {A.}~\bibnamefont
  {Bazavov}} \emph {et~al.},\ }\href {\doibase 10.1103/PhysRevD.95.054504}
  {\bibfield  {journal} {\bibinfo  {journal} {Phys. Rev.}\ }\textbf {\bibinfo
  {volume} {D95}},\ \bibinfo {pages} {054504} (\bibinfo {year} {2017})},\
  \Eprint {http://arxiv.org/abs/1701.04325} {arXiv:1701.04325 [hep-lat]}
  \BibitemShut {NoStop}%
\bibitem [{\citenamefont {Bellwied}\ \emph {et~al.}(2015)\citenamefont
  {Bellwied}, \citenamefont {Borsanyi}, \citenamefont {Fodor}, \citenamefont
  {Katz}, \citenamefont {Pasztor}, \citenamefont {Ratti},\ and\ \citenamefont
  {Szabo}}]{Bellwied:2015lba}%
  \BibitemOpen
  \bibfield  {author} {\bibinfo {author} {\bibfnamefont {R.}~\bibnamefont
  {Bellwied}}, \bibinfo {author} {\bibfnamefont {S.}~\bibnamefont {Borsanyi}},
  \bibinfo {author} {\bibfnamefont {Z.}~\bibnamefont {Fodor}}, \bibinfo
  {author} {\bibfnamefont {S.~D.}\ \bibnamefont {Katz}}, \bibinfo {author}
  {\bibfnamefont {A.}~\bibnamefont {Pasztor}}, \bibinfo {author} {\bibfnamefont
  {C.}~\bibnamefont {Ratti}}, \ and\ \bibinfo {author} {\bibfnamefont {K.~K.}\
  \bibnamefont {Szabo}},\ }\href {\doibase 10.1103/PhysRevD.92.114505}
  {\bibfield  {journal} {\bibinfo  {journal} {Phys. Rev.}\ }\textbf {\bibinfo
  {volume} {D92}},\ \bibinfo {pages} {114505} (\bibinfo {year} {2015})},\
  \Eprint {http://arxiv.org/abs/1507.04627} {arXiv:1507.04627 [hep-lat]}
  \BibitemShut {NoStop}%
\bibitem [{\citenamefont {Kovtun}\ \emph {et~al.}(2005)\citenamefont {Kovtun},
  \citenamefont {Son},\ and\ \citenamefont {Starinets}}]{Kovtun:2004de}%
  \BibitemOpen
  \bibfield  {author} {\bibinfo {author} {\bibfnamefont {P.}~\bibnamefont
  {Kovtun}}, \bibinfo {author} {\bibfnamefont {D.~T.}\ \bibnamefont {Son}}, \
  and\ \bibinfo {author} {\bibfnamefont {A.~O.}\ \bibnamefont {Starinets}},\
  }\href {\doibase 10.1103/PhysRevLett.94.111601} {\bibfield  {journal}
  {\bibinfo  {journal} {Phys. Rev. Lett.}\ }\textbf {\bibinfo {volume} {94}},\
  \bibinfo {pages} {111601} (\bibinfo {year} {2005})},\ \Eprint
  {http://arxiv.org/abs/hep-th/0405231} {arXiv:hep-th/0405231 [hep-th]}
  \BibitemShut {NoStop}%
\bibitem [{\citenamefont {Karsch}\ \emph {et~al.}(2008)\citenamefont {Karsch},
  \citenamefont {Kharzeev},\ and\ \citenamefont {Tuchin}}]{Karsch:2007jc}%
  \BibitemOpen
  \bibfield  {author} {\bibinfo {author} {\bibfnamefont {F.}~\bibnamefont
  {Karsch}}, \bibinfo {author} {\bibfnamefont {D.}~\bibnamefont {Kharzeev}}, \
  and\ \bibinfo {author} {\bibfnamefont {K.}~\bibnamefont {Tuchin}},\ }\href
  {\doibase 10.1016/j.physletb.2008.01.080} {\bibfield  {journal} {\bibinfo
  {journal} {Phys. Lett.}\ }\textbf {\bibinfo {volume} {B663}},\ \bibinfo
  {pages} {217} (\bibinfo {year} {2008})},\ \Eprint
  {http://arxiv.org/abs/0711.0914} {arXiv:0711.0914 [hep-ph]} \BibitemShut
  {NoStop}%
\bibitem [{\citenamefont {Noronha-Hostler}\ \emph {et~al.}(2009)\citenamefont
  {Noronha-Hostler}, \citenamefont {Noronha},\ and\ \citenamefont
  {Greiner}}]{NoronhaHostler:2008ju}%
  \BibitemOpen
  \bibfield  {author} {\bibinfo {author} {\bibfnamefont {J.}~\bibnamefont
  {Noronha-Hostler}}, \bibinfo {author} {\bibfnamefont {J.}~\bibnamefont
  {Noronha}}, \ and\ \bibinfo {author} {\bibfnamefont {C.}~\bibnamefont
  {Greiner}},\ }\href {\doibase 10.1103/PhysRevLett.103.172302} {\bibfield
  {journal} {\bibinfo  {journal} {Phys. Rev. Lett.}\ }\textbf {\bibinfo
  {volume} {103}},\ \bibinfo {pages} {172302} (\bibinfo {year} {2009})},\
  \Eprint {http://arxiv.org/abs/0811.1571} {arXiv:0811.1571 [nucl-th]}
  \BibitemShut {NoStop}%
\bibitem [{\citenamefont {Meehan}(2017)}]{Meehan:2017cum}%
  \BibitemOpen
  \bibfield  {author} {\bibinfo {author} {\bibfnamefont {K.}~\bibnamefont
  {Meehan}} (\bibinfo {collaboration} {STAR}),\ }\bibfield  {booktitle} {\emph
  {\bibinfo {booktitle} {{Proceedings, 26th International Conference on
  Ultra-relativistic Nucleus-Nucleus Collisions (Quark Matter 2017): Chicago,
  Illinois, USA, February 5-11, 2017}}},\ }\href {\doibase
  10.1016/j.nuclphysa.2017.06.007} {\bibfield  {journal} {\bibinfo  {journal}
  {Nucl. Phys.}\ }\textbf {\bibinfo {volume} {A967}},\ \bibinfo {pages} {808}
  (\bibinfo {year} {2017})},\ \Eprint {http://arxiv.org/abs/1704.06342}
  {arXiv:1704.06342 [nucl-ex]} \BibitemShut {NoStop}%
\bibitem [{\citenamefont {Ablyazimov}\ \emph {et~al.}(2017)\citenamefont
  {Ablyazimov} \emph {et~al.}}]{Ablyazimov:2017guv}%
  \BibitemOpen
  \bibfield  {author} {\bibinfo {author} {\bibfnamefont {T.}~\bibnamefont
  {Ablyazimov}} \emph {et~al.} (\bibinfo {collaboration} {CBM}),\ }\href
  {\doibase 10.1140/epja/i2017-12248-y} {\bibfield  {journal} {\bibinfo
  {journal} {Eur. Phys. J.}\ }\textbf {\bibinfo {volume} {A53}},\ \bibinfo
  {pages} {60} (\bibinfo {year} {2017})},\ \Eprint
  {http://arxiv.org/abs/1607.01487} {arXiv:1607.01487 [nucl-ex]} \BibitemShut
  {NoStop}%
\bibitem [{\citenamefont {Kovtun}\ and\ \citenamefont
  {Starinets}(2005)}]{Kovtun:2005ev}%
  \BibitemOpen
  \bibfield  {author} {\bibinfo {author} {\bibfnamefont {P.~K.}\ \bibnamefont
  {Kovtun}}\ and\ \bibinfo {author} {\bibfnamefont {A.~O.}\ \bibnamefont
  {Starinets}},\ }\href {\doibase 10.1103/PhysRevD.72.086009} {\bibfield
  {journal} {\bibinfo  {journal} {Phys. Rev.}\ }\textbf {\bibinfo {volume}
  {D72}},\ \bibinfo {pages} {086009} (\bibinfo {year} {2005})},\ \Eprint
  {http://arxiv.org/abs/hep-th/0506184} {arXiv:hep-th/0506184 [hep-th]}
  \BibitemShut {NoStop}%
\bibitem [{\citenamefont {Berti}\ \emph {et~al.}(2009)\citenamefont {Berti},
  \citenamefont {Cardoso},\ and\ \citenamefont {Starinets}}]{Berti:2009kk}%
  \BibitemOpen
  \bibfield  {author} {\bibinfo {author} {\bibfnamefont {E.}~\bibnamefont
  {Berti}}, \bibinfo {author} {\bibfnamefont {V.}~\bibnamefont {Cardoso}}, \
  and\ \bibinfo {author} {\bibfnamefont {A.~O.}\ \bibnamefont {Starinets}},\
  }\href {\doibase 10.1088/0264-9381/26/16/163001} {\bibfield  {journal}
  {\bibinfo  {journal} {Class. Quant. Grav.}\ }\textbf {\bibinfo {volume}
  {26}},\ \bibinfo {pages} {163001} (\bibinfo {year} {2009})},\ \Eprint
  {http://arxiv.org/abs/0905.2975} {arXiv:0905.2975 [gr-qc]} \BibitemShut
  {NoStop}%
\bibitem [{\citenamefont {Horowitz}\ and\ \citenamefont
  {Hubeny}(2000)}]{Horowitz:1999jd}%
  \BibitemOpen
  \bibfield  {author} {\bibinfo {author} {\bibfnamefont {G.~T.}\ \bibnamefont
  {Horowitz}}\ and\ \bibinfo {author} {\bibfnamefont {V.~E.}\ \bibnamefont
  {Hubeny}},\ }\href {\doibase 10.1103/PhysRevD.62.024027} {\bibfield
  {journal} {\bibinfo  {journal} {Phys. Rev.}\ }\textbf {\bibinfo {volume}
  {D62}},\ \bibinfo {pages} {024027} (\bibinfo {year} {2000})},\ \Eprint
  {http://arxiv.org/abs/hep-th/9909056} {arXiv:hep-th/9909056 [hep-th]}
  \BibitemShut {NoStop}%
\bibitem [{\citenamefont {Critelli}\ \emph
  {et~al.}(2017{\natexlab{b}})\citenamefont {Critelli}, \citenamefont
  {Rougemont},\ and\ \citenamefont {Noronha}}]{Critelli:2017euk}%
  \BibitemOpen
  \bibfield  {author} {\bibinfo {author} {\bibfnamefont {R.}~\bibnamefont
  {Critelli}}, \bibinfo {author} {\bibfnamefont {R.}~\bibnamefont {Rougemont}},
  \ and\ \bibinfo {author} {\bibfnamefont {J.}~\bibnamefont {Noronha}},\ }\href
  {\doibase 10.1007/JHEP12(2017)029} {\bibfield  {journal} {\bibinfo  {journal}
  {JHEP}\ }\textbf {\bibinfo {volume} {12}},\ \bibinfo {pages} {029} (\bibinfo
  {year} {2017}{\natexlab{b}})},\ \Eprint {http://arxiv.org/abs/1709.03131}
  {arXiv:1709.03131 [hep-th]} \BibitemShut {NoStop}%
\bibitem [{\citenamefont {Hohenberg}\ and\ \citenamefont
  {Halperin}(1977)}]{Hohenberg:1977ym}%
  \BibitemOpen
  \bibfield  {author} {\bibinfo {author} {\bibfnamefont {P.~C.}\ \bibnamefont
  {Hohenberg}}\ and\ \bibinfo {author} {\bibfnamefont {B.~I.}\ \bibnamefont
  {Halperin}},\ }\href {\doibase 10.1103/RevModPhys.49.435} {\bibfield
  {journal} {\bibinfo  {journal} {Rev. Mod. Phys.}\ }\textbf {\bibinfo {volume}
  {49}},\ \bibinfo {pages} {435} (\bibinfo {year} {1977})}\BibitemShut
  {NoStop}%
\bibitem [{\citenamefont {Natsuume}\ and\ \citenamefont
  {Okamura}(2011)}]{Natsuume:2010bs}%
  \BibitemOpen
  \bibfield  {author} {\bibinfo {author} {\bibfnamefont {M.}~\bibnamefont
  {Natsuume}}\ and\ \bibinfo {author} {\bibfnamefont {T.}~\bibnamefont
  {Okamura}},\ }\href {\doibase 10.1103/PhysRevD.83.046008} {\bibfield
  {journal} {\bibinfo  {journal} {Phys. Rev.}\ }\textbf {\bibinfo {volume}
  {D83}},\ \bibinfo {pages} {046008} (\bibinfo {year} {2011})},\ \Eprint
  {http://arxiv.org/abs/1012.0575} {arXiv:1012.0575 [hep-th]} \BibitemShut
  {NoStop}%
\bibitem [{\citenamefont {Son}\ and\ \citenamefont
  {Stephanov}(2004)}]{Son:2004iv}%
  \BibitemOpen
  \bibfield  {author} {\bibinfo {author} {\bibfnamefont {D.~T.}\ \bibnamefont
  {Son}}\ and\ \bibinfo {author} {\bibfnamefont {M.~A.}\ \bibnamefont
  {Stephanov}},\ }\href {\doibase 10.1103/PhysRevD.70.056001} {\bibfield
  {journal} {\bibinfo  {journal} {Phys. Rev.}\ }\textbf {\bibinfo {volume}
  {D70}},\ \bibinfo {pages} {056001} (\bibinfo {year} {2004})},\ \Eprint
  {http://arxiv.org/abs/hep-ph/0401052} {arXiv:hep-ph/0401052 [hep-ph]}
  \BibitemShut {NoStop}%
\bibitem [{\citenamefont {Gubser}\ and\ \citenamefont
  {Nellore}(2008)}]{Gubser:2008ny}%
  \BibitemOpen
  \bibfield  {author} {\bibinfo {author} {\bibfnamefont {S.~S.}\ \bibnamefont
  {Gubser}}\ and\ \bibinfo {author} {\bibfnamefont {A.}~\bibnamefont
  {Nellore}},\ }\href {\doibase 10.1103/PhysRevD.78.086007} {\bibfield
  {journal} {\bibinfo  {journal} {Phys. Rev.}\ }\textbf {\bibinfo {volume}
  {D78}},\ \bibinfo {pages} {086007} (\bibinfo {year} {2008})},\ \Eprint
  {http://arxiv.org/abs/0804.0434} {arXiv:0804.0434 [hep-th]} \BibitemShut
  {NoStop}%
\bibitem [{\citenamefont {DeWolfe}\ \emph
  {et~al.}(2011{\natexlab{a}})\citenamefont {DeWolfe}, \citenamefont {Gubser},\
  and\ \citenamefont {Rosen}}]{DeWolfe:2010he}%
  \BibitemOpen
  \bibfield  {author} {\bibinfo {author} {\bibfnamefont {O.}~\bibnamefont
  {DeWolfe}}, \bibinfo {author} {\bibfnamefont {S.~S.}\ \bibnamefont {Gubser}},
  \ and\ \bibinfo {author} {\bibfnamefont {C.}~\bibnamefont {Rosen}},\ }\href
  {\doibase 10.1103/PhysRevD.83.086005} {\bibfield  {journal} {\bibinfo
  {journal} {Phys. Rev.}\ }\textbf {\bibinfo {volume} {D83}},\ \bibinfo {pages}
  {086005} (\bibinfo {year} {2011}{\natexlab{a}})},\ \Eprint
  {http://arxiv.org/abs/1012.1864} {arXiv:1012.1864 [hep-th]} \BibitemShut
  {NoStop}%
\bibitem [{\citenamefont {DeWolfe}\ \emph
  {et~al.}(2011{\natexlab{b}})\citenamefont {DeWolfe}, \citenamefont {Gubser},\
  and\ \citenamefont {Rosen}}]{DeWolfe:2011ts}%
  \BibitemOpen
  \bibfield  {author} {\bibinfo {author} {\bibfnamefont {O.}~\bibnamefont
  {DeWolfe}}, \bibinfo {author} {\bibfnamefont {S.~S.}\ \bibnamefont {Gubser}},
  \ and\ \bibinfo {author} {\bibfnamefont {C.}~\bibnamefont {Rosen}},\ }\href
  {\doibase 10.1103/PhysRevD.84.126014} {\bibfield  {journal} {\bibinfo
  {journal} {Phys. Rev.}\ }\textbf {\bibinfo {volume} {D84}},\ \bibinfo {pages}
  {126014} (\bibinfo {year} {2011}{\natexlab{b}})},\ \Eprint
  {http://arxiv.org/abs/1108.2029} {arXiv:1108.2029 [hep-th]} \BibitemShut
  {NoStop}%
\bibitem [{\citenamefont {Maldacena}(1999)}]{Maldacena:1997re}%
  \BibitemOpen
  \bibfield  {author} {\bibinfo {author} {\bibfnamefont {J.~M.}\ \bibnamefont
  {Maldacena}},\ }\href {\doibase 10.1023/A:1026654312961} {\bibfield
  {journal} {\bibinfo  {journal} {Int. J. Theor. Phys.}\ }\textbf {\bibinfo
  {volume} {38}},\ \bibinfo {pages} {1113} (\bibinfo {year} {1999})},\ \bibinfo
  {note} {[Adv. Theor. Math. Phys.2,231(1998)]},\ \Eprint
  {http://arxiv.org/abs/hep-th/9711200} {arXiv:hep-th/9711200 [hep-th]}
  \BibitemShut {NoStop}%
\bibitem [{\citenamefont {Gubser}\ \emph {et~al.}(1998)\citenamefont {Gubser},
  \citenamefont {Klebanov},\ and\ \citenamefont {Polyakov}}]{Gubser:1998bc}%
  \BibitemOpen
  \bibfield  {author} {\bibinfo {author} {\bibfnamefont {S.~S.}\ \bibnamefont
  {Gubser}}, \bibinfo {author} {\bibfnamefont {I.~R.}\ \bibnamefont
  {Klebanov}}, \ and\ \bibinfo {author} {\bibfnamefont {A.~M.}\ \bibnamefont
  {Polyakov}},\ }\href {\doibase 10.1016/S0370-2693(98)00377-3} {\bibfield
  {journal} {\bibinfo  {journal} {Phys. Lett.}\ }\textbf {\bibinfo {volume}
  {B428}},\ \bibinfo {pages} {105} (\bibinfo {year} {1998})},\ \Eprint
  {http://arxiv.org/abs/hep-th/9802109} {arXiv:hep-th/9802109 [hep-th]}
  \BibitemShut {NoStop}%
\bibitem [{\citenamefont {Witten}(1998{\natexlab{a}})}]{Witten:1998qj}%
  \BibitemOpen
  \bibfield  {author} {\bibinfo {author} {\bibfnamefont {E.}~\bibnamefont
  {Witten}},\ }\href@noop {} {\bibfield  {journal} {\bibinfo  {journal} {Adv.
  Theor. Math. Phys.}\ }\textbf {\bibinfo {volume} {2}},\ \bibinfo {pages}
  {253} (\bibinfo {year} {1998}{\natexlab{a}})},\ \Eprint
  {http://arxiv.org/abs/hep-th/9802150} {arXiv:hep-th/9802150 [hep-th]}
  \BibitemShut {NoStop}%
\bibitem [{\citenamefont {Witten}(1998{\natexlab{b}})}]{Witten:1998zw}%
  \BibitemOpen
  \bibfield  {author} {\bibinfo {author} {\bibfnamefont {E.}~\bibnamefont
  {Witten}},\ }\href@noop {} {\bibfield  {journal} {\bibinfo  {journal} {Adv.
  Theor. Math. Phys.}\ }\textbf {\bibinfo {volume} {2}},\ \bibinfo {pages}
  {505} (\bibinfo {year} {1998}{\natexlab{b}})},\ \Eprint
  {http://arxiv.org/abs/hep-th/9803131} {arXiv:hep-th/9803131 [hep-th]}
  \BibitemShut {NoStop}%
\bibitem [{\citenamefont {Gubser}\ \emph {et~al.}(2008)\citenamefont {Gubser},
  \citenamefont {Nellore}, \citenamefont {Pufu},\ and\ \citenamefont
  {Rocha}}]{Gubser:2008yx}%
  \BibitemOpen
  \bibfield  {author} {\bibinfo {author} {\bibfnamefont {S.~S.}\ \bibnamefont
  {Gubser}}, \bibinfo {author} {\bibfnamefont {A.}~\bibnamefont {Nellore}},
  \bibinfo {author} {\bibfnamefont {S.~S.}\ \bibnamefont {Pufu}}, \ and\
  \bibinfo {author} {\bibfnamefont {F.~D.}\ \bibnamefont {Rocha}},\ }\href
  {\doibase 10.1103/PhysRevLett.101.131601} {\bibfield  {journal} {\bibinfo
  {journal} {Phys. Rev. Lett.}\ }\textbf {\bibinfo {volume} {101}},\ \bibinfo
  {pages} {131601} (\bibinfo {year} {2008})},\ \Eprint
  {http://arxiv.org/abs/0804.1950} {arXiv:0804.1950 [hep-th]} \BibitemShut
  {NoStop}%
\bibitem [{\citenamefont {Finazzo}\ and\ \citenamefont
  {Noronha}(2014)}]{Finazzo:2013efa}%
  \BibitemOpen
  \bibfield  {author} {\bibinfo {author} {\bibfnamefont {S.~I.}\ \bibnamefont
  {Finazzo}}\ and\ \bibinfo {author} {\bibfnamefont {J.}~\bibnamefont
  {Noronha}},\ }\href {\doibase 10.1103/PhysRevD.89.106008} {\bibfield
  {journal} {\bibinfo  {journal} {Phys. Rev.}\ }\textbf {\bibinfo {volume}
  {D89}},\ \bibinfo {pages} {106008} (\bibinfo {year} {2014})},\ \Eprint
  {http://arxiv.org/abs/1311.6675} {arXiv:1311.6675 [hep-th]} \BibitemShut
  {NoStop}%
\bibitem [{\citenamefont {Finazzo}\ \emph {et~al.}(2015)\citenamefont
  {Finazzo}, \citenamefont {Rougemont}, \citenamefont {Marrochio},\ and\
  \citenamefont {Noronha}}]{Finazzo:2014cna}%
  \BibitemOpen
  \bibfield  {author} {\bibinfo {author} {\bibfnamefont {S.~I.}\ \bibnamefont
  {Finazzo}}, \bibinfo {author} {\bibfnamefont {R.}~\bibnamefont {Rougemont}},
  \bibinfo {author} {\bibfnamefont {H.}~\bibnamefont {Marrochio}}, \ and\
  \bibinfo {author} {\bibfnamefont {J.}~\bibnamefont {Noronha}},\ }\href
  {\doibase 10.1007/JHEP02(2015)051} {\bibfield  {journal} {\bibinfo  {journal}
  {JHEP}\ }\textbf {\bibinfo {volume} {02}},\ \bibinfo {pages} {051} (\bibinfo
  {year} {2015})},\ \Eprint {http://arxiv.org/abs/1412.2968} {arXiv:1412.2968
  [hep-ph]} \BibitemShut {NoStop}%
\bibitem [{\citenamefont {Rougemont}\ \emph {et~al.}(2015)\citenamefont
  {Rougemont}, \citenamefont {Noronha},\ and\ \citenamefont
  {Noronha-Hostler}}]{Rougemont:2015ona}%
  \BibitemOpen
  \bibfield  {author} {\bibinfo {author} {\bibfnamefont {R.}~\bibnamefont
  {Rougemont}}, \bibinfo {author} {\bibfnamefont {J.}~\bibnamefont {Noronha}},
  \ and\ \bibinfo {author} {\bibfnamefont {J.}~\bibnamefont
  {Noronha-Hostler}},\ }\href {\doibase 10.1103/PhysRevLett.115.202301}
  {\bibfield  {journal} {\bibinfo  {journal} {Phys. Rev. Lett.}\ }\textbf
  {\bibinfo {volume} {115}},\ \bibinfo {pages} {202301} (\bibinfo {year}
  {2015})},\ \Eprint {http://arxiv.org/abs/1507.06972} {arXiv:1507.06972
  [hep-ph]} \BibitemShut {NoStop}%
\bibitem [{\citenamefont {Finazzo}\ and\ \citenamefont
  {Rougemont}(2016)}]{Finazzo:2015xwa}%
  \BibitemOpen
  \bibfield  {author} {\bibinfo {author} {\bibfnamefont {S.~I.}\ \bibnamefont
  {Finazzo}}\ and\ \bibinfo {author} {\bibfnamefont {R.}~\bibnamefont
  {Rougemont}},\ }\href {\doibase 10.1103/PhysRevD.93.034017} {\bibfield
  {journal} {\bibinfo  {journal} {Phys. Rev.}\ }\textbf {\bibinfo {volume}
  {D93}},\ \bibinfo {pages} {034017} (\bibinfo {year} {2016})},\ \Eprint
  {http://arxiv.org/abs/1510.03321} {arXiv:1510.03321 [hep-ph]} \BibitemShut
  {NoStop}%
\bibitem [{\citenamefont {Rougemont}\ \emph
  {et~al.}(2016{\natexlab{b}})\citenamefont {Rougemont}, \citenamefont
  {Critelli},\ and\ \citenamefont {Noronha}}]{Rougemont:2015oea}%
  \BibitemOpen
  \bibfield  {author} {\bibinfo {author} {\bibfnamefont {R.}~\bibnamefont
  {Rougemont}}, \bibinfo {author} {\bibfnamefont {R.}~\bibnamefont {Critelli}},
  \ and\ \bibinfo {author} {\bibfnamefont {J.}~\bibnamefont {Noronha}},\ }\href
  {\doibase 10.1103/PhysRevD.93.045013} {\bibfield  {journal} {\bibinfo
  {journal} {Phys. Rev.}\ }\textbf {\bibinfo {volume} {D93}},\ \bibinfo {pages}
  {045013} (\bibinfo {year} {2016}{\natexlab{b}})},\ \Eprint
  {http://arxiv.org/abs/1505.07894} {arXiv:1505.07894 [hep-th]} \BibitemShut
  {NoStop}%
\bibitem [{\citenamefont {Finazzo}\ \emph {et~al.}(2016)\citenamefont
  {Finazzo}, \citenamefont {Critelli}, \citenamefont {Rougemont},\ and\
  \citenamefont {Noronha}}]{Finazzo:2016mhm}%
  \BibitemOpen
  \bibfield  {author} {\bibinfo {author} {\bibfnamefont {S.~I.}\ \bibnamefont
  {Finazzo}}, \bibinfo {author} {\bibfnamefont {R.}~\bibnamefont {Critelli}},
  \bibinfo {author} {\bibfnamefont {R.}~\bibnamefont {Rougemont}}, \ and\
  \bibinfo {author} {\bibfnamefont {J.}~\bibnamefont {Noronha}},\ }\href
  {\doibase 10.1103/PhysRevD.94.054020, 10.1103/PhysRevD.96.019903} {\bibfield
  {journal} {\bibinfo  {journal} {Phys. Rev.}\ }\textbf {\bibinfo {volume}
  {D94}},\ \bibinfo {pages} {054020} (\bibinfo {year} {2016})},\ \bibinfo
  {note} {[Erratum: Phys. Rev.D96,no.1,019903(2017)]},\ \Eprint
  {http://arxiv.org/abs/1605.06061} {arXiv:1605.06061 [hep-ph]} \BibitemShut
  {NoStop}%
\bibitem [{\citenamefont {Critelli}\ \emph {et~al.}(2016)\citenamefont
  {Critelli}, \citenamefont {Rougemont}, \citenamefont {Finazzo},\ and\
  \citenamefont {Noronha}}]{Critelli:2016cvq}%
  \BibitemOpen
  \bibfield  {author} {\bibinfo {author} {\bibfnamefont {R.}~\bibnamefont
  {Critelli}}, \bibinfo {author} {\bibfnamefont {R.}~\bibnamefont {Rougemont}},
  \bibinfo {author} {\bibfnamefont {S.~I.}\ \bibnamefont {Finazzo}}, \ and\
  \bibinfo {author} {\bibfnamefont {J.}~\bibnamefont {Noronha}},\ }\href
  {\doibase 10.1103/PhysRevD.94.125019} {\bibfield  {journal} {\bibinfo
  {journal} {Phys. Rev.}\ }\textbf {\bibinfo {volume} {D94}},\ \bibinfo {pages}
  {125019} (\bibinfo {year} {2016})},\ \Eprint
  {http://arxiv.org/abs/1606.09484} {arXiv:1606.09484 [hep-ph]} \BibitemShut
  {NoStop}%
\bibitem [{\citenamefont {Borsanyi}\ \emph {et~al.}(2012)\citenamefont
  {Borsanyi}, \citenamefont {Fodor}, \citenamefont {Katz}, \citenamefont
  {Krieg}, \citenamefont {Ratti},\ and\ \citenamefont
  {Szabo}}]{Borsanyi:2011sw}%
  \BibitemOpen
  \bibfield  {author} {\bibinfo {author} {\bibfnamefont {S.}~\bibnamefont
  {Borsanyi}}, \bibinfo {author} {\bibfnamefont {Z.}~\bibnamefont {Fodor}},
  \bibinfo {author} {\bibfnamefont {S.~D.}\ \bibnamefont {Katz}}, \bibinfo
  {author} {\bibfnamefont {S.}~\bibnamefont {Krieg}}, \bibinfo {author}
  {\bibfnamefont {C.}~\bibnamefont {Ratti}}, \ and\ \bibinfo {author}
  {\bibfnamefont {K.}~\bibnamefont {Szabo}},\ }\href {\doibase
  10.1007/JHEP01(2012)138} {\bibfield  {journal} {\bibinfo  {journal} {JHEP}\
  }\textbf {\bibinfo {volume} {01}},\ \bibinfo {pages} {138} (\bibinfo {year}
  {2012})},\ \Eprint {http://arxiv.org/abs/1112.4416} {arXiv:1112.4416
  [hep-lat]} \BibitemShut {NoStop}%
\bibitem [{\citenamefont {Borsanyi}\ \emph {et~al.}(2014)\citenamefont
  {Borsanyi}, \citenamefont {Fodor}, \citenamefont {Hoelbling}, \citenamefont
  {Katz}, \citenamefont {Krieg},\ and\ \citenamefont
  {Szabo}}]{Borsanyi:2013bia}%
  \BibitemOpen
  \bibfield  {author} {\bibinfo {author} {\bibfnamefont {S.}~\bibnamefont
  {Borsanyi}}, \bibinfo {author} {\bibfnamefont {Z.}~\bibnamefont {Fodor}},
  \bibinfo {author} {\bibfnamefont {C.}~\bibnamefont {Hoelbling}}, \bibinfo
  {author} {\bibfnamefont {S.~D.}\ \bibnamefont {Katz}}, \bibinfo {author}
  {\bibfnamefont {S.}~\bibnamefont {Krieg}}, \ and\ \bibinfo {author}
  {\bibfnamefont {K.~K.}\ \bibnamefont {Szabo}},\ }\href {\doibase
  10.1016/j.physletb.2014.01.007} {\bibfield  {journal} {\bibinfo  {journal}
  {Phys. Lett.}\ }\textbf {\bibinfo {volume} {B730}},\ \bibinfo {pages} {99}
  (\bibinfo {year} {2014})},\ \Eprint {http://arxiv.org/abs/1309.5258}
  {arXiv:1309.5258 [hep-lat]} \BibitemShut {NoStop}%
\bibitem [{\citenamefont {Borsanyi}\ \emph {et~al.}(2018)\citenamefont
  {Borsanyi}, \citenamefont {Fodor}, \citenamefont {Guenther}, \citenamefont
  {Katz}, \citenamefont {Szabó}, \citenamefont {Pasztor}, \citenamefont
  {Portillo},\ and\ \citenamefont {Ratti}}]{Borsanyi:2018grb}%
  \BibitemOpen
  \bibfield  {author} {\bibinfo {author} {\bibfnamefont {S.}~\bibnamefont
  {Borsanyi}}, \bibinfo {author} {\bibfnamefont {Z.}~\bibnamefont {Fodor}},
  \bibinfo {author} {\bibfnamefont {J.~N.}\ \bibnamefont {Guenther}}, \bibinfo
  {author} {\bibfnamefont {S.~K.}\ \bibnamefont {Katz}}, \bibinfo {author}
  {\bibfnamefont {K.~K.}\ \bibnamefont {Szabó}}, \bibinfo {author}
  {\bibfnamefont {A.}~\bibnamefont {Pasztor}}, \bibinfo {author} {\bibfnamefont
  {I.}~\bibnamefont {Portillo}}, \ and\ \bibinfo {author} {\bibfnamefont
  {C.}~\bibnamefont {Ratti}},\ }\href@noop {} {\  (\bibinfo {year} {2018})},\
  \Eprint {http://arxiv.org/abs/1805.04445} {arXiv:1805.04445 [hep-lat]}
  \BibitemShut {NoStop}%
\bibitem [{\citenamefont {Vovchenko}\ \emph {et~al.}(2018)\citenamefont
  {Vovchenko}, \citenamefont {Steinheimer}, \citenamefont {Philipsen},\ and\
  \citenamefont {Stoecker}}]{Vovchenko:2017gkg}%
  \BibitemOpen
  \bibfield  {author} {\bibinfo {author} {\bibfnamefont {V.}~\bibnamefont
  {Vovchenko}}, \bibinfo {author} {\bibfnamefont {J.}~\bibnamefont
  {Steinheimer}}, \bibinfo {author} {\bibfnamefont {O.}~\bibnamefont
  {Philipsen}}, \ and\ \bibinfo {author} {\bibfnamefont {H.}~\bibnamefont
  {Stoecker}},\ }\href {\doibase 10.1103/PhysRevD.97.114030} {\bibfield
  {journal} {\bibinfo  {journal} {Phys. Rev.}\ }\textbf {\bibinfo {volume}
  {D97}},\ \bibinfo {pages} {114030} (\bibinfo {year} {2018})},\ \Eprint
  {http://arxiv.org/abs/1711.01261} {arXiv:1711.01261 [hep-ph]} \BibitemShut
  {NoStop}%
\bibitem [{\citenamefont {Haque}\ \emph {et~al.}(2014)\citenamefont {Haque},
  \citenamefont {Bandyopadhyay}, \citenamefont {Andersen}, \citenamefont
  {Mustafa}, \citenamefont {Strickland},\ and\ \citenamefont
  {Su}}]{Haque:2014rua}%
  \BibitemOpen
  \bibfield  {author} {\bibinfo {author} {\bibfnamefont {N.}~\bibnamefont
  {Haque}}, \bibinfo {author} {\bibfnamefont {A.}~\bibnamefont
  {Bandyopadhyay}}, \bibinfo {author} {\bibfnamefont {J.~O.}\ \bibnamefont
  {Andersen}}, \bibinfo {author} {\bibfnamefont {M.~G.}\ \bibnamefont
  {Mustafa}}, \bibinfo {author} {\bibfnamefont {M.}~\bibnamefont {Strickland}},
  \ and\ \bibinfo {author} {\bibfnamefont {N.}~\bibnamefont {Su}},\ }\href
  {\doibase 10.1007/JHEP05(2014)027} {\bibfield  {journal} {\bibinfo  {journal}
  {JHEP}\ }\textbf {\bibinfo {volume} {05}},\ \bibinfo {pages} {027} (\bibinfo
  {year} {2014})},\ \Eprint {http://arxiv.org/abs/1402.6907} {arXiv:1402.6907
  [hep-ph]} \BibitemShut {NoStop}%
\bibitem [{\citenamefont {Kokkotas}\ and\ \citenamefont
  {Schmidt}(1999)}]{Kokkotas:1999bd}%
  \BibitemOpen
  \bibfield  {author} {\bibinfo {author} {\bibfnamefont {K.~D.}\ \bibnamefont
  {Kokkotas}}\ and\ \bibinfo {author} {\bibfnamefont {B.~G.}\ \bibnamefont
  {Schmidt}},\ }\href {\doibase 10.12942/lrr-1999-2} {\bibfield  {journal}
  {\bibinfo  {journal} {Living Rev. Rel.}\ }\textbf {\bibinfo {volume} {2}},\
  \bibinfo {pages} {2} (\bibinfo {year} {1999})},\ \Eprint
  {http://arxiv.org/abs/gr-qc/9909058} {arXiv:gr-qc/9909058 [gr-qc]}
  \BibitemShut {NoStop}%
\bibitem [{\citenamefont {Konoplya}\ and\ \citenamefont
  {Zhidenko}(2011)}]{Konoplya:2011qq}%
  \BibitemOpen
  \bibfield  {author} {\bibinfo {author} {\bibfnamefont {R.~A.}\ \bibnamefont
  {Konoplya}}\ and\ \bibinfo {author} {\bibfnamefont {A.}~\bibnamefont
  {Zhidenko}},\ }\href {\doibase 10.1103/RevModPhys.83.793} {\bibfield
  {journal} {\bibinfo  {journal} {Rev. Mod. Phys.}\ }\textbf {\bibinfo {volume}
  {83}},\ \bibinfo {pages} {793} (\bibinfo {year} {2011})},\ \Eprint
  {http://arxiv.org/abs/1102.4014} {arXiv:1102.4014 [gr-qc]} \BibitemShut
  {NoStop}%
\bibitem [{\citenamefont {Buchel}\ \emph {et~al.}(2015)\citenamefont {Buchel},
  \citenamefont {Heller},\ and\ \citenamefont {Myers}}]{Buchel:2015saa}%
  \BibitemOpen
  \bibfield  {author} {\bibinfo {author} {\bibfnamefont {A.}~\bibnamefont
  {Buchel}}, \bibinfo {author} {\bibfnamefont {M.~P.}\ \bibnamefont {Heller}},
  \ and\ \bibinfo {author} {\bibfnamefont {R.~C.}\ \bibnamefont {Myers}},\
  }\href {\doibase 10.1103/PhysRevLett.114.251601} {\bibfield  {journal}
  {\bibinfo  {journal} {Phys. Rev. Lett.}\ }\textbf {\bibinfo {volume} {114}},\
  \bibinfo {pages} {251601} (\bibinfo {year} {2015})},\ \Eprint
  {http://arxiv.org/abs/1503.07114} {arXiv:1503.07114 [hep-th]} \BibitemShut
  {NoStop}%
\bibitem [{\citenamefont {Janik}\ \emph {et~al.}(2015)\citenamefont {Janik},
  \citenamefont {Plewa}, \citenamefont {Soltanpanahi},\ and\ \citenamefont
  {Spalinski}}]{Janik:2015waa}%
  \BibitemOpen
  \bibfield  {author} {\bibinfo {author} {\bibfnamefont {R.~A.}\ \bibnamefont
  {Janik}}, \bibinfo {author} {\bibfnamefont {G.}~\bibnamefont {Plewa}},
  \bibinfo {author} {\bibfnamefont {H.}~\bibnamefont {Soltanpanahi}}, \ and\
  \bibinfo {author} {\bibfnamefont {M.}~\bibnamefont {Spalinski}},\ }\href
  {\doibase 10.1103/PhysRevD.91.126013} {\bibfield  {journal} {\bibinfo
  {journal} {Phys. Rev.}\ }\textbf {\bibinfo {volume} {D91}},\ \bibinfo {pages}
  {126013} (\bibinfo {year} {2015})},\ \Eprint
  {http://arxiv.org/abs/1503.07149} {arXiv:1503.07149 [hep-th]} \BibitemShut
  {NoStop}%
\bibitem [{\citenamefont {Janik}\ \emph
  {et~al.}(2016{\natexlab{a}})\citenamefont {Janik}, \citenamefont
  {Jankowski},\ and\ \citenamefont {Soltanpanahi}}]{Janik:2015iry}%
  \BibitemOpen
  \bibfield  {author} {\bibinfo {author} {\bibfnamefont {R.~A.}\ \bibnamefont
  {Janik}}, \bibinfo {author} {\bibfnamefont {J.}~\bibnamefont {Jankowski}}, \
  and\ \bibinfo {author} {\bibfnamefont {H.}~\bibnamefont {Soltanpanahi}},\
  }\href {\doibase 10.1103/PhysRevLett.117.091603} {\bibfield  {journal}
  {\bibinfo  {journal} {Phys. Rev. Lett.}\ }\textbf {\bibinfo {volume} {117}},\
  \bibinfo {pages} {091603} (\bibinfo {year} {2016}{\natexlab{a}})},\ \Eprint
  {http://arxiv.org/abs/1512.06871} {arXiv:1512.06871 [hep-th]} \BibitemShut
  {NoStop}%
\bibitem [{\citenamefont {Janik}\ \emph
  {et~al.}(2016{\natexlab{b}})\citenamefont {Janik}, \citenamefont
  {Jankowski},\ and\ \citenamefont {Soltanpanahi}}]{Janik:2016btb}%
  \BibitemOpen
  \bibfield  {author} {\bibinfo {author} {\bibfnamefont {R.~A.}\ \bibnamefont
  {Janik}}, \bibinfo {author} {\bibfnamefont {J.}~\bibnamefont {Jankowski}}, \
  and\ \bibinfo {author} {\bibfnamefont {H.}~\bibnamefont {Soltanpanahi}},\
  }\href {\doibase 10.1007/JHEP06(2016)047} {\bibfield  {journal} {\bibinfo
  {journal} {JHEP}\ }\textbf {\bibinfo {volume} {06}},\ \bibinfo {pages} {047}
  (\bibinfo {year} {2016}{\natexlab{b}})},\ \Eprint
  {http://arxiv.org/abs/1603.05950} {arXiv:1603.05950 [hep-th]} \BibitemShut
  {NoStop}%
\bibitem [{\citenamefont {Attems}\ \emph {et~al.}(2016)\citenamefont {Attems},
  \citenamefont {Casalderrey-Solana}, \citenamefont {Mateos}, \citenamefont
  {Papadimitriou}, \citenamefont {Santos-Olivan}, \citenamefont {Sopuerta},
  \citenamefont {Triana},\ and\ \citenamefont {Zilhao}}]{Attems:2016ugt}%
  \BibitemOpen
  \bibfield  {author} {\bibinfo {author} {\bibfnamefont {M.}~\bibnamefont
  {Attems}}, \bibinfo {author} {\bibfnamefont {J.}~\bibnamefont
  {Casalderrey-Solana}}, \bibinfo {author} {\bibfnamefont {D.}~\bibnamefont
  {Mateos}}, \bibinfo {author} {\bibfnamefont {I.}~\bibnamefont
  {Papadimitriou}}, \bibinfo {author} {\bibfnamefont {D.}~\bibnamefont
  {Santos-Olivan}}, \bibinfo {author} {\bibfnamefont {C.~F.}\ \bibnamefont
  {Sopuerta}}, \bibinfo {author} {\bibfnamefont {M.}~\bibnamefont {Triana}}, \
  and\ \bibinfo {author} {\bibfnamefont {M.}~\bibnamefont {Zilhao}},\ }\href
  {\doibase 10.1007/JHEP10(2016)155} {\bibfield  {journal} {\bibinfo  {journal}
  {JHEP}\ }\textbf {\bibinfo {volume} {10}},\ \bibinfo {pages} {155} (\bibinfo
  {year} {2016})},\ \Eprint {http://arxiv.org/abs/1603.01254} {arXiv:1603.01254
  [hep-th]} \BibitemShut {NoStop}%
\bibitem [{\citenamefont {Gursoy}\ \emph {et~al.}(2016)\citenamefont {Gursoy},
  \citenamefont {Jansen},\ and\ \citenamefont {van~der
  Schee}}]{Gursoy:2016ggq}%
  \BibitemOpen
  \bibfield  {author} {\bibinfo {author} {\bibfnamefont {U.}~\bibnamefont
  {Gursoy}}, \bibinfo {author} {\bibfnamefont {A.}~\bibnamefont {Jansen}}, \
  and\ \bibinfo {author} {\bibfnamefont {W.}~\bibnamefont {van~der Schee}},\
  }\href {\doibase 10.1103/PhysRevD.94.061901} {\bibfield  {journal} {\bibinfo
  {journal} {Phys. Rev.}\ }\textbf {\bibinfo {volume} {D94}},\ \bibinfo {pages}
  {061901} (\bibinfo {year} {2016})},\ \Eprint
  {http://arxiv.org/abs/1603.07724} {arXiv:1603.07724 [hep-th]} \BibitemShut
  {NoStop}%
\bibitem [{\citenamefont {Demircik}\ and\ \citenamefont
  {Gursoy}(2017)}]{Demircik:2016nhr}%
  \BibitemOpen
  \bibfield  {author} {\bibinfo {author} {\bibfnamefont {T.}~\bibnamefont
  {Demircik}}\ and\ \bibinfo {author} {\bibfnamefont {U.}~\bibnamefont
  {Gursoy}},\ }\href {\doibase 10.1016/j.nuclphysb.2017.03.020} {\bibfield
  {journal} {\bibinfo  {journal} {Nucl. Phys.}\ }\textbf {\bibinfo {volume}
  {B919}},\ \bibinfo {pages} {384} (\bibinfo {year} {2017})},\ \Eprint
  {http://arxiv.org/abs/1605.08118} {arXiv:1605.08118 [hep-th]} \BibitemShut
  {NoStop}%
\bibitem [{\citenamefont {Betzios}\ \emph {et~al.}(2017)\citenamefont
  {Betzios}, \citenamefont {Gursoy}, \citenamefont {Jarvinen},\ and\
  \citenamefont {Policastro}}]{Betzios:2017dol}%
  \BibitemOpen
  \bibfield  {author} {\bibinfo {author} {\bibfnamefont {P.}~\bibnamefont
  {Betzios}}, \bibinfo {author} {\bibfnamefont {U.}~\bibnamefont {Gursoy}},
  \bibinfo {author} {\bibfnamefont {M.}~\bibnamefont {Jarvinen}}, \ and\
  \bibinfo {author} {\bibfnamefont {G.}~\bibnamefont {Policastro}},\
  }\href@noop {} {\  (\bibinfo {year} {2017})},\ \Eprint
  {http://arxiv.org/abs/1708.02252} {arXiv:1708.02252 [hep-th]} \BibitemShut
  {NoStop}%
\bibitem [{\citenamefont {Baier}\ \emph {et~al.}(2008)\citenamefont {Baier},
  \citenamefont {Romatschke}, \citenamefont {Son}, \citenamefont {Starinets},\
  and\ \citenamefont {Stephanov}}]{Baier:2007ix}%
  \BibitemOpen
  \bibfield  {author} {\bibinfo {author} {\bibfnamefont {R.}~\bibnamefont
  {Baier}}, \bibinfo {author} {\bibfnamefont {P.}~\bibnamefont {Romatschke}},
  \bibinfo {author} {\bibfnamefont {D.~T.}\ \bibnamefont {Son}}, \bibinfo
  {author} {\bibfnamefont {A.~O.}\ \bibnamefont {Starinets}}, \ and\ \bibinfo
  {author} {\bibfnamefont {M.~A.}\ \bibnamefont {Stephanov}},\ }\href {\doibase
  10.1088/1126-6708/2008/04/100} {\bibfield  {journal} {\bibinfo  {journal}
  {JHEP}\ }\textbf {\bibinfo {volume} {04}},\ \bibinfo {pages} {100} (\bibinfo
  {year} {2008})},\ \Eprint {http://arxiv.org/abs/0712.2451} {arXiv:0712.2451
  [hep-th]} \BibitemShut {NoStop}%
\bibitem [{\citenamefont {Son}\ and\ \citenamefont
  {Starinets}(2002)}]{Son:2002sd}%
  \BibitemOpen
  \bibfield  {author} {\bibinfo {author} {\bibfnamefont {D.~T.}\ \bibnamefont
  {Son}}\ and\ \bibinfo {author} {\bibfnamefont {A.~O.}\ \bibnamefont
  {Starinets}},\ }\href {\doibase 10.1088/1126-6708/2002/09/042} {\bibfield
  {journal} {\bibinfo  {journal} {JHEP}\ }\textbf {\bibinfo {volume} {09}},\
  \bibinfo {pages} {042} (\bibinfo {year} {2002})},\ \Eprint
  {http://arxiv.org/abs/hep-th/0205051} {arXiv:hep-th/0205051 [hep-th]}
  \BibitemShut {NoStop}%
\bibitem [{\citenamefont {Policastro}\ \emph
  {et~al.}(2002{\natexlab{a}})\citenamefont {Policastro}, \citenamefont {Son},\
  and\ \citenamefont {Starinets}}]{Policastro:2002se}%
  \BibitemOpen
  \bibfield  {author} {\bibinfo {author} {\bibfnamefont {G.}~\bibnamefont
  {Policastro}}, \bibinfo {author} {\bibfnamefont {D.~T.}\ \bibnamefont {Son}},
  \ and\ \bibinfo {author} {\bibfnamefont {A.~O.}\ \bibnamefont {Starinets}},\
  }\href {\doibase 10.1088/1126-6708/2002/09/043} {\bibfield  {journal}
  {\bibinfo  {journal} {JHEP}\ }\textbf {\bibinfo {volume} {09}},\ \bibinfo
  {pages} {043} (\bibinfo {year} {2002}{\natexlab{a}})},\ \Eprint
  {http://arxiv.org/abs/hep-th/0205052} {arXiv:hep-th/0205052 [hep-th]}
  \BibitemShut {NoStop}%
\bibitem [{\citenamefont {Policastro}\ \emph
  {et~al.}(2002{\natexlab{b}})\citenamefont {Policastro}, \citenamefont {Son},\
  and\ \citenamefont {Starinets}}]{Policastro:2002tn}%
  \BibitemOpen
  \bibfield  {author} {\bibinfo {author} {\bibfnamefont {G.}~\bibnamefont
  {Policastro}}, \bibinfo {author} {\bibfnamefont {D.~T.}\ \bibnamefont {Son}},
  \ and\ \bibinfo {author} {\bibfnamefont {A.~O.}\ \bibnamefont {Starinets}},\
  }\href {\doibase 10.1088/1126-6708/2002/12/054} {\bibfield  {journal}
  {\bibinfo  {journal} {JHEP}\ }\textbf {\bibinfo {volume} {12}},\ \bibinfo
  {pages} {054} (\bibinfo {year} {2002}{\natexlab{b}})},\ \Eprint
  {http://arxiv.org/abs/hep-th/0210220} {arXiv:hep-th/0210220 [hep-th]}
  \BibitemShut {NoStop}%
\bibitem [{\citenamefont {Heller}\ \emph {et~al.}(2013)\citenamefont {Heller},
  \citenamefont {Janik},\ and\ \citenamefont {Witaszczyk}}]{Heller:2013fn}%
  \BibitemOpen
  \bibfield  {author} {\bibinfo {author} {\bibfnamefont {M.~P.}\ \bibnamefont
  {Heller}}, \bibinfo {author} {\bibfnamefont {R.~A.}\ \bibnamefont {Janik}}, \
  and\ \bibinfo {author} {\bibfnamefont {P.}~\bibnamefont {Witaszczyk}},\
  }\href {\doibase 10.1103/PhysRevLett.110.211602} {\bibfield  {journal}
  {\bibinfo  {journal} {Phys. Rev. Lett.}\ }\textbf {\bibinfo {volume} {110}},\
  \bibinfo {pages} {211602} (\bibinfo {year} {2013})},\ \Eprint
  {http://arxiv.org/abs/1302.0697} {arXiv:1302.0697 [hep-th]} \BibitemShut
  {NoStop}%
\bibitem [{\citenamefont {Finazzo}\ \emph {et~al.}(2017)\citenamefont
  {Finazzo}, \citenamefont {Rougemont}, \citenamefont {Zaniboni}, \citenamefont
  {Critelli},\ and\ \citenamefont {Noronha}}]{Finazzo:2016psx}%
  \BibitemOpen
  \bibfield  {author} {\bibinfo {author} {\bibfnamefont {S.~I.}\ \bibnamefont
  {Finazzo}}, \bibinfo {author} {\bibfnamefont {R.}~\bibnamefont {Rougemont}},
  \bibinfo {author} {\bibfnamefont {M.}~\bibnamefont {Zaniboni}}, \bibinfo
  {author} {\bibfnamefont {R.}~\bibnamefont {Critelli}}, \ and\ \bibinfo
  {author} {\bibfnamefont {J.}~\bibnamefont {Noronha}},\ }\href {\doibase
  10.1007/JHEP01(2017)137} {\bibfield  {journal} {\bibinfo  {journal} {JHEP}\
  }\textbf {\bibinfo {volume} {01}},\ \bibinfo {pages} {137} (\bibinfo {year}
  {2017})},\ \Eprint {http://arxiv.org/abs/1610.01519} {arXiv:1610.01519
  [hep-th]} \BibitemShut {NoStop}%
\bibitem [{\citenamefont {Attems}\ \emph
  {et~al.}(2017{\natexlab{a}})\citenamefont {Attems}, \citenamefont
  {Casalderrey-Solana}, \citenamefont {Mateos}, \citenamefont {Santos-Oliván},
  \citenamefont {Sopuerta}, \citenamefont {Triana},\ and\ \citenamefont
  {Zilhão}}]{Attems:2017zam}%
  \BibitemOpen
  \bibfield  {author} {\bibinfo {author} {\bibfnamefont {M.}~\bibnamefont
  {Attems}}, \bibinfo {author} {\bibfnamefont {J.}~\bibnamefont
  {Casalderrey-Solana}}, \bibinfo {author} {\bibfnamefont {D.}~\bibnamefont
  {Mateos}}, \bibinfo {author} {\bibfnamefont {D.}~\bibnamefont
  {Santos-Oliván}}, \bibinfo {author} {\bibfnamefont {C.~F.}\ \bibnamefont
  {Sopuerta}}, \bibinfo {author} {\bibfnamefont {M.}~\bibnamefont {Triana}}, \
  and\ \bibinfo {author} {\bibfnamefont {M.}~\bibnamefont {Zilhão}},\ }\href
  {\doibase 10.1007/JHEP06(2017)154} {\bibfield  {journal} {\bibinfo  {journal}
  {JHEP}\ }\textbf {\bibinfo {volume} {06}},\ \bibinfo {pages} {154} (\bibinfo
  {year} {2017}{\natexlab{a}})},\ \Eprint {http://arxiv.org/abs/1703.09681}
  {arXiv:1703.09681 [hep-th]} \BibitemShut {NoStop}%
\bibitem [{\citenamefont {Stephanov}\ and\ \citenamefont
  {Yin}(2017)}]{Stephanov:2017ghc}%
  \BibitemOpen
  \bibfield  {author} {\bibinfo {author} {\bibfnamefont {M.}~\bibnamefont
  {Stephanov}}\ and\ \bibinfo {author} {\bibfnamefont {Y.}~\bibnamefont
  {Yin}},\ }\href@noop {} {\  (\bibinfo {year} {2017})},\ \Eprint
  {http://arxiv.org/abs/1712.10305} {arXiv:1712.10305 [nucl-th]} \BibitemShut
  {NoStop}%
\bibitem [{\citenamefont {Critelli}\ \emph {et~al.}(2018)\citenamefont
  {Critelli}, \citenamefont {Rougemont},\ and\ \citenamefont
  {Noronha}}]{Critelli:2018osu}%
  \BibitemOpen
  \bibfield  {author} {\bibinfo {author} {\bibfnamefont {R.}~\bibnamefont
  {Critelli}}, \bibinfo {author} {\bibfnamefont {R.}~\bibnamefont {Rougemont}},
  \ and\ \bibinfo {author} {\bibfnamefont {J.}~\bibnamefont {Noronha}},\
  }\href@noop {} {\  (\bibinfo {year} {2018})},\ \Eprint
  {http://arxiv.org/abs/1805.00882} {arXiv:1805.00882 [hep-th]} \BibitemShut
  {NoStop}%
\bibitem [{\citenamefont {Bjorken}(1983)}]{Bjorken:1982qr}%
  \BibitemOpen
  \bibfield  {author} {\bibinfo {author} {\bibfnamefont {J.~D.}\ \bibnamefont
  {Bjorken}},\ }\href {\doibase 10.1103/PhysRevD.27.140} {\bibfield  {journal}
  {\bibinfo  {journal} {Phys. Rev.}\ }\textbf {\bibinfo {volume} {D27}},\
  \bibinfo {pages} {140} (\bibinfo {year} {1983})}\BibitemShut {NoStop}%
\bibitem [{\citenamefont {Chesler}\ and\ \citenamefont
  {Yaffe}(2010)}]{Chesler:2009cy}%
  \BibitemOpen
  \bibfield  {author} {\bibinfo {author} {\bibfnamefont {P.~M.}\ \bibnamefont
  {Chesler}}\ and\ \bibinfo {author} {\bibfnamefont {L.~G.}\ \bibnamefont
  {Yaffe}},\ }\href {\doibase 10.1103/PhysRevD.82.026006} {\bibfield  {journal}
  {\bibinfo  {journal} {Phys. Rev.}\ }\textbf {\bibinfo {volume} {D82}},\
  \bibinfo {pages} {026006} (\bibinfo {year} {2010})},\ \Eprint
  {http://arxiv.org/abs/0906.4426} {arXiv:0906.4426 [hep-th]} \BibitemShut
  {NoStop}%
\bibitem [{\citenamefont {Heller}\ \emph {et~al.}(2012)\citenamefont {Heller},
  \citenamefont {Janik},\ and\ \citenamefont {Witaszczyk}}]{Heller:2011ju}%
  \BibitemOpen
  \bibfield  {author} {\bibinfo {author} {\bibfnamefont {M.~P.}\ \bibnamefont
  {Heller}}, \bibinfo {author} {\bibfnamefont {R.~A.}\ \bibnamefont {Janik}}, \
  and\ \bibinfo {author} {\bibfnamefont {P.}~\bibnamefont {Witaszczyk}},\
  }\href {\doibase 10.1103/PhysRevLett.108.201602} {\bibfield  {journal}
  {\bibinfo  {journal} {Phys. Rev. Lett.}\ }\textbf {\bibinfo {volume} {108}},\
  \bibinfo {pages} {201602} (\bibinfo {year} {2012})},\ \Eprint
  {http://arxiv.org/abs/1103.3452} {arXiv:1103.3452 [hep-th]} \BibitemShut
  {NoStop}%
\bibitem [{\citenamefont {Jankowski}\ \emph {et~al.}(2014)\citenamefont
  {Jankowski}, \citenamefont {Plewa},\ and\ \citenamefont
  {Spalinski}}]{Jankowski:2014lna}%
  \BibitemOpen
  \bibfield  {author} {\bibinfo {author} {\bibfnamefont {J.}~\bibnamefont
  {Jankowski}}, \bibinfo {author} {\bibfnamefont {G.}~\bibnamefont {Plewa}}, \
  and\ \bibinfo {author} {\bibfnamefont {M.}~\bibnamefont {Spalinski}},\ }\href
  {\doibase 10.1007/JHEP12(2014)105} {\bibfield  {journal} {\bibinfo  {journal}
  {JHEP}\ }\textbf {\bibinfo {volume} {12}},\ \bibinfo {pages} {105} (\bibinfo
  {year} {2014})},\ \Eprint {http://arxiv.org/abs/1411.1969} {arXiv:1411.1969
  [hep-th]} \BibitemShut {NoStop}%
\bibitem [{\citenamefont {Romatschke}(2018)}]{Romatschke:2017vte}%
  \BibitemOpen
  \bibfield  {author} {\bibinfo {author} {\bibfnamefont {P.}~\bibnamefont
  {Romatschke}},\ }\href {\doibase 10.1103/PhysRevLett.120.012301} {\bibfield
  {journal} {\bibinfo  {journal} {Phys. Rev. Lett.}\ }\textbf {\bibinfo
  {volume} {120}},\ \bibinfo {pages} {012301} (\bibinfo {year} {2018})},\
  \Eprint {http://arxiv.org/abs/1704.08699} {arXiv:1704.08699 [hep-th]}
  \BibitemShut {NoStop}%
\bibitem [{\citenamefont {Spalinski}(2018)}]{Spalinski:2017mel}%
  \BibitemOpen
  \bibfield  {author} {\bibinfo {author} {\bibfnamefont {M.}~\bibnamefont
  {Spalinski}},\ }\href {\doibase 10.1016/j.physletb.2017.11.059} {\bibfield
  {journal} {\bibinfo  {journal} {Phys. Lett.}\ }\textbf {\bibinfo {volume}
  {B776}},\ \bibinfo {pages} {468} (\bibinfo {year} {2018})},\ \Eprint
  {http://arxiv.org/abs/1708.01921} {arXiv:1708.01921 [hep-th]} \BibitemShut
  {NoStop}%
\bibitem [{\citenamefont {Chesler}\ and\ \citenamefont
  {Yaffe}(2011)}]{Chesler:2010bi}%
  \BibitemOpen
  \bibfield  {author} {\bibinfo {author} {\bibfnamefont {P.~M.}\ \bibnamefont
  {Chesler}}\ and\ \bibinfo {author} {\bibfnamefont {L.~G.}\ \bibnamefont
  {Yaffe}},\ }\href {\doibase 10.1103/PhysRevLett.106.021601} {\bibfield
  {journal} {\bibinfo  {journal} {Phys. Rev. Lett.}\ }\textbf {\bibinfo
  {volume} {106}},\ \bibinfo {pages} {021601} (\bibinfo {year} {2011})},\
  \Eprint {http://arxiv.org/abs/1011.3562} {arXiv:1011.3562 [hep-th]}
  \BibitemShut {NoStop}%
\bibitem [{\citenamefont {Casalderrey-Solana}\ \emph
  {et~al.}(2013)\citenamefont {Casalderrey-Solana}, \citenamefont {Heller},
  \citenamefont {Mateos},\ and\ \citenamefont {van~der
  Schee}}]{Casalderrey-Solana:2013aba}%
  \BibitemOpen
  \bibfield  {author} {\bibinfo {author} {\bibfnamefont {J.}~\bibnamefont
  {Casalderrey-Solana}}, \bibinfo {author} {\bibfnamefont {M.~P.}\ \bibnamefont
  {Heller}}, \bibinfo {author} {\bibfnamefont {D.}~\bibnamefont {Mateos}}, \
  and\ \bibinfo {author} {\bibfnamefont {W.}~\bibnamefont {van~der Schee}},\
  }\href {\doibase 10.1103/PhysRevLett.111.181601} {\bibfield  {journal}
  {\bibinfo  {journal} {Phys. Rev. Lett.}\ }\textbf {\bibinfo {volume} {111}},\
  \bibinfo {pages} {181601} (\bibinfo {year} {2013})},\ \Eprint
  {http://arxiv.org/abs/1305.4919} {arXiv:1305.4919 [hep-th]} \BibitemShut
  {NoStop}%
\bibitem [{\citenamefont {van~der Schee}\ \emph {et~al.}(2013)\citenamefont
  {van~der Schee}, \citenamefont {Romatschke},\ and\ \citenamefont
  {Pratt}}]{vanderSchee:2013pia}%
  \BibitemOpen
  \bibfield  {author} {\bibinfo {author} {\bibfnamefont {W.}~\bibnamefont
  {van~der Schee}}, \bibinfo {author} {\bibfnamefont {P.}~\bibnamefont
  {Romatschke}}, \ and\ \bibinfo {author} {\bibfnamefont {S.}~\bibnamefont
  {Pratt}},\ }\href {\doibase 10.1103/PhysRevLett.111.222302} {\bibfield
  {journal} {\bibinfo  {journal} {Phys. Rev. Lett.}\ }\textbf {\bibinfo
  {volume} {111}},\ \bibinfo {pages} {222302} (\bibinfo {year} {2013})},\
  \Eprint {http://arxiv.org/abs/1307.2539} {arXiv:1307.2539 [nucl-th]}
  \BibitemShut {NoStop}%
\bibitem [{\citenamefont {Chesler}(2015)}]{Chesler:2015bba}%
  \BibitemOpen
  \bibfield  {author} {\bibinfo {author} {\bibfnamefont {P.~M.}\ \bibnamefont
  {Chesler}},\ }\href {\doibase 10.1103/PhysRevLett.115.241602} {\bibfield
  {journal} {\bibinfo  {journal} {Phys. Rev. Lett.}\ }\textbf {\bibinfo
  {volume} {115}},\ \bibinfo {pages} {241602} (\bibinfo {year} {2015})},\
  \Eprint {http://arxiv.org/abs/1506.02209} {arXiv:1506.02209 [hep-th]}
  \BibitemShut {NoStop}%
\bibitem [{\citenamefont {Chesler}(2016)}]{Chesler:2016ceu}%
  \BibitemOpen
  \bibfield  {author} {\bibinfo {author} {\bibfnamefont {P.~M.}\ \bibnamefont
  {Chesler}},\ }\href {\doibase 10.1007/JHEP03(2016)146} {\bibfield  {journal}
  {\bibinfo  {journal} {JHEP}\ }\textbf {\bibinfo {volume} {03}},\ \bibinfo
  {pages} {146} (\bibinfo {year} {2016})},\ \Eprint
  {http://arxiv.org/abs/1601.01583} {arXiv:1601.01583 [hep-th]} \BibitemShut
  {NoStop}%
\bibitem [{\citenamefont {Attems}\ \emph
  {et~al.}(2017{\natexlab{b}})\citenamefont {Attems}, \citenamefont
  {Casalderrey-Solana}, \citenamefont {Mateos}, \citenamefont {Santos-Olivan},
  \citenamefont {Sopuerta}, \citenamefont {Triana},\ and\ \citenamefont
  {Zilhao}}]{Attems:2016tby}%
  \BibitemOpen
  \bibfield  {author} {\bibinfo {author} {\bibfnamefont {M.}~\bibnamefont
  {Attems}}, \bibinfo {author} {\bibfnamefont {J.}~\bibnamefont
  {Casalderrey-Solana}}, \bibinfo {author} {\bibfnamefont {D.}~\bibnamefont
  {Mateos}}, \bibinfo {author} {\bibfnamefont {D.}~\bibnamefont
  {Santos-Olivan}}, \bibinfo {author} {\bibfnamefont {C.~F.}\ \bibnamefont
  {Sopuerta}}, \bibinfo {author} {\bibfnamefont {M.}~\bibnamefont {Triana}}, \
  and\ \bibinfo {author} {\bibfnamefont {M.}~\bibnamefont {Zilhao}},\ }\href
  {\doibase 10.1007/JHEP01(2017)026} {\bibfield  {journal} {\bibinfo  {journal}
  {JHEP}\ }\textbf {\bibinfo {volume} {01}},\ \bibinfo {pages} {026} (\bibinfo
  {year} {2017}{\natexlab{b}})},\ \Eprint {http://arxiv.org/abs/1604.06439}
  {arXiv:1604.06439 [hep-th]} \BibitemShut {NoStop}%
\bibitem [{\citenamefont {Casalderrey-Solana}\ \emph
  {et~al.}(2016)\citenamefont {Casalderrey-Solana}, \citenamefont {Mateos},
  \citenamefont {van~der Schee},\ and\ \citenamefont
  {Triana}}]{Casalderrey-Solana:2016xfq}%
  \BibitemOpen
  \bibfield  {author} {\bibinfo {author} {\bibfnamefont {J.}~\bibnamefont
  {Casalderrey-Solana}}, \bibinfo {author} {\bibfnamefont {D.}~\bibnamefont
  {Mateos}}, \bibinfo {author} {\bibfnamefont {W.}~\bibnamefont {van~der
  Schee}}, \ and\ \bibinfo {author} {\bibfnamefont {M.}~\bibnamefont
  {Triana}},\ }\href {\doibase 10.1007/JHEP09(2016)108} {\bibfield  {journal}
  {\bibinfo  {journal} {JHEP}\ }\textbf {\bibinfo {volume} {09}},\ \bibinfo
  {pages} {108} (\bibinfo {year} {2016})},\ \Eprint
  {http://arxiv.org/abs/1607.05273} {arXiv:1607.05273 [hep-th]} \BibitemShut
  {NoStop}%
\bibitem [{\citenamefont {Skokov}\ \emph {et~al.}(2009)\citenamefont {Skokov},
  \citenamefont {Illarionov},\ and\ \citenamefont {Toneev}}]{Skokov:2009qp}%
  \BibitemOpen
  \bibfield  {author} {\bibinfo {author} {\bibfnamefont {V.}~\bibnamefont
  {Skokov}}, \bibinfo {author} {\bibfnamefont {A.~{\relax Yu}.}\ \bibnamefont
  {Illarionov}}, \ and\ \bibinfo {author} {\bibfnamefont {V.}~\bibnamefont
  {Toneev}},\ }\href {\doibase 10.1142/S0217751X09047570} {\bibfield  {journal}
  {\bibinfo  {journal} {Int. J. Mod. Phys.}\ }\textbf {\bibinfo {volume}
  {A24}},\ \bibinfo {pages} {5925} (\bibinfo {year} {2009})},\ \Eprint
  {http://arxiv.org/abs/0907.1396} {arXiv:0907.1396 [nucl-th]} \BibitemShut
  {NoStop}%
\end{thebibliography}%

\end{document}